\documentclass[12pt]{article}

\usepackage[dvips]{graphicx}
\usepackage{epsfig}
\usepackage{amsmath,amsfonts,amssymb,amsthm}
\usepackage{mathrsfs,mathtools}
\usepackage{verbatim}
\usepackage{psfrag}
\usepackage{bm}
\usepackage{bbm}
\usepackage[utf8]{inputenc}
\usepackage[square,comma,sort&compress,numbers]{natbib}
\usepackage[dvipsnames]{xcolor}
\usepackage{slashed}
\usepackage{upgreek}
\usepackage{enumitem}
\usepackage{float}
\usepackage[T1]{fontenc}
\usepackage{soul}
\usepackage{pgfplots}
\usepackage{extarrows}
\pgfplotsset{compat=1.16}
\usetikzlibrary{external}
\usepackage{cases}

\usepackage[T1]{fontenc}

\tikzset{
  external/only named=true,
  thick/.style={line width=.5pt},
  approximation/.style={line width=1.2pt},
  numerics/.style={black, dotted, line width=.8pt},
  amplitude/.style={dashed},
  estimate/.style={dashed, line width=.8pt},
  normal plot/.style={line width=.8pt},
}

\usepackage[colorlinks=true,urlcolor=blue,anchorcolor=blue,citecolor=blue,filecolor=blue
,linkcolor=blue,menucolor=blue,linktocpage=true,pdfproducer=medialab,pdfa=true
]{hyperref}

\usepackage{epsf,epsfig}
\usepackage{graphics}
\usepackage{subcaption}

\newcommand\blfootnote[1]{%
  \begingroup
  \renewcommand\thefootnote{}\footnote{#1}%
  \addtocounter{footnote}{-1}%
  \endgroup
}

\def\d{\mathrm{d}}

\def\C{\mathcal{C}}
\def\O{\mathcal{O}}
\let\vec\mathbf
\def\i{\mathrm{i}}
\def\e{\mathrm{e}}

\def\G{\mathcal{G}}

\def\sign{\rm sign}

\def\veck{\vec{k}}
\def\vecp{\vec{p}}

\usepackage[errorstop]{feynmp}
\newcommand{\lsim}
{\;\raisebox{-.3em}{$\stackrel{\displaystyle <}{\sim}$}\;}
\newcommand{\gsim}
{\;\raisebox{-.3em}{$\stackrel{\displaystyle >}{\sim}$}\;}

\newcommand{\rom}[1]{\uppercase\expandafter{\romannumeral #1\relax}}

\begin{document}                                                                                                 
\thispagestyle{empty}

\begin{flushright}
{
\small
KCL-PH-TH/2023-41\\
}
\end{flushright}

\vspace{-0.5cm}

\begin{center}
\Large\bf\boldmath
Fate of oscillating homogeneous $\mathbb{Z}_2$-symmetric scalar condensates in the early Universe
\unboldmath
\end{center}

\vspace{-0.2cm}

\begin{center}
Wen-Yuan Ai$^{*1}$\blfootnote{$^*$~wenyuan.ai@kcl.ac.uk} and Zi-Liang Wang$^{\dagger 2}$\blfootnote{$^\dagger$~ziliang.wang@just.edu.cn} \\
\vskip0.4cm

{\it $^1$Theoretical Particle Physics and Cosmology, King’s College London,\\ Strand, London WC2R 2LS, UK}\par
{\it $^2$Department of Physics, School of Science,\\
Jiangsu University of Science and Technology,
Zhenjiang, 212003, China} 

\vskip1.cm
\end{center}

\begin{abstract}
Dark matter, if represented by a $\mathbb{Z}_2$-symmetric scalar field, can manifest as both particles and condensates. In this paper, we study the evolution of an oscillating homogeneous condensate of a $\mathbb{Z}_2$-symmetric scalar field in a thermal plasma in an FLRW universe. We focus on the perturbative regime where the oscillation amplitude is sufficiently small so that parametric resonance is inefficient. This perturbative regime necessarily comprises the late stage of the condensate decay and determines its fate. The coupled coarse-grained equations of motion for the condensate, radiation, and spacetime are derived from first principles using nonequilibrium quantum field theory. We obtain analytical expressions for the relevant microscopic quantities that enter the equations of motion and solve the latter numerically. We find that there is always a nonvanishing relic abundance for a condensate with a $\mathbb{Z}_2$ symmetry that is not spontaneously broken. This is because its decay rate decreases faster than the Hubble parameter at late times due to either the amplitude dependence or the temperature dependence in the condensate decay rate. Consequently, accounting for the condensate contribution to the overall dark matter relic density is essential for $\mathbb{Z}_2$ scalar singlet dark matter.    
\end{abstract}

\newpage

\hrule
\tableofcontents
\vskip.85cm
\hrule


\section{Introduction}

Scalar fields appear in many
theories beyond the Standard Model (SM) of particle physics. For instance, in theories with extra dimensions, there are numerous moduli scalar fields~\cite{Douglas:2006es,Denef:2007pq}. The inflationary paradigm, which describes the very early stage of the Universe, commonly assumes the inflaton to be a scalar field \cite{Starobinsky:1980te,Guth:1980zm,Linde:1981mu,Albrecht:1982wi,Martin:2013tda}.  Scalar fields can also explain the strong CP problem~\cite{Peccei:2006as,Kim:2008hd}\footnote{See Refs.~\cite{Ai:2020ptm,Ai:2022htq} for a different perspective on the strong CP problem.}, be dark matter (DM)~\cite{Silveira:1985rk,McDonald:1993ex,Burgess:2000yq,Bento:2000ah,Cline:2013gha,Marsh:2015xka} and dark energy~\cite{Wetterich:1987fm,Armendariz-Picon:2000nqq,Copeland:2006wr}. 

Scalar fields with $\mathbb{Z}_2$ symmetry are natural DM candidates~\cite{Silveira:1985rk,McDonald:1993ex,Burgess:2000yq}. In this scenario, it is typically assumed that DM exists in the form of particles. However, a scalar field may also form condensate in the early Universe. For example, the scalar field may be pushed away from its equilibrium value during inflation due to the dramatic universe expansion~\cite{Linde:1982uu,Affleck:1984fy,Starobinsky:1994bd}. After the inflation and when the Hubble parameter is smaller than its mass, it starts oscillating coherently, forming an oscillating condensate. It is known that the dynamics of condensates is very different from that of particles. The $\mathbb{Z}_2$ symmetry cannot prohibit the coherently oscillating condensate from decaying. 

In this paper, we study the evolution of a $\mathbb{Z}_2$-symmetric scalar condensate in the early Universe, taking into account fully the interactions between the condensate, plasma, and spacetime. We focus on the late stage where the oscillation amplitude is sufficiently small such that nonperturbative processes are inefficient. This stage determines the fate of the condensate. Our analysis is based on the Closed-Time-Path (CTP) formalism~\cite{Schwinger:1960qe,Keldysh:1964ud}.  Earlier studies on the dissipation of scalar backgrounds using the same formalism can be found in Refs.~\cite{Calzetta:1989vs,Paz:1990sd,Boyanovsky:1994me,Greiner:1996dx,Yokoyama:2004pf,Bastero-Gil:2010dgy,Bastero-Gil:2012akf,Mukaida:2012qn}. Different from these references, we adopt the multiple-scale analysis~\cite{Bender,Holmes} to derive the coupled Markovian equations of motion from first principles, closely following the recent work~\cite{Ai:2021gtg}. We present an explicit computation for the relevant quantities that enter the coupled equations of motion (EoMs), allowing us to solve the latter numerically. 

From the obtained equations, the decay rate of the oscillating condensate is shown to be both temperature-dependent and amplitude-dependent. This kind of time-dependent dissipation rate is usually considered only phenomenologically~\cite{Drewes:2014pfa,Co:2020xaf,Ahmed:2021fvt,Barman:2022tzk,Banerjee:2022fiw,Chowdhury:2023jft}. 
In this work, we go beyond the high-temperature approximation for the condensate decay rate. This is crucial because it leads to a conclusion different from that given in Refs.~\cite{Mukaida:2013xxa,Wang:2022mvv}. If one uses the high-temperature approximation, it is possible to have a decay rate that is always larger than the Hubble parameter, and one may conclude that the condensate can transfer all its energy to radiation for some values of the coupling and mass parameters. However, going beyond the high-temperature approximation we find that the decay rate always decreases faster than the Hubble parameter at low temperatures and thus the condensate must ``freeze out'' at some point, leading to a nonvanishing relic density provided the $\mathbb{Z}_2$ symmetry is not spontaneously broken.

The coupled EoMs derived in this work could be used for investigating perturbative production of DM from an oscillating scalar field~\cite{Garcia:2020eof} and perturbative reheating~\cite{Chung:1998rq,Giudice:2000ex,Garcia:2020wiy,Aoki:2022dzd}. However, it is important to note that our derivation assumes all particles in the plasma are in thermal equilibrium. In scenarios such as freeze-in~\cite{Hall:2009bx,Bernal:2017kxu}, where certain particle species remain out of equilibrium, it becomes necessary to track the evolution of their particle distribution functions. A derivation of the coupled EoMs from nonequilibrium quantum field theory in such cases is provided in Ref.~\cite{Ai:2023qnr}.

The paper is organised as follows. In Sec.~\ref{sec:model} we introduce the model and the non-local condensate EoM derived from the CTP formalism. The non-local terms are due to self-energy corrections. In order to describe the condensate's evolution in the early Universe, it is sufficient to derive the EoM for its envelope function. In Sec.~\ref{sec:coarse-grained EoMs} we therefore derive the coupled coarse-grained EoMs for the condensate, radiation, and spacetime. This requires the Markovianization of the original non-local condensate EoM. In Sec.~\ref{sec:freeze-out} we analyse the freeze-out behaviour of the condensate analytically. The obtained coupled coarse-grained equations are then solved numerically in Sec.~\ref{sec:numerics}. Sec.~\ref{sec:Conc} is left for conclusion and discussion. We collect several technical details in appendices. Appendix~\ref{app:CTP} is a brief introduction to the CTP formalism. In Appendix~\ref{app:self-energy_and_four-vertex} we present an explicit computation of the self-energies. In Appendix~\ref{app:sol_eom}, we solve the non-local condensate EoM with a time-dependent mass term using multiple-scale analysis.

\section{Model and the condensate equation of motion}

\label{sec:model}

\subsection{Model}
We consider the following model 
\begin{align}
\label{eq:model}
S[\Phi, \chi]=\int \d^{4}x \,\sqrt{-g} \biggl[
	\frac{1}{2} (\partial_\mu \Phi) (\partial^\mu \Phi)
	+ \frac{1}{2} (\partial_\mu \chi) (\partial^\mu \chi)
	-V(\Phi,\chi)
	\biggr] \, ,
\end{align}
with
\begin{equation}
\label{eq:pot}
V(\Phi,\chi)= \frac{m_\phi^2}{2} \Phi^2
	+ \frac{m_\chi^2}{2} \chi^2
	+ \frac{\lambda_\phi}{4!} \Phi^4
	+ \frac{\lambda_\chi}{4!} \chi^4
	+ \frac{g}{4} \Phi^2 \chi^2\,.
\end{equation}
Here $\Phi$ and $\chi$ are two real scalars. We assume that $\Phi$ forms condensate, possessing a nonvanishing expectation value $\langle \Phi\rangle\equiv \varphi$. The other scalar field $\chi$ may interact with the SM fields. For example, $\chi$ could be the Higgs but in this case, one needs to be careful with the electroweak phase transition that may occur during the decay of the condensate. To avoid additional complications from spontaneous symmetry breaking, in this work we assume that both $m_\phi^2$ and $m_\chi^2$ are positive. For the case that $\chi$ is the Higgs field and there is the electroweak symmetry breaking, our main conclusions are not changed. We will comment on this in the Conclusion section. For a spatially flat FLRW universe,
\begin{equation}
\d{s}^2 = \d{t}^2 - a^2(t) \d{\vec x}^2\,,
\end{equation}
and $\sqrt{-g}=a^3(t)$. 

Expanding $\Phi = \varphi + \phi$, one can view $\varphi$ as a background field and $\phi$, $\chi$ as fluctuations about this background. Particles are then defined as excitations of these fluctuation fields. Discarding the linear terms,\footnote{The linear terms in fluctuations, e.g., $m^2_\phi\varphi\phi$, would not contribute to the perturbative diagrammatic expansion of the effective action basically due to $\langle\phi\rangle =\langle \chi\rangle=0$ and will not induce any dynamics. For a more detailed explanation see, e.g., Ref.~\cite{Peskin:1995ev}.} the action $S[\Phi,\chi]$ can be written as $S_\varphi[\varphi]+\hat{S}[\phi, \chi;\varphi]$ where
\begin{subequations}
\begin{align}
&S_\varphi[\varphi]  =\int\d^4 x\, a^3(t) \left[\frac{1}{2}(\partial_\mu\varphi)(\partial^\mu\varphi)-\frac{1}{2}m^2_\phi\varphi^2-\frac{\lambda_\phi}{4!}\varphi^4\right]\,, \\
&\hat{S}[\phi,\chi;\varphi]  =\int\d^4 x\,a^3(t)\left[\frac{1}{2} (\partial_\mu \phi) (\partial^\mu \phi)
	+ \frac{1}{2} (\partial_\mu \chi) (\partial^\mu \chi)
	-V(\phi,\chi;\varphi)\right]\,, \label{eq:Sphichi}
\end{align}
\end{subequations}
with
\begin{align}
\label{eq:Vhphi}
V(\phi, \chi;\varphi) &=
\frac{1}{2} \left(m_\phi^2 + \frac{\lambda_{\phi}}{2}\varphi^2\right) \phi^2+\frac{1}{2} \left(m^2_\chi +  \frac{g}{2} \varphi^2\right) \chi^2+\frac{\lambda_\phi}{3!}\varphi\phi^3+\frac{g}{2}\varphi\phi \chi^2
+ \frac{\lambda_\phi}{4!} \phi^4\notag\\
&\quad+ \frac{\lambda_\chi}{4!} \chi^4 +\frac{g}{4}\phi^2 \chi^2\,.
\end{align}
The oscillation of $\varphi$ could induce particle production for both $\phi$ and $\chi$, either through the time-dependent mass terms or the interactions $\varphi\phi^3$, $\varphi\phi \chi^2$.

\subsection{Coupled fundamental equations of motion for the condensate and plasma.}

In addition to the condensate $\varphi$, our system also includes a plasma. The plasma can be described by the two-point functions of the fluctuation fields, $\langle \phi\phi\rangle\equiv \Delta_\phi$, $\langle\chi\chi\rangle\equiv\Delta_\chi$. In the present subsection, for simplicity, we avoid distinguishing from various types of two-point functions. For more details, see Appendix~\ref{app:CTP}. In principle, we have coupled EoMs for these one- and two-point functions. Such coupled EoMs can be most generally derived from the two-particle-irreducible (2PI) effective action~\cite{Cornwall:1974vz,Berges:2004yj}. The latter is a functional of the one- and two-point functions, $\Gamma_{\rm 2PI}[\varphi,\Delta_\phi,\Delta_\chi]$, constructed from the Legendre transform of the generating functional with local one-point and non-local two-point sources (Appendix~\ref{app:CTP}). The EoMs are given by
\begin{subequations}
\label{eq:eom}
\begin{align}
\frac{\delta\Gamma_{\rm 2PI}[\varphi,\Delta_\phi,\Delta_\chi]}{\delta\varphi(x)}&=0\,,\\
\frac{\delta\Gamma_{\rm 2PI}[\varphi,\Delta_\phi,\Delta_\chi]}{\delta \Delta_\phi(x,x')}&=0\,,\\
\frac{\delta\Gamma_{\rm 2PI}[\varphi,\Delta_\phi,\Delta_\chi]}{\delta \Delta_\chi(x,x')}&=0\,.
\end{align}
\end{subequations}
The last two equations, typically studied in the absence of a background field, can be reduced to Boltzmann equations with a certain localization procedure~\cite{Calzetta:1986cq,Ivanov:1999tj,Buchmuller:2000nd,Blaizot:2001nr,Prokopec:2003pj,Berges:2005md,DeSimone:2007gkc,Cirigliano:2009yt,Beneke:2010wd,Beneke:2010dz,Drewes:2012qw,Berges:2014xea}.\footnote{More specifically, the Boltzmann equation is obtained from the EoM for the Wightman functions (cf. Eq.~\eqref{eq:wigthman}, the same for $\chi$).} The presence of the condensate introduces both $\varphi$-dependent masses and vertices.

The 2PI effective action $\Gamma_{\rm 2PI}$ can be expanded in terms of Feynman diagrams. In such diagrams, the propagators are the {\it full} two-point functions $\Delta_\phi$ and $\Delta_\chi$. Of particular interest is the condensate EoM at the one-loop\footnote{We emphasise that here ``one-loop'' is said in the sense of the 2PI effective action. If we use free propagators, the full two-point functions $\langle \phi\phi\rangle$ and $\langle\chi\chi\rangle$ can be expressed as a sum of infinitely many diagrams with arbitrarily higher
loops.}
\begin{align}
    \left(\Box+m_\phi^2\right)\varphi(x)+\frac{\lambda_\phi}{6}\varphi^3(x)+\frac{\lambda_\phi}{2}\varphi(x)\Delta_\phi(x,x)+\frac{g}{2}\varphi(x) \Delta_\chi(x,x)=0\,,\label{eq:condensate_eq_one_loop}
\end{align}
which is local. At this level, one can already have damping for the condensate if the plasma is not in thermal equilibrium, i.e., if
\begin{align}
    \Delta_\phi\neq \Delta_\phi^{\rm eq}\,,\qquad \Delta_\chi\neq \Delta_\chi^{\rm eq}\,,
\end{align}
where the superscript ``eq'' indicates thermal equilibrium. Writing $\Delta_{\phi/\chi}=\Delta_{\phi/\chi}^{\rm eq}+\delta\Delta_{\phi/\chi}$, the equilibrium parts $\Delta^{\rm eq}_{\phi/\chi}(x,x)$ simply correspond to thermal corrections to the squared mass of the field $\Phi$ (and hence also of the condensate). The resulting damping term is proportional to $\delta\Delta_{\phi/\chi}$, and the latter can be determined by solving the Boltzmann equations that are derived from the last two equations in~\eqref{eq:eom}. At high temperatures $T\gg M_\phi$ where $M_\phi$ is the oscillation frequency, the damping coefficient can be related to the bulk viscosity~\cite{Bodeker:2022ihg}. Such a map must break down at low temperatures $T\lsim M_\phi$ because the bulk viscosity diverges for $T\rightarrow 0$~\cite{Jeon:1994if,Jeon:1995zm} while the damping coefficient due to the deviation from equilibrium should decrease to zero for $T\rightarrow 0$ as in that limit the condensate oscillates simply in vacuum.

\paragraph{The small-field regime.}
The plasma can be driven out of equilibrium essentially because of the $\varphi$-dependent masses (see e.g.,~\cite{Bodeker:2022ihg}). In this paper, we are interested in the evolution of the condensate at a sufficiently late stage so that 
\begin{align}
\label{eq:conditions-small-field}
    \frac{\lambda_\phi}{2}\varphi^2 \ll  M^2_\phi \equiv m_\phi^2+ \frac{\lambda_\phi}{2}\varphi^2+m^2_{\phi,\rm th}\,,\quad \frac{g}{2}\varphi^2  \ll  M^2_\chi \equiv m_\chi^2+\frac{g}{2}\varphi^2+m_{\chi,\rm th}^2 \,,
\end{align}
where $m_{\phi,\rm th}$ and $m_{\chi,\rm th}$ are thermal corrections to the masses of $\phi$ and $\chi$ particles. The first condition also indicates that the oscillation is quasi-harmonic. This late stage is still interesting because it determines the fate of the oscillating condensate, i.e., whether the condensate can have a nonvanishing relic abundance. The conditions in Eq.~\eqref{eq:conditions-small-field} define the so-called small-field regime. In this case, one can perform a perturbative expansion in $\varphi$ in $\Delta_{\phi,\chi}[\varphi]$ and at the leading order, the two-point functions are independent of $\varphi$. In the small-field regime, it is expected that the damping due to the out-of-equilibrium effects mentioned above becomes inefficient as the particle masses are not sensitive to the change of $\varphi$ anymore. Furthermore, the small-field regime is precisely the regime in which nonperturbative particle production from condensate oscillations can be neglected and the condensate decays mainly via perturbative processes~\cite{Drewes:2019rxn}.

In the small-field regime, we have to consider other types of damping for the condensate oscillations. Going beyond the one-loop of the 2PI effective action, one has non-local terms in the condensate EoM which describe direct interactions between the condensate and particles induced by the $\varphi$-dependent vertices. Such interactions cause damping even if the plasma is in thermal equilibrium. This is the damping more often studied in the literature~\cite{Calzetta:1989vs,Paz:1990sd,Boyanovsky:1994me,Greiner:1996dx,Yokoyama:2004pf,Bastero-Gil:2010dgy,Bastero-Gil:2012akf,Mukaida:2012qn,Ai:2021gtg,Wang:2022mvv} and is the focus of the present work.

The above observation also applies to bubble wall dynamics in cosmological first-order phase transitions. The local EoM~\eqref{eq:condensate_eq_one_loop} is usually used for studying the bubble expansion. In the ultrarelativistic regime, the third and fourth terms, usually written as
\begin{align}
    \sum_i\frac{\d m^2_i(\phi)}{\d\phi}\int \frac{\d^3{\bf p}}{(2\pi)^32E_i}\,f_i(p,z)\,,
\end{align}
give rise to the B\"{o}deker-Moore thermal friction~\cite{Bodeker:2009qy,Ai:2024shx}. However, there is also additional friction caused by particle-production processes induced by condensate-dependent vertices, see e.g. Refs.~\cite{Bodeker:2017cim,Ai:2023suz}. This force is not captured by the local terms in Eq.~\eqref{eq:condensate_eq_one_loop} but rather is encoded in the non-local terms when one goes beyond the one-loop of the 2PI effective action~\cite{Ai:2024X}.

\subsection{Condensate equation of motion in the small-field regime}

We make the approximation that the plasma is in thermal equilibrium during the evolution of the condensate under study. Therefore, the leading-order two-point functions $\Delta_{\phi,\chi}[\varphi=0]$ in the small-field expansion are fixed to be the thermal equilibrium propagators and essentially, one only needs to solve the condensate EoM~\cite{Ai:2021gtg}. This dramatically simplifies the problem. The condensate EoM can be derived in the CTP formalism~\cite{Schwinger:1960qe,Keldysh:1964ud} using the effective action, see e.g., Ref.~\cite{Ai:2021gtg}. To be self-contained, we give a brief introduction to the CTP formalism in Appendix~\ref{app:CTP}.

Properly truncating the effective action and taking into account the homogeneity of the condensate and plasma, one obtains the condensate EoM as~\cite{Ai:2021gtg,Wang:2022mvv}
\begin{align}
\ddot{\varphi}(t) +  M_\phi^2 \varphi(t)+3H\dot{\varphi}(t)+ \frac{\lambda_\phi \varphi^3(t) }{6}+ \int_{t_i}^{t}
	    \d t' \, \pi^{\rm R}(t-t') \varphi(t')
	+ \frac{\varphi(t) }{6}
	\int_{t_i}^{t} \d t' \,
	    v^{\rm R}(t-t') \varphi^2(t') = 0 \,,
\label{eq:condensate eom}
\end{align}
where~$t_i$ is the initial time at which initial conditions~$\varphi(t_i)$ and~$\dot{\varphi}(t_i)$
are specified. $\pi^{\rm R}$ and $v^{\rm R}$ are the retarded {\it self-energy} and {\it proper four-vertex function} integrated over space, respectively. The proper four-vertex function is also called self-energy in a broad sense. They correspond to the following diagrams in the effective action,
\begin{subequations}
\label{eq:diagrams}
\begin{align}
\label{eq:diag-self-energy}
&\pi^{\rm R}\quad \leftrightarrow \quad
    \begin{tikzpicture}[baseline={0cm-0.5*height("$=$")}]
\draw[thick] (0,0) circle (0.5) ;
\draw[thick] (-0.5,0) -- (0.5,0) ;
\filldraw (-0.5,0) circle (1.5pt) node {} ;
\draw[thick] (0.5,0) -- (1,0) ;
\draw[thick] (1.15,0) circle (0.15) ;
\draw[thick] (1.15-0.707*0.15,0.707*0.15) -- (1.15+0.707*0.15,-0.707*0.15) ;
\draw[thick] (1.15-0.707*0.15,-0.707*0.15) -- (1.15+0.707*0.15,+0.707*0.15) ;
\filldraw (0.5,0) circle (1.5pt) node {} ;
\draw[thick] (-1,0) -- (-0.5,0) ;
\draw[thick] (-1.15,0) circle (0.15) ;
\draw[thick] (-1.15-0.707*0.15,0.707*0.15) -- (-1.15+0.707*0.15,-0.707*0.15) ;
\draw[thick] (-1.15-0.707*0.15,-0.707*0.15) -- (-1.15+0.707*0.15,+0.707*0.15) ;
\end{tikzpicture}\,,
\quad
\begin{tikzpicture}[baseline={0cm-0.5*height("$=$")}]
\draw[thick,dashed] (0,0) circle (0.5) ;
\draw[thick] (-0.5,0) -- (0.5,0) ;
\draw[thick] (0.5,0) -- (1,0) ;
\draw[thick] (1.15,0) circle (0.15) ;
\draw[thick] (1.15-0.707*0.15,0.707*0.15) -- (1.15+0.707*0.15,-0.707*0.15) ;
\draw[thick] (1.15-0.707*0.15,-0.707*0.15) -- (1.15+0.707*0.15,+0.707*0.15) ;
\filldraw[fill=black] (0.5,0) circle (1.5pt) node {} ;
\draw[thick] (-1,0) -- (-0.5,0) ;
\draw[thick] (-1.15,0) circle (0.15) ;
\draw[thick] (-1.15-0.707*0.15,0.707*0.15) -- (-1.15+0.707*0.15,-0.707*0.15) ;
\draw[thick] (-1.15-0.707*0.15,-0.707*0.15) -- (-1.15+0.707*0.15,+0.707*0.15) ;
\filldraw[fill=black] (-0.5,0) circle (1.5pt) node {} ;
\end{tikzpicture}\,,\\
& v^{\rm R} \quad \leftrightarrow \quad 
\begin{tikzpicture}[baseline={0cm-0.5*height("$=$")}]
\draw[thick] (0,0) circle (0.5) ;
\filldraw (0.5,0) circle (1.5pt) node {} ;
\filldraw (-0.5,0) circle (1.5pt) node {} ;
\draw[thick] (0.5,0) -- (0.6+0.3*0.707,0.1+0.3*0.707) ;
\draw[thick] (0.6+0.45*0.707,0.1+0.45*0.707) circle (0.15) ;
\draw[thick] (0.6+0.3*0.707,0.1+0.3*0.707) -- (0.6+0.6*0.707,0.1+0.6*0.707) ;
\draw[thick] (0.6+0.6*0.707,0.1+0.3*0.707) -- (0.6+0.3*0.707,0.1+0.6*0.707) ;
\draw[thick] (0.5,0) -- (0.6+0.3*0.707,-0.1-0.3*0.707) ;
\draw[thick] (0.6+0.45*0.707,-0.1-0.45*0.707) circle (0.15) ;
\draw[thick] (0.6+0.3*0.707,-0.1-0.3*0.707) -- (0.6+0.6*0.707,-0.1-0.6*0.707) ;
\draw[thick] (0.6+0.6*0.707,-0.1-0.3*0.707) -- (0.6+0.3*0.707,-0.1-0.6*0.707) ;
\draw[thick] (-0.5,0) -- (-0.6-0.3*0.707,-0.1-0.3*0.707) ;
\draw[thick] (-0.6-0.45*0.707,-0.1-0.45*0.707) circle (0.15) ;
\draw[thick] (-0.6-0.3*0.707,-0.1-0.3*0.707) -- (-0.6-0.6*0.707,-0.1-0.6*0.707) ;
\draw[thick] (-0.6-0.6*0.707,-0.1-0.3*0.707) -- (-0.6-0.3*0.707,-0.1-0.6*0.707) ;
\draw[thick] (-0.5,0) -- (-0.6-0.3*0.707,0.1+0.3*0.707) ;
\draw[thick] (-0.6-0.45*0.707,0.1+0.45*0.707) circle (0.15) ;
\draw[thick] (-0.6-0.3*0.707,0.1+0.3*0.707) -- (-0.6-0.6*0.707,0.1+0.6*0.707) ;
\draw[thick] (-0.6-0.6*0.707,0.1+0.3*0.707) -- (-0.6-0.3*0.707,0.1+0.6*0.707) ;
\end{tikzpicture}\,,
\quad
\begin{tikzpicture}[baseline={0cm-0.5*height("$=$")}]
\draw[thick, dashed] (0,0) circle (0.5) ;
\filldraw (0.5,0) circle (1.5pt) node {} ;
\filldraw (-0.5,0) circle (1.5pt) node {} ;
\draw[thick] (0.5,0) -- (0.6+0.3*0.707,0.1+0.3*0.707) ;
\draw[thick] (0.6+0.45*0.707,0.1+0.45*0.707) circle (0.15) ;
\draw[thick] (0.6+0.3*0.707,0.1+0.3*0.707) -- (0.6+0.6*0.707,0.1+0.6*0.707) ;
\draw[thick] (0.6+0.6*0.707,0.1+0.3*0.707) -- (0.6+0.3*0.707,0.1+0.6*0.707) ;
\draw[thick] (0.5,0) -- (0.6+0.3*0.707,-0.1-0.3*0.707) ;
\draw[thick] (0.6+0.45*0.707,-0.1-0.45*0.707) circle (0.15) ;
\draw[thick] (0.6+0.3*0.707,-0.1-0.3*0.707) -- (0.6+0.6*0.707,-0.1-0.6*0.707) ;
\draw[thick] (0.6+0.6*0.707,-0.1-0.3*0.707) -- (0.6+0.3*0.707,-0.1-0.6*0.707) ;
\draw[thick] (-0.5,0) -- (-0.6-0.3*0.707,-0.1-0.3*0.707) ;
\draw[thick] (-0.6-0.45*0.707,-0.1-0.45*0.707) circle (0.15) ;
\draw[thick] (-0.6-0.3*0.707,-0.1-0.3*0.707) -- (-0.6-0.6*0.707,-0.1-0.6*0.707) ;
\draw[thick] (-0.6-0.6*0.707,-0.1-0.3*0.707) -- (-0.6-0.3*0.707,-0.1-0.6*0.707) ;
\draw[thick] (-0.5,0) -- (-0.6-0.3*0.707,0.1+0.3*0.707) ;
\draw[thick] (-0.6-0.45*0.707,0.1+0.45*0.707) circle (0.15) ;
\draw[thick] (-0.6-0.3*0.707,0.1+0.3*0.707) -- (-0.6-0.6*0.707,0.1+0.6*0.707) ;
\draw[thick] (-0.6-0.6*0.707,0.1+0.3*0.707) -- (-0.6-0.3*0.707,0.1+0.6*0.707) ;
\end{tikzpicture}\,,
\end{align}
\end{subequations}
where a solid/dashed line denotes the free thermal equilibrium propagator for $\phi/\chi$ (cf. Eq.~\eqref{thermal propagators}), and a solid line ended with a wheel cross indicates a quantum of the condensate.
$\pi^{\rm R}$ and $v^{\rm R}$ are explicitly defined in Appendix~\ref{app:self-energy_and_four-vertex}. The self-energies are discussed in Appendix~\ref{app:self-energy_and_four-vertex} and the expressions for the relevant quantities that enter the coarse-grained EoMs will be given below. Note that the diagrams of $v^{\rm R}$ are generated from the one loop diagrams in $\Gamma_{\rm 2PI}$ by the small-field expansion while $\pi^{\rm R}$ is obtained from two loop diagrams in $\Gamma_{\rm 2PI}$~\cite{Ai:2021gtg}.

The last two terms in Eq.~\eqref{eq:condensate eom} exhibit non-locality in time and therefore contain memory effects, making it challenging to obtain a rigorous solution for the EoM. However, one can reasonably expect the first two terms in Eq.~\eqref{eq:condensate eom} to dominate. By treating the remaining terms as perturbations, one can solve the condensate EoM using multiple-scale analysis~\cite{Bender,Holmes} as described in Refs.~\cite{Ai:2021gtg,Wang:2022mvv}. There, constant temperature and $M_\phi$ are assumed. In the present work, we account for the evolution of temperature which makes $M_\phi$ time dependent. In Appendix~\ref{app:sol_eom}, we extend the analysis of Refs.~\cite{Ai:2021gtg,Wang:2022mvv} to the case with time-dependent but adiabatic $M_\phi$. The solution takes the following form:
\begin{align}
    \varphi(t)\approx A(t) \cos\left[\int ^{t} M_\phi (t')\d t'+f(t)\right]\,,
\end{align}
where $A(t)$ is the envelope function of the oscillation and $f(t)$ represents corrections to the local frequency $M_\phi$. The EoMs for $A(t)$ and $f(t)$ are found to be
\begin{subequations}
\label{eq:eom_A and f(tau)}
\begin{align}
\label{eq:eom_A(tau)}
\frac{\d A (t)}{\d t}+\left(\gamma +\frac{3}{2}H(t)+\frac{1}{2 M_\phi(t)}\frac{\d M_\phi(t)}{\d t}\right) A(t) +\frac{\sigma}{2} [A(t)]^3&=0\,,\\
\frac{\d f(t)}{\d t}-\mu -\left(\frac{\lambda_{\phi}}{16 M_{\phi}}+\alpha \right)[A(t)]^2&=0\,,
\end{align}
\end{subequations}
where\footnote{Note that the $\gamma$ defined in Ref.~\cite{Wang:2022mvv} is half of that defined in Ref.~\cite{Ai:2021gtg} and here we follow the former.} 
\begin{align}
\label{eq:important_quantities}
\mu&\equiv \frac{\text{Re}[\widetilde{\pi}^{\rm R} (M_\phi )]}{2M_{\phi}}\,, \; \gamma \equiv -\frac{\text{Im} [\widetilde{\pi}^{\rm R} (M_\phi )]}{2M_{\phi}}\,, \; \sigma \equiv -\frac{\text{Im} [\widetilde{v}^{\rm R} (2M_\phi )]}{24 M_{\phi}}
\,,\notag\\
\alpha_1 &\equiv \frac{\text{Re} [\widetilde{v}^{\rm R} (0)]}{24 M_{\phi}}\,,\ \ \, \alpha _2\equiv \frac{\text{Re} [\widetilde{v}^{\rm R} (2M_\phi )]}{48 M_{\phi}}\,,\ \; \alpha \equiv \alpha_1+\alpha_2\,,
\end{align}
with
\begin{equation}
\label{PiTildeDef}
\widetilde{\pi}^{\rm R}(\omega) = \int_{-\infty}^{\infty} \d t'\, \e^{\i\omega(t-t')} \pi^{\rm R}(t-t') \, ,
\qquad 
\widetilde{v}^{\rm R}(\omega) = \int_{-\infty}^{\infty} \d t' \, \e^{\i\omega(t-t')} v^{\rm R}(t-t') \, .
\end{equation}
Above, one actually has neglected the early-time transient effects related to the initial conditions in the non-local terms. This is why the EoMs for $A(t)$ and $f(t)$ above depend on the normal Fourier transforms of the retarded self-energy and proper four-vertex function.

\section{Coupled coarse-grained equations of motion}
\label{sec:coarse-grained EoMs}

In order to understand the energy transfer between the condensate and plasma, it is sufficient to study the EoM of the envelope function, Eq.~\eqref{eq:eom_A(tau)}. Indeed, the energy of the condensate can be approximated as
\begin{align}
       \rho _{\varphi}= \frac{1}{2} \dot{\varphi}^2 +\frac{1}{2} M^2_{\phi} \varphi^2 \approx \frac{1}{2}M^2_\phi A(t)^2\,,
\end{align}
where we have assumed that
\begin{align}\label{eq:static universe assumption_0}
   \frac{\dot{A}}{A} \ll M_{\phi} \,,\quad \frac{\d f}{\d t}\ll M_{\phi}\,.
\end{align}
The above conditions imply that the amplitude and the frequency of oscillations do not change much during one oscillation. Therefore, the evolution of $A(t)$ contains information on the dissipation of the condensate energy. From Eq.~\eqref{eq:eom_A(tau)} we obtain
\begin{align}
\label{eq:eom_energy_density}
\dot{\rho}_\varphi+3H\rho_\varphi+\left(2\gamma-\frac{\dot{M}_\phi}{M_\phi}\right)\rho_\varphi+\frac{2\sigma \rho_\varphi^2}{M_{\phi}^2}=0\,.
\end{align}
The $\gamma$ and $\sigma$ depend on the temperature and thus are time dependent. Eq.~\eqref{eq:eom_energy_density} is accompanied by the EoM for the radiation energy density and the Friedmann equation
\begin{align}
\label{eq:eom_energy_density_radiation}
   & \dot{\rho}_{\rm R}+4H\rho_{\rm R}-\left(2\gamma-\frac{\dot{M}_\phi}{M_\phi}\right)\rho_\varphi-\frac{2\sigma \rho_\varphi^2}{M_{\phi}^2}=0\,,\\
\label{eq:eom_Friedmann}
  & H^2=\frac{8\pi(\rho_\varphi+\rho_{\rm R})}{3M^2_{\rm P}}\,,
\end{align}
where $M_{\rm P}$ is the Planck mass.
The radiation energy density is related to the temperature through
\begin{align}
    \rho_{\rm R}=\frac{\pi^2}{30} g_* T^4\,,
\end{align}
where $g_*$ is the effective number of relativistic degrees of freedom.
Here we have assumed that during the dissipation of the condensate, there are no other forms of energy densities.

If we define the number density of the condensate quanta $n_\varphi$ through $n_\varphi=\rho_\varphi/M_\phi$,
then Eq.~\eqref{eq:eom_energy_density} becomes
    \begin{align}
    \dot{n}_\varphi+3H n_\varphi=-2\gamma n_\varphi-\frac{2\sigma}{M_\phi} n_\varphi^2\,.
\end{align}
This implies that the comoving number density of the condensate quanta is conserved for vanishing $\gamma$ and $\sigma$.

We note that the evolution of the envelope function (and of $\rho_\varphi$) depends on two particular microscopic quantities, $\gamma$ and $\sigma$, the imaginary parts of the retarded self-energy and retarded proper four-vertex function evaluated at particular frequencies. These quantities have an interpretation in terms of particle production via the cutting rules at finite temperature~\cite{Cutkosky:1960sp,Weldon:1983jn,Kobes:1985kc,Kobes:1986za,Landshoff:1996ta,Gelis:1997zv,Bedaque:1996af}, which explains the origin of the dissipation. For more details see Ref.~\cite{Wang:2022mvv} and Appendix~\ref{app:self-energy_and_four-vertex}. In summary, nonvanishing $\gamma$ describes the following processes
\begin{align}
\label{eq:decaychannel-one}
    \gamma{\rm \ channels:}\quad 
    \varphi\phi\leftrightarrow \phi\phi\,,\quad
    \varphi\chi\leftrightarrow \phi\chi\,,\quad \varphi\phi\leftrightarrow\chi\chi\,,
\end{align}
while $\sigma$ describes
\begin{align}
\label{eq:decaychannel-two}
\sigma\ {\rm channel:}\quad
\quad (\varphi\varphi)\leftrightarrow \chi\chi\quad  {\rm for\ } M_\phi> M_\chi\,.
\end{align}
Eq.~\eqref{eq:decaychannel-one} are scattering processes that are absent at zero temperature and are called {\it Landau damping}. 
We leave the details of an explicit computation of $\gamma$ and $\sigma$ in Appendix~\ref{app:self-energy_and_four-vertex}.

The imaginary part of the retarded proper four-vertex is known in the literature~\cite{Boyanovsky:2004dj}, and reads
\begin{align}
\sigma=\theta\left(M_\phi - 
	    M_\chi \right)
		\frac{g^2 }{ 256\pi M_\phi }
		\sqrt{ 1 -\frac{M_\chi^2}{M_\phi^2} }
		\,\left[ 1 + 2f_{\rm B} \left(M_\phi \right) \right] \,.
\label{Im vR omega T}
\end{align}
In the limit of $T=0$, one has 
\begin{align}\label{eq:sigma0}
    \sigma_0\equiv \left.\sigma\right|_{T=0}=\theta\left(m_\phi - 
	    m_\chi \right)
		\frac{g^2 }{ 256\pi m_\phi}
		\sqrt{ 1 -\frac{m_\chi^2}{ m_\phi^2}}\,.
\end{align}

The $\gamma$ has contributions from the self-interactions of $\Phi$ and the interaction between $\Phi$ and $\chi$: $\gamma=\gamma_\phi +\gamma_{\phi \chi}$. The imaginary part of the self-energy from the $\Phi$ self-interaction has been obtained in Refs.~\cite{Jeon:1994if,Wang:1995qg,Wang:1995qf} and reads\footnote{Note a sign difference in the definition of the imaginary part of Ref.~\cite{Jeon:1994if}. See Eq.~(2.35) of that reference.}
\begin{align}
    \left.{\rm Im}[\widetilde{\pi}^{\rm R}(M_\phi)]\right|_{\rm self\ interaction}=-\frac{\lambda_\phi^2 T^2}{128\pi^3}{\rm Li}_2\left(\e^{-M_\phi/T}\right)\,.
\end{align}
where ${\rm Li}_2 (z)$ is the dilogarithm function defined as
\begin{align}
    {\rm Li}_2(z) =-\int ^{1} _{0} \frac{\d x}{x} \ln (1-z\,x)\,,
\end{align}
for $z\in (-\infty, 1)$. The dilogarithm function has the following asymptotic behaviours
\begin{align}
     &\text{Li}_2(\e ^{- M_{\phi}/T}) = \frac{\pi ^2}{6} + O\left[\frac{M_{\phi}}{T}\ln(\frac{M_{\phi}}{T})\right]\quad {\rm for\ }T\gg M_\phi\,,\\
     \label{eq:Li2-lowT}
     &\text{Li}_2(\e ^{- M_{\phi}/T}) = \e ^{- M_{\phi}/T} + O\left(\e ^{-2 M_{\phi}/T}\right)\quad {\rm for\ } T\ll M_\phi\,.
\end{align}
Substituting the above into the definition of $\gamma$, we obtain
\begin{align}
\label{eq:Im-pi-phi}
  \gamma_{\rm \phi}=\frac{\lambda_\phi^2 T^2}{256 M_\phi\pi^3}{\rm Li}_2\left(\e^{-M_\phi/T}\right)\,.
\end{align}
The high-temperature limit of Eq.~\eqref{eq:Im-pi-phi} is consistent with the result given in Refs.~\cite{Parwani:1991gq,Drewes:2013iaa}. For $T\rightarrow 0$, $\gamma_\phi$ vanishes. This is reasonable because at zero temperature all the $\gamma$ channels are not possible.

While we cannot provide a closed form for $\gamma_{\phi \chi}$, we do obtain an analytical fit for it (cf. Appendix~\ref{app:sefl-energy}):
\begin{align}
\label{eq:gammafit}
    \gamma_{\phi \chi}\sim \gamma_{\phi \chi,\rm fit}&\equiv 4.5\times\frac{g^2 T^2}{128 M_\phi\pi^5}{\rm Li}_2\left(\frac{4}{(M_\chi/M_\phi)^3+4}\right){\rm Li}_2\left(\e^{-\frac{M_\chi+2M_\phi}{3T}}\right)\notag\\
    &+6.8\times\frac{g^2 T^2}{64 M_\phi\pi^5}{\rm Li}_2\left(\frac{1.6}{(M_\chi/M_\phi)+1.6}\right){\rm Li}_2\left(\e^{-\frac{2M_\chi+M_\phi}{3T}}\right)\,.
\end{align}
One can use the above fit to solve the coupled dynamics.  For $T\rightarrow 0$, $\gamma_{\phi\chi,\rm fit}$ vanishes as one would expect.

\section{``Freeze-out'' of the \texorpdfstring{$\mathbb{Z}_2$}{TEXT} condensate}
\label{sec:freeze-out}

Before we present some exact numerical results, let us first estimate the condition for the ``freeze-out'' of the $\mathbb{Z}_2$ condensate. From Eq.~\eqref{eq:eom_energy_density}, one can read the condensate ``decay rate'' as
\begin{align}
\label{eq:decay-rate}
    \Gamma_\varphi=2\gamma-\frac{\dot{M}_\phi}{M_\phi}+\frac{2\sigma\rho_\varphi}{M^2_\phi}\,.
\end{align}
Note that this decay rate is not necessarily equal to the damping rate defined from the envelop function $A(t)$~\cite{Ai:2021gtg}. Nevertheless, it can still characterise how fast the condensate depletes its energy into the plasma. An efficient decay requires $\Gamma_\varphi \gsim H$ so that one can use $\Gamma_\varphi\sim H$ as an estimate of the ``freeze-out'' time of the condensate. The full evolution is complicated because there could be some reheating effects due to the energy transfer from the condensate to the plasma. However, the ``freeze-out'' of the condensate typically occurs after the reheating. At that stage, one would have $\rho_{\rm R}\gg \rho_{\varphi}$,\footnote{If the initial conditions already satisfy this, then reheating effects are negligible.} and
\begin{align}
    H\sim 1.66\times\sqrt{g_*} T^2\sim \O(10)\, T^2 \sim \frac{1}{t}\,.
\end{align}

\paragraph{$\gamma$ channels.}  
In the high-temperature limit, $\gamma$ exhibits a decrease proportional to the temperature $T$ as the Universe cools down. When the temperature drops to the regime where $M_\phi \sim m_\phi$, $\gamma$ decreases with a quadratic dependence on $T$, thus scaling as $T^2$. At lower temperatures, $\gamma$ experiences an exponential suppression. As a consequence, when the temperature drops to a sufficiently low value, $\gamma$ decreases faster than $H$.

\paragraph{$\sigma$ channel.} While $\sigma$ is not vanishing even for $T\rightarrow 0$ provided $m_\phi > m_\chi$, the condensate decay rate due to the sigma channel is proportional to $\rho_\varphi$ which decreases fast. For the case of $\gamma=0$ and $\dot{M}_\phi=0$, $\rho_\varphi$ drops as $t^{-3/2}$~\cite{Wang:2022mvv} while $H$ drops as $1/t$. With nonvanishing $\gamma$ and $\dot{M}_\phi$, $\rho_\varphi$ should decrease even faster. 

\paragraph{Damping due to the change of $M_\phi$.} In the high-temperature limit, one has \begin{align}
    \frac{\dot{M}_\phi}{M_\phi}=\frac{(\lambda_\phi + g) T\dot{T}/24}{{m^2_\phi+(\lambda_\phi +g) T^2/24 }} \sim \frac{\dot{T}}{T}
\end{align}
which scales as $H$. Going beyond the high temperature limit, $\dot{M_\phi}/M_\phi$ decreases faster and quickly approaches zero when $(\lambda_\phi+g)T^2/24< m_\phi^{2}$. Actually, in the parameter region for our numerical calculations below, $\dot{M}_\phi/M_\phi$ is negligible.   

In summary, $\Gamma_\varphi$ would decrease faster than $H$ as the Universe cools down and therefore the condensate must ``freeze out'' at some point. This will be confirmed further with our numerical results below. We emphasize that it is crucial to go beyond the high-temperature approximation for $\gamma$. Otherwise, one can have, for some chosen parameters, a $\gamma$ that is larger than $H$ all the time and conclude that the condensate energy can be fully transferred to radiation~\cite{Mukaida:2013xxa,Wang:2022mvv}. The above analysis does not draw any conclusion about the relic density of the condensate. The latter strongly depends on the parameters in the theory as well as the initial conditions.

\section{Numerical results}
\label{sec:numerics}

In this section, we numerically solve the coupled equations \eqref{eq:eom_energy_density}, \eqref{eq:eom_energy_density_radiation}, and \eqref{eq:eom_Friedmann}. The purpose of this section is to numerically demonstrate the ``freeze-out'' behaviour of the $\mathbb{Z}_2$ condensate analysed in the last section. Since we do not aim for a rigorous phenomenological study, the mass and coupling parameters will be to some extent chosen randomly without good phenomenological motivations. A complete phenomenological study on the condensate contribution to the $\mathbb{Z}_2$ singlet scalar DM will be left for future work. Above, analytical expressions for $\sigma$ and $\gamma$ (approximated as $\gamma_{\rm fit}\equiv \gamma_{\phi} + \gamma_{\phi \chi,\rm fit}$) have already been presented. For the thermal corrections to the masses, we shall use the high-temperature approximation~\eqref{eq:thermal-masses-highT}. We assume that the thermal corrections to the $\chi$ mass will be dominated by the term $\kappa^2 T^2$ and we take $\kappa^2\approx 0.1$. We note however that in the numerical calculation below, the high-temperature limit may not hold. However, for perturbatively small couplings, the thermal mass corrections away from the high-temperature regime are immaterial anyway (cf. Appendix~\ref{app:self-energy_and_four-vertex}). Therefore whether or not one uses the high-temperature approximation for the thermal mass corrections does not change the results quantitatively. However, as we emphasised in the Introduction, it is crucial to use the complete $T$-dependence for the $\gamma$ parameter as its high-temperature approximation would lead to incorrect conclusions.

\begin{figure}[ht]
    \centering
    \includegraphics[scale=1.1]{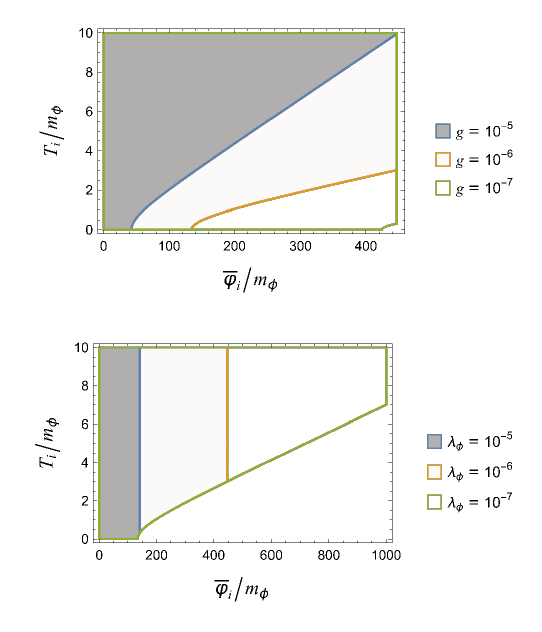}
    \caption{Initial conditions that fall within the perturbative regime with fixed masses $m_{\phi}=500{\rm GeV}$, $m_{\chi}=150{\rm GeV}$.  $\lambda_{\phi}=10^{-6}$ for the upper panel and $g=10^{-6}$ for the lower panel. }
    \label{fig:constraint1}
\end{figure}

We will not be concerned with the formation of the condensate and its early nonperturbative decay through parametric resonance. Rather, we assume that these processes produce some initial condition $\{T_i, \bar{\varphi}_i\}$, where $T_i$ is the initial temperature and $\bar{\varphi}_i$ is the initial oscillation amplitude, for the perturbative regime at the late stage. In this work, we randomly choose the initial condition in performing the numerical calculation, provided that it falls within the perturbative regime of particle production, satisfying $(i)\ M^2_\phi\gg \lambda_\phi \bar{\varphi}_i^2/2$ and $(ii)\ M^2_\chi\gg g \bar{\varphi}_i^2/2$. We will consider different sets of $\{T_i,\bar{\varphi}_i\}$ to explore the sensitivity of the results to the initial condition. For our analysis, we set $g_*=100$, $m_\chi= 150{\rm GeV}$ and consider two cases for $m_\phi$, $m_\phi=100 {\rm GeV}<m_\chi$ and $m_\phi=500 {\rm GeV}>m_\chi$. An example of the constraints on the initial conditions around $g=10^{-6}$ and $\lambda_\phi=10^{-6}$ is given in Fig.~\ref{fig:constraint1} where the boundary is obtained with $M_\phi^2=10\times \lambda_\phi\bar{\varphi}_i^2/2$, $M_\chi^2=10\times g\bar{\varphi}_i^2/2$. One can also find the constraints on $\{T_i,\bar{\varphi}_i\}$ for other chosen couplings. When fixing $\lambda_\phi$ but changing $g$, the regions are mostly affected by the condition $(ii)$ (see the upper panel). When fixing $g$ but changing $\lambda_\phi$, the regions are mostly affected by the condition $(i)$ (see the lower panel). Note that the seemingly vertical lines are not strictly vertical; they just have a very large slope.

\subsection{\texorpdfstring{$m_\phi> m_\chi$}{TEXT}}

Let us first consider the case $m_\phi=500 {\rm GeV} > m_\chi =150 {\rm GeV}$.

\begin{figure}[H]
    \centering
    \includegraphics[scale=0.75]{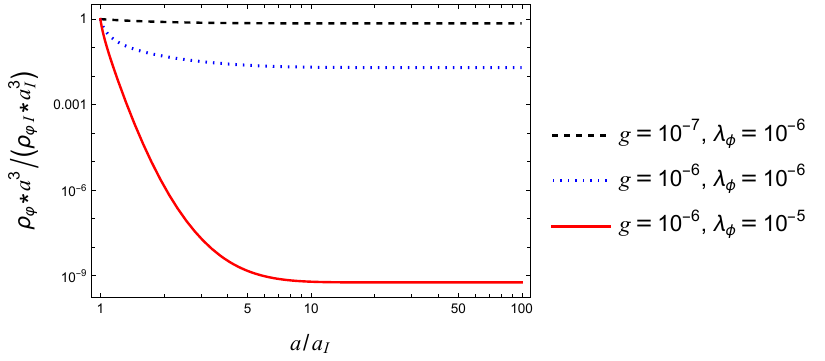}
    \caption{Relic comoving energy densities of the condensate with various coupling constants. The initial condition is chosen as $\bar{\varphi}_i=5m_{\phi}$, and $T_i=2m_{\phi}$.  }
    \label{fig:Comvingenergycomparingv2}
\end{figure}

\begin{figure}[H]
    \centering
    \includegraphics[scale=0.75]{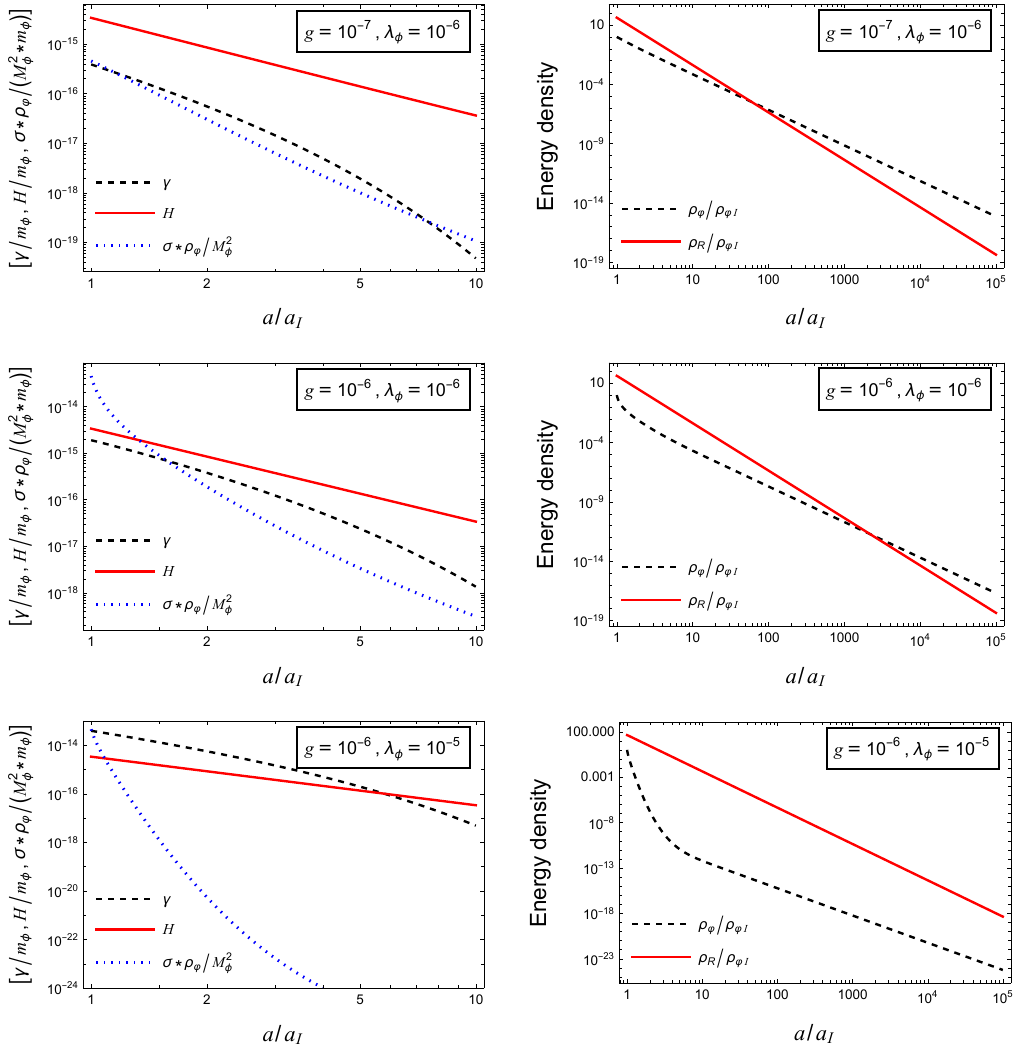}
    \caption{$\gamma$, $\sigma\rho_\varphi/M_\phi^2$ and the Hubble parameter $H$ (the first column) and the energy densities (the second column) for different couplings but with fixed initial condition $\bar{\varphi}_i=5m_{\phi}$, and $T_i=2m_{\phi}$.}
    \label{fig:numericalsolGridv2}
\end{figure}

\paragraph{Different couplings.} 
We fix the initial condition as $\bar{\varphi}_i=5 m_\phi$, $T_i=2 m_\phi$ and consider three sets of couplings: $\{g=10^{-7}, \lambda_\phi=10^{-6}\}$,  $\{g=10^{-6}, \lambda_\phi=10^{-6}\}$, and $\{g=10^{-6}, \lambda_\phi=10^{-5}\}$. Fig.~\ref{fig:Comvingenergycomparingv2} shows the relic comoving energy density $\rho_{\varphi} a^3$ for these three cases while Fig.~\ref{fig:numericalsolGridv2} shows the details about the evolution. For $g=10^{-7}$, $\lambda_\phi=10^{-6}$, the condensate decay is rather incomplete. This is because both $\gamma$ and $\sigma\rho_\varphi/M_\phi^2$ are smaller than $H$ all the time and no efficient decay can occur (see the first row of Fig.~\ref{fig:numericalsolGridv2}). For the second case, $\sigma \rho_\varphi/M_\phi^2$ is larger than $H$ initially while $\gamma$ is still smaller than $H$ all the time. Hence the decay is dominated by the $\sigma$ channel. However, due to dependence on $\bar{\varphi}^2$, $\sigma\rho_\varphi/M_\phi^2$ decreases quickly and the channel is efficient only for a short time, leading to again an incomplete decay. In the last case, $g=10^{-6}$, $\lambda_\phi=10^{-5}$, $\gamma$ is larger than $H$ for a period of the universe expansion and the decay is quite complete. In the last two cases, one clearly sees a ``freeze-out'' behaviour of the condensate which is characterised by the crossing between the evolution line of either $\gamma$ or $\sigma \rho_\varphi/M_\phi^2$ and that of $H$. We note that even if the relic {\it comoving} energy density of the condensate is very small compared with the radiation {\it comoving} energy density, its actual relic energy density decreases slower than the radiation energy density and can catch up with the latter after a sufficiently long time of universe expansion. This should lead to additional phenomenological constraints on such $\mathbb{Z}_2$-symmetric scalar fields. For example, for the chosen initial temperature $T_i=2m_\phi=1000 {\rm GeV}$, the temperature decreases to $10{\rm MeV}$ for $a/a_I\sim 10^5$. Since the Universe must be radiation-dominated at the Big Bang nucleosynthesis (BBN), the first two cases in Fig.~\ref{fig:numericalsolGridv2} would then be excluded as a phenomenological model.\footnote{We thank the anonymous referee for pointing out this to us.} However, our initial conditions are chosen randomly. To have concrete conclusions, one has to carefully analyse the nonperturbative process prior to the perturbative condensate decay studied here. We leave a detailed study on this for future work.  

\paragraph{Different initial conditions.} 

\begin{figure}[H]
    \centering
    \includegraphics[scale=0.69]{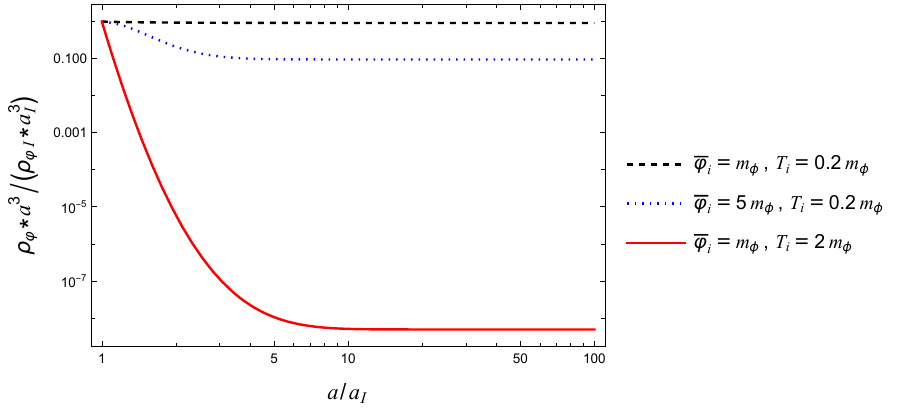}
    \caption{Comoving energy densities of the condensate for various initial conditions but with fixed couplings $g=10^{-7}$ and $\lambda_{\phi}=10^{-5}$.}
    \label{fig:Comvingenergycomparing2}
\end{figure}

\begin{figure}[H]
    \centering
    \includegraphics[scale=0.75]{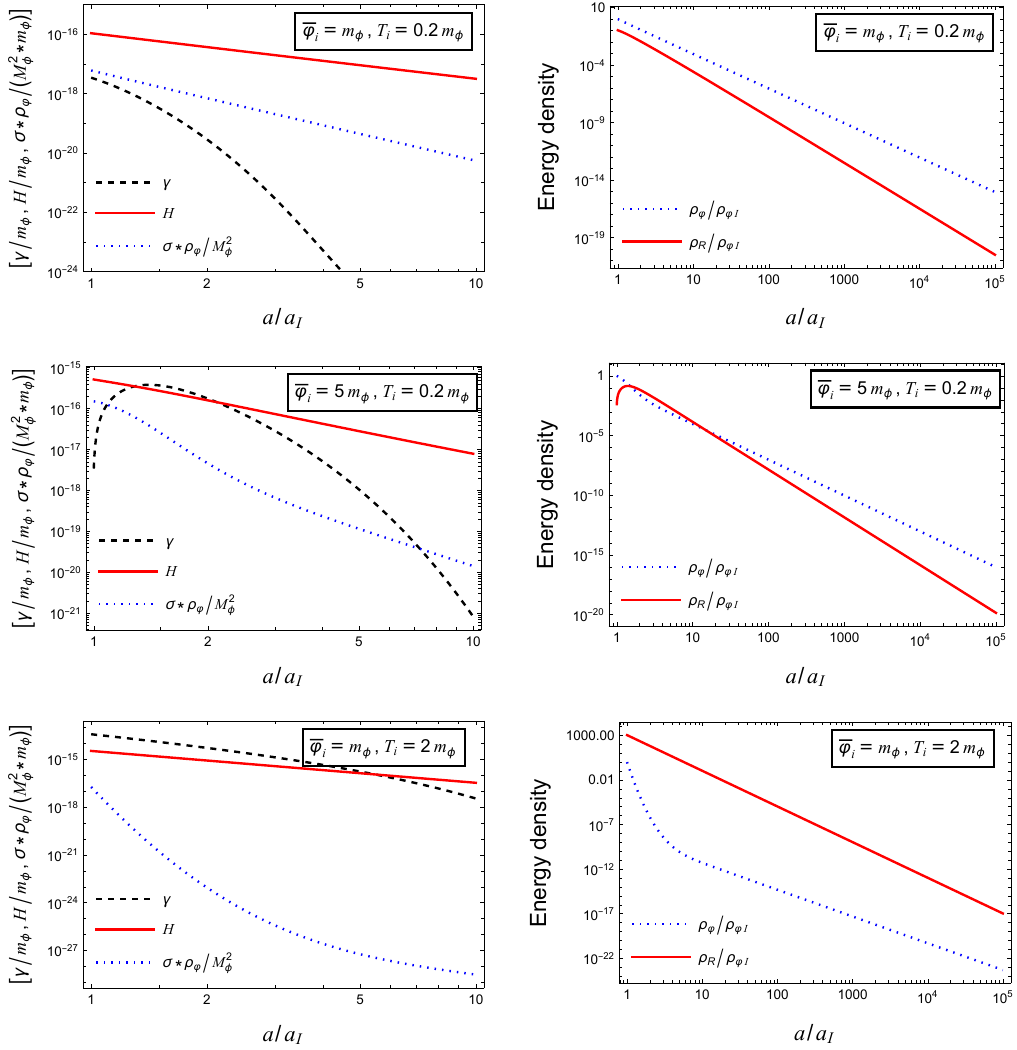}
    \caption{$\gamma$, $\sigma\rho_\varphi/M_\phi^2$ and the Hubble parameter $H$ (the first column) and the energy densities (the second column) for different initial conditions but fixed couplings $g=10^{-7}$ and $\lambda_\phi=10^{-5}$.}
    \label{fig:numericalsolGrid2}
\end{figure}

\begin{figure}[H]
    \centering
    \includegraphics[scale=0.69]{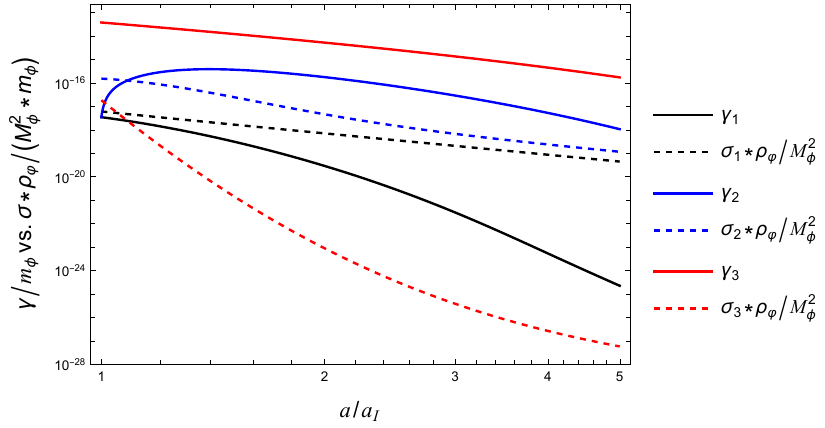}
    \caption{Early evolution of $\gamma$ and $\sigma \rho_\varphi/M_\phi^2$ with various initial conditions: $\{\bar{\varphi}_i=m_{\phi}, T_i=0.2m_{\phi}\}$ (black), $\{\bar{\varphi}_i=5m_{\phi}, T_i=0.2m_{\phi}\}$ (blue),  $\{\bar{\varphi}_i=m_{\phi}, T_i=2m_{\phi}\}$ (red) with fixed couplings $g=10^{-7}$, $\lambda_\phi=10^{-5}$. }
    \label{fig:comparingsmallavs2}
\end{figure}

Fig.~\ref{fig:Comvingenergycomparing2} and Fig.~\ref{fig:numericalsolGrid2} present the numerical solutions for different initial conditions, $\{\bar{\varphi}_i=m_\phi,T_i=0.2 m_\phi\}$, $\{\bar{\varphi}_i=5m_\phi,T_i=0.2 m_\phi\}$, and $\{\bar{\varphi}_i=m_\phi,T_i=2m_\phi\}$, but with fixed couplings $g=10^{-7}$, $\lambda_\phi=10^{-5}$. For the first initial condition, the decay is rather incomplete because both $\gamma$ and $\sigma \rho_\varphi/M_\phi^2$ are smaller than $H$ all the time. A larger value of either $\bar{\varphi}_i$ or $T_i$ can enhance the decay efficiency. A larger $\bar{\varphi}_i$ gives a larger initial decay rate of the $\sigma$ channel. Particle production from the $\sigma$ channel then leads to reheating and increases the temperature. As a consequence, the decay rate from the $\gamma$ channels increases at early times, see the left panel of the second row in Fig.~\ref{fig:numericalsolGrid2}. Increasing $T_i$ would directly increase $\gamma$ at the beginning, see the left panel of the third row in Fig.~\ref{fig:numericalsolGrid2}. Fig.~\ref{fig:comparingsmallavs2} presents the evolution of $\gamma$, $\sigma\rho_\varphi/M_\phi^2$, and $H$ at early times for all the three cases. Again, the first two cases would lead to a matter-dominated universe already before BBN.

\subsection{\texorpdfstring{$m_\phi<m_\chi$}{TEXT}}

Now we consider the case $m_\phi=100 {\rm GeV}< m_\chi= 150 {\rm GeV}$. In this case, one typically has $M_\chi> M_\phi$ unless $\lambda_\phi$ is too large which would then probably lead to unitarity problems. Therefore, for this case, the $\sigma$ channel is closed.

\begin{figure}[H]
    \centering
    \includegraphics[scale=0.75]{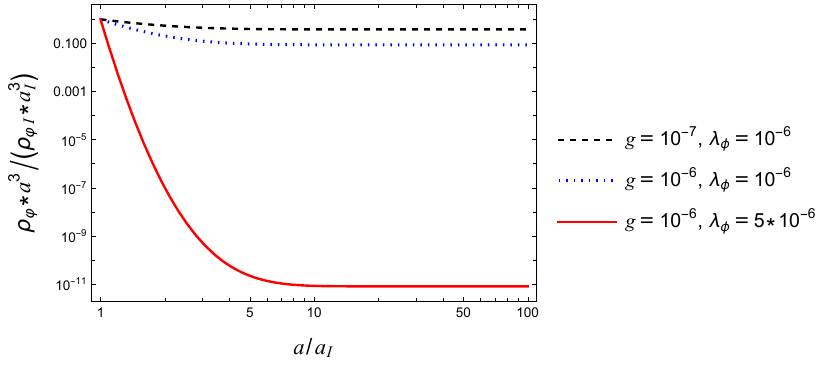}
    \caption{Comoving energy densities of the condensate for various coupling constants but with fixed initial condition $\bar{\varphi}_i=5m_{\phi}$, and $T_i=2m_{\phi}$.  }
    \label{fig:Comvingenergycomparing3}
\end{figure}

\begin{figure}[H]
    \centering
    \includegraphics[scale=0.75]{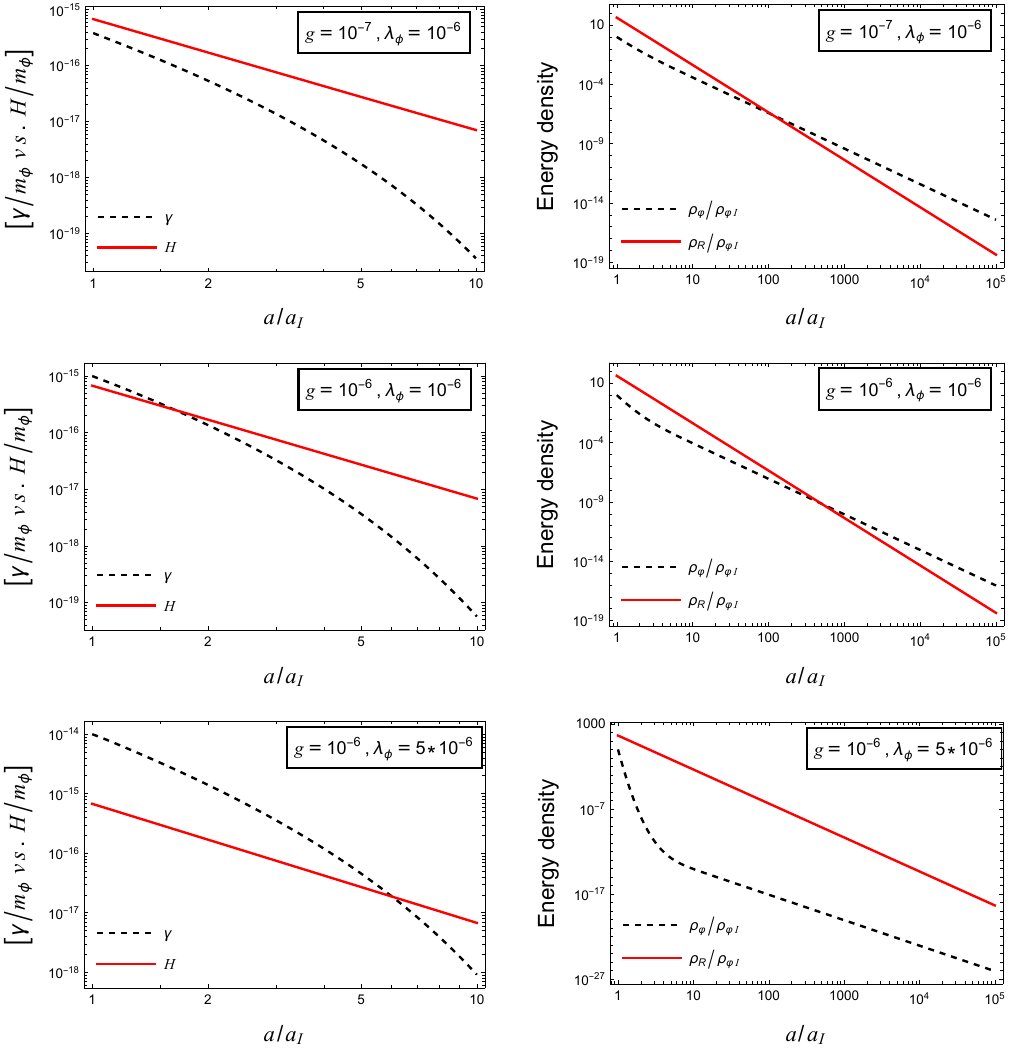}
    \caption{$\gamma$ and the Hubble parameter $H$ (the first column) and the energy densities (the second column) for different couplings but fixed initial condition $\bar{\varphi}_i=5m_{\phi}$, and $T_i=2m_{\phi}$.}
    \label{fig:numericalsolGrid3}
\end{figure}

\paragraph{Different couplings.} Again we fix the initial condition to $\bar{\varphi}_i=5m_\phi$ and $T_i=2m_\phi$. Fig.~\ref{fig:Comvingenergycomparing3} presents the obtained comoving energy densities of the condensate for $\{g=10^{-7},\lambda_\phi=10^{-6}\}$, $\{g=10^{-6},\lambda_\phi=10^{-6}\}$ and $\{g=10^{-6},\lambda_\phi=5*10^{-6}\}$. The ``freeze-out'' behaviour can be explained by the evolution of $\gamma$ and $H$ shown in Fig.~\ref{fig:numericalsolGrid3}.

\begin{figure}[H]
    \centering
    \includegraphics[scale=0.66]{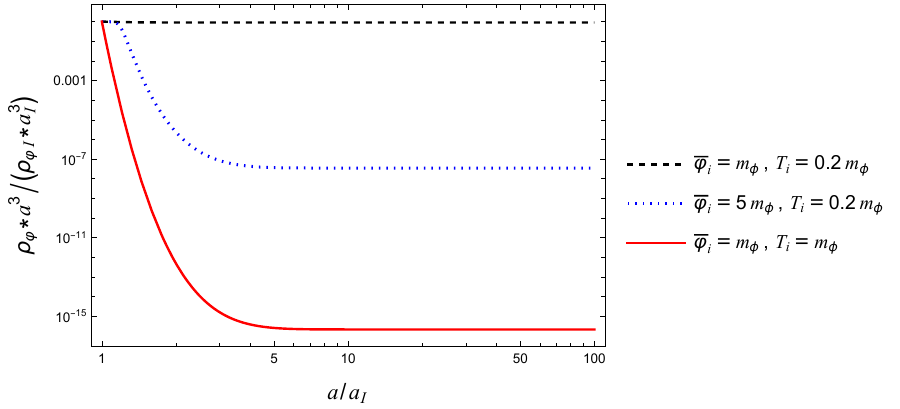}
    \caption{Comoving energy densities of the condensate for various initial conditions with fixed couplings $g=10^{-6}$ and $\lambda_{\phi}=10^{-5}$.}
    \label{fig:Comvingenergycomparing4}
\end{figure}

\begin{figure}[H]
    \centering
    \includegraphics[scale=0.73]{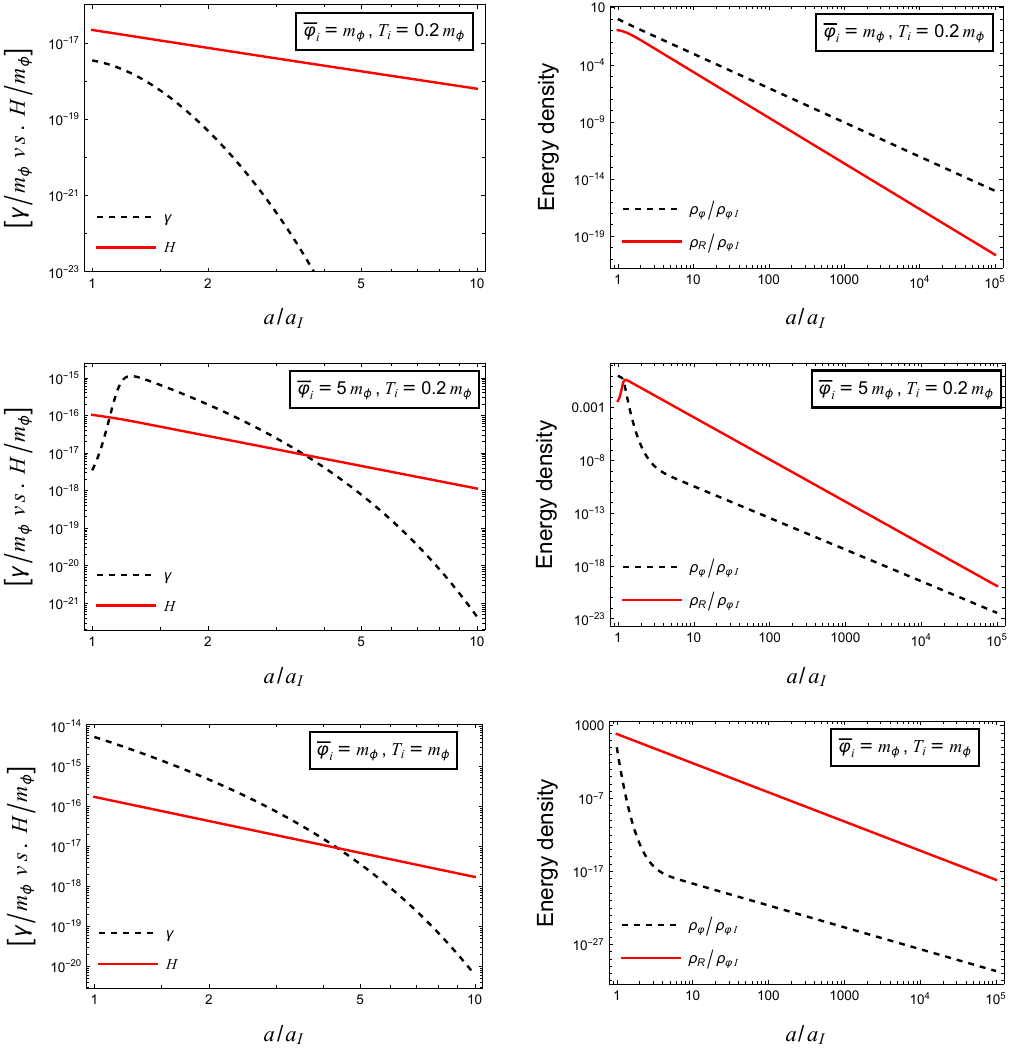}
    \caption{Numerical solutions for coupled equations \eqref{eq:eom_energy_density}, \eqref{eq:eom_energy_density_radiation}, and \eqref{eq:eom_Friedmann}. The first column shows the evolution of $\gamma$ and the Hubble parameter $H$. The second column shows the evolution of energy densities.}
    \label{fig:numericalsolGrid4}
\end{figure}

\begin{figure}[ht]
    \centering
    \includegraphics[scale=0.75]{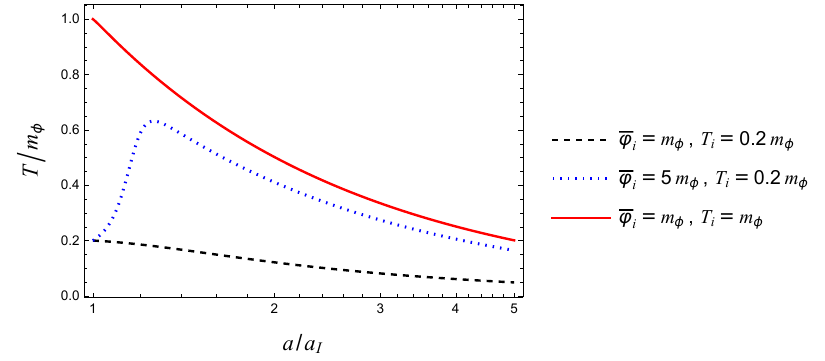}
    \caption{Evolution of radiation temperature for various initial conditions with fixed couplings $g=10^{-6}$ and $\lambda_{\phi}=10^{-5}$.}
    \label{fig:Temvs4}
\end{figure}

\paragraph{Different initial conditions.}
Figs.~\ref{fig:Comvingenergycomparing4} and~\ref{fig:numericalsolGrid4} present the numerical solutions for different initial conditions, $\{\bar{\varphi}_i=m_\phi,T_i=0.2 m_\phi\}$, $\{\bar{\varphi}_i=5m_\phi,T_i=0.2 m_\phi\}$, and  $\{\bar{\varphi}_i=m_\phi,T_i=m_\phi\}$, but fixed couplings $g=10^{-6}$, $\lambda_\phi=10^{-5}$. Similar to the case $m_\phi> m_\chi$, a larger $T_i$ would directly increase $\gamma$ at the beginning, hence improving the decay efficiency. However, we still observe an enhancement of the decay rate with a larger initial oscillation amplitude even though the $\sigma$ channel is closed. This is again due to reheating effects from particle production but via the $\gamma$ channels now. In the middle panel of Fig.~\ref{fig:numericalsolGrid4}, one can see an early increase of $\gamma$ (reheating), which is followed by a slower decrease. The evolution of the temperature in three cases is shown in Fig.~\ref{fig:Temvs4}.

\section{Conclusion and discussion}
\label{sec:Conc}

In this work, we have studied the evolution of homogeneous $\mathbb{Z}_2$-symmetric scalar condensates in the early Universe in the perturbative regime where the oscillation amplitude is sufficiently small so that dissipation occurs via perturbative processes. This also indicates that the oscillation is quasi-harmonic. Otherwise, self-interactions of the scalar field would lead to significant parametric resonance. This perturbative regime necessarily comprises the late stage of the condensate evolution and determines the fate of the condensate. Our analysis is based on the CTP formalism which naturally accommodates quantum statistical effects of the plasma
as well as the backreaction effects from particle production. 

We have derived, in the perturbative regime, the coupled coarse-grained EoMs for the condensate energy, radiation, and spacetime from the original non-local condensate EoM obtained from nonequilibrium quantum field theory. This derivation makes use of multiple-scale analysis that was first introduced in Ref.~\cite{Ai:2021gtg} to study the dissipation of scalar background fields. In this work, we have extended the analysis in Ref.~\cite{Ai:2021gtg} to take into account both the expanding spacetime and the time dependence in the temperature. The coarse-grained EoMs depend on two microscopic quantities $\gamma$ and $\sigma$, defined in Eq.~\eqref{eq:important_quantities}, which have an interpretation in terms of particle production. We have presented a very detailed computation of these two quantities. The analytic result for $\gamma$ contributed from the sunset diagram generated by the $\varphi\phi\chi^2$ interaction (the second diagram from Eq.~\eqref{eq:diag-self-energy}) is a new result. The coupled EoMs have been solved numerically.

The key finding in this work is that a scalar condensate with $\mathbb{Z}_2$ symmetry cannot transfer all its energy to the plasma. The condensate decay rate is given in Eq.~\eqref{eq:decay-rate} and depends on both the temperature and the oscillation amplitude (through $\rho_\varphi$). As a result, the condensate decay rate always decreases faster than the Hubble parameter at late times (see the analysis in Sec.~\ref{sec:freeze-out}). Therefore, similar to the freeze-out process for particles, the condensate comoving energy density approaches a constant once the condensate decay rate drops below the Hubble parameter. A homogeneous $\mathbb{Z}_2$-symmetric scalar condensate must have a nonvanishing relic density. This should put additional constraints on $\mathbb{Z}_2$-symmetric scalar fields as DM. 

In our model, we did not consider spontaneous symmetry breaking. If $m_\chi^2<0$, as it would be when $\chi$ is identified as the Higgs, there may be spontaneous symmetry breaking during the evolution of the condensate. If such spontaneous symmetry breaking occurs, we should have a cubic interaction term $v\Phi^2\chi$ where $v$ is the symmetry broken value of the $\chi$ field. In the presence of the condensate, one has the vertex $v \varphi \phi \chi$ which leads to the following additional diagram in the effective action that contributes to the condensate decay,
\begin{align}
    \begin{tikzpicture}[baseline={-0.025cm*height("$=$")}]
\draw[draw=black,thick] (-0.5,0) circle (0.15) ;
\draw[draw=black,thick] (-0.5-0.12,0+0.12) -- (-0.5+0.12,0-0.12);
\draw[draw=black,thick] (-0.5-0.12,0-0.12) -- (-0.5+0.12,0+0.12);
\draw[draw=black,thick] (-0.35,0) -- (0.25,0);
\draw[draw=black,dashed] (0.25,0) arc (180:0:0.5);
\draw[draw=black,thick] (0.25,0) -- (1.25,0);
\draw[draw=black,thick] (1.25,0) -- (1.85,0);
\draw[draw=black,thick] (2,0) circle (0.15) ;
\draw[draw=black,thick] (2-0.12,0+0.12) -- (2+0.12,0-0.12);
\draw[draw=black,thick] (2-0.12,0-0.12) -- (2+0.12,0+0.12);
\end{tikzpicture}\,.
\end{align}
This diagram would contribute to $\gamma$ because it induces a linear term in the EoM of $\varphi$. By the cutting rules~\cite{Cutkosky:1960sp,Weldon:1983jn,Kobes:1985kc,Kobes:1986za,Landshoff:1996ta,Gelis:1997zv,Bedaque:1996af}, this diagram describes the scattering process $\varphi\phi \leftrightarrow \chi$ [the other processes ($\varphi\leftrightarrow \phi \chi$, $\varphi\chi\leftrightarrow \phi$) cannot satisfy the on-shell condition]. The process $\varphi\phi\leftrightarrow \chi$ again belongs to Landau damping and is suppressed at low temperatures. Therefore, $\gamma$ would still drop below $H$ at later times. The key finding discussed above is thus not affected by a spontaneous symmetry breaking from the $\chi$ field but the actual evolution of the condensate may be quite different. 

We emphasize that in this work, $m_\phi^2>0$ is assumed, i.e., the $\mathbb{Z}_2$ symmetry of $\Phi$ is not spontaneously broken. Different from the portal field $\chi$, a negative $m^2_\phi$ would completely change our conclusion. This is because such a negative squared-mass of $\Phi$ would induce symmetric broken minima for field $\Phi$, say $v_\phi$ and $-v_\phi$. Then the condensate would oscillate about one of them. In this case, one has to substitute $\Phi=v_{\phi}+\varphi+\phi$ (assuming oscillations about the minimum $v_\phi$ without loss of generality) into the Lagrangian and now one has an interaction term $v_\phi\varphi \chi^2$. This term gives the decay $\varphi\rightarrow \chi\chi$ (provided $M_\phi>2M_\chi$) which is not suppressed at low temperatures and meanwhile would contribute to the non-local term linear in $\varphi$ in the EoM~\eqref{eq:condensate eom}. This means that such a decay channel is suppressed neither by low temperature nor by small oscillation amplitude at late times. Therefore, it would lead to a complete decay of the condensate. 

At last, since our analysis is based on the one and two-loop diagrams given in Eqs.~\eqref{eq:diagrams}, one may wonder whether higher loop processes would spoil our conclusion or not. In particular, it is known that in computing correlation functions in the zero-momentum, zero-frequency limit, e.g., in studying hydrodynamic transport coefficients~\cite{Jeon:1994if} or damping coefficients in the slow-roll regime or at very high temperatures~\cite{Bodeker:2022ihg}, resummation of an infinity set of diagrams is required. This is because a near on-shell singularity appears whenever there is a product of two equal four-momentum propagators. However, such near on-shell singularity only appears in the zero-momentum, zero-frequency limit of the correlation functions. Since we do not assume $T\gg M_\phi$ (high temperature) nor $\Gamma_\varphi\gg M_\phi$ (over-damped or slow-roll), the self-energies in our problem are always evaluated at finite frequencies related to $M_\phi$ depending on the powers of $\varphi$ associated with the external vertices (e.g., $\gamma$ and $\sigma$ in Eq.~\eqref{eq:important_quantities} are evaluated at $M_\phi$ and $2M_\phi$, respectively). Therefore, the near on-shell singularity does not appear in our case. For scalar field theory at finite temperature, there is the famous problem with infrared divergences, essentially related to the Matsubara zero mode~\cite{Laine:2016hma}. Such an infrared problem is cured if one does a daisy resummation. This has already been done with the thermal corrections being included in the propagators used in this work.

\section*{Acknowledgments}

WYA is supported by the UK Engineering and Physical Sciences Research Council (EPSRC), under Research Grant No. EP/V002821/1. ZLW is supported by the  Natural Science Foundation of Jiangsu Province (BK20220642). We are grateful to John Ellis and Jian Wang for helpful discussions.

\newpage

\begin{appendix}
\renewcommand{\theequation}{\Alph{section}\arabic{equation}}

\section{Closed-Time-Path formalism}
\label{app:CTP}

In this section, we give a brief introduction to the CTP formalism, with the main purpose of setting up the notation. For a more detailed introduction, we refer to Refs.~\cite{Chou:1984es,Calzetta:1986cq,Berges:2004yj}.\footnote{See Ref.~\cite{Lundberg:2020mwu} for a recent review.} The CTP formalism has been widely applied to the studies of baryogenesis, especially in the form of leptogenesis (see, e.g., Refs.~\cite{Buchmuller:2000nd,Prokopec:2003pj,Prokopec:2004ic,Konstandin:2004gy,Lee:2004we,Konstandin:2005cd,DeSimone:2007gkc,Garny:2009rv,Garny:2009qn,Anisimov:2010aq,Beneke:2010wd,Beneke:2010dz,Anisimov:2010dk,BhupalDev:2014oar,Postma:2022dbr}).

\begin{figure}[ht]
    \centering
    \includegraphics[scale=0.3]{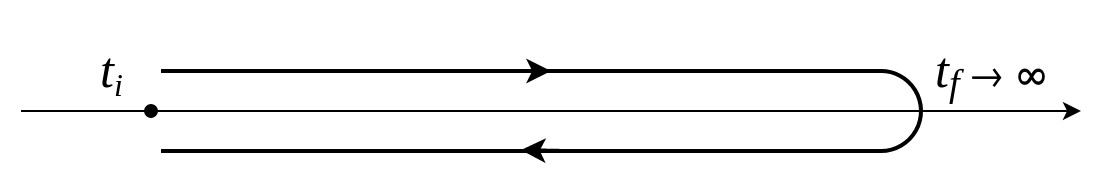}
    \caption{The Keldysh contour $\mathcal{C}$ for the generating functional in the CTP formalism. In the plot, the forward and backward time contours are slightly shifted off the real line only for illustration; both contours should be understood as lying exactly on the real line. }
    \label{fig:keldyshcontour}
\end{figure}

Out-of-equilibrium dynamics is an initial-value problem with the initial conditions given by a density matrix at a given time, $\rho_D(t_i)$. The expectation value of an observable $\mathcal{O}$ at time $t$ is given by
\begin{align}
\label{eq:expetation}
    \langle \mathcal{O}(t) \rangle ={\rm Tr} \left\{\rho_D(t_i)\mathcal{O}(t)\right\}\,,
\end{align}
where $\mathcal{O}(t)=\exp(\i H(t-t_i))\mathcal{O}(t_i)\exp(-\i H(t-t_i))$. The operator is sandwiched between the same state. For this reason, the CTP formalism is also called in-in formalism.  Viewed from the Schr\"{o}dinger picture, the state evolves first forward in time from $t_i$ to $t$ and then backward in time to $t_i$.\footnote{In the conventional zero-temperature quantum field theory studying scattering amplitudes, $t_i=-\infty$ and the backward evolution from $t$ to $-\infty$ can be reversed to that from $t$ to $\infty$ by inserting a $S$-matrix, $S(\infty,-\infty)$, into the leftmost in the brackets in Eq.~\eqref{eq:expetation}, which induces only a total phase.} Expectation values can thus be obtained by a generating functional formulated on a closed time contour $\mathcal{C}$, as illustrated in Fig.~\ref{fig:keldyshcontour}.\footnote{If the density matrix at the initial time is thermal, it can be encoded with an additional contour segment of length $1/T$ to the negative imaginary direction of time, glued to the end point of the lower branch of $\C$ at $t_i$.} For the quantity given in Eq.~\eqref{eq:expetation}, $t_f=t$, but one can also choose any $t_f\geq t$ by inserting an operator $\exp(\i H(t-t_f))\exp(-\i H (t_f-t))=\mathbf{1}$ on the left of $\O(t)$ in Eq.~\eqref{eq:expetation}. In particular, one can choose $t_f\rightarrow\infty$ so that the generating functional is independent of the quantities to be computed.

All the information in a quantum field system is encoded in its correlation functions. For most purposes, it is sufficient to study the one- and two-point functions. For the scalar field $\Phi$, there are two independent connected two-point functions
\begin{subequations}
\label{eq:wigthman}
\begin{align}
   \langle \Phi(x_1)\Phi(x_2)\rangle_c &=\langle \phi(x_1)\phi(x_2)\rangle\equiv  \Delta_{\phi}^{>}(x_1,x_2)\,, \\
    \langle \Phi(x_2)\Phi(x_1)\rangle_c &=\langle \phi(x_2)\phi(x_1)\rangle\equiv \Delta_{\phi}^{<}(x_1,x_2)\,.
\end{align}
\end{subequations}
They are called {\it Wightman functions}.
Here and in what follows, analogous expressions also apply for the $\chi$ field and corresponding quantities are denoted similarly with the subscript $\phi$ replaced by $\chi$. Since the generating functional is formulated on the closed time contour $\C$, we would have two-point functions time-ordered on $\C$, 
\begin{align}
\label{eq:Delta}
    \Delta^{\mathcal{C}}_{\phi}(x_1,x_2)=\theta_{\mathcal{C}}(t_1,t_2)\Delta_{\phi}^{>}(x_1,x_2)+\theta_{\mathcal{C}}(t_2,t_1)\Delta_{\phi}^{<}(x_1,x_2)\,.
\end{align}
To distinguish the times on the forward and backward contours in $\Delta^\C_\phi$, one can denote the forward time with a superscript ``$+$'' and the backward time ``$-$'': $x^+=\{t^+,\vec{x}\}$, $x^-=\{t^-,\vec{x}\}$. According to the $\C$-ordering, one would have
\begin{subequations}
\label{eq:C-odering-propagators}
\begin{align}
    \Delta^\mathcal{C}_\phi(x_1^+,x_2^-)&=\Delta _\phi^<(x_1,x_2)\,,\\ \Delta^\mathcal{C}_\phi(x_1^-,x_2^+)&=\Delta _\phi^>(x_1,x_2)\,,\\ \Delta^\mathcal{C}_\phi(x_1^+,x_2^+)&=\langle T\phi(x_1)\phi(x_2)\rangle\,, \\ \Delta^\mathcal{C}_\phi(x_1^-,x_2^-)&=\langle \overline{T}\phi(x_1) \phi(x_2)\rangle\,,
\end{align}
\end{subequations}
where $T$ and $\overline{T}$ denote the normal time-ordering and anti-time-ordering operators, respectively.

Compared to the closed time variable, a more convenient notation is the ``double-field'' formulation where one distinguishes the fields on the forward contour from those on the backward contour while using a common time variable. For example, one denotes $\Phi(x^+)$ as $\Phi^+(x)$, $\Phi(x^-)$ as $\Phi^-(x)$, and similarly for $\chi$. Then the action integrated over the contour $\C$ can be written as
\begin{align}
    S_\C[\Phi,\chi]=S[\Phi^+,\chi^+]-S[\Phi^-,\chi^-]\,,
\end{align}
where the minus sign is due to the backward direction in the integral on the backward contour in $S_\C[\Phi, \chi]$. In this notation, Eq.~\eqref{eq:C-odering-propagators} can be rewritten as
\begin{subequations}
\begin{align}
    \Delta^{+-}_\phi(x_1,x_2)&=\Delta _\phi^<(x_1,x_2)\,,\\ \Delta^{-+}_\phi(x_1,x_2)&=\Delta _\phi^>(x_1,x_2)\,,\\ \Delta^{++}_\phi(x_1,x_2)&=\langle T\phi(x_1)\phi(x_2)\rangle\,, \\ \Delta^{--}_\phi(x_1,x_2)&=\langle \overline{T}\phi(x_1) \phi(x_2)\rangle\,.
\end{align}
\end{subequations}
The indices ``$\pm$'' are called the Schwinger-Keldysh polarity indices.

Usually, it is convenient to introduce the spectral and statistical correlation functions defined as, respectively, 
\begin{subequations}
\label{eq:DeltaplusDeltaminus}
\begin{align}
 \Delta_{\phi}^-(x_1,x_2)&=\i\langle [\Phi(x_1),\Phi(x_2)]\rangle_c=\i\left(\Delta_{\phi}^{-+}(x_1,x_2)-\Delta_{\phi}^{+-}(x_1,x_2)\right)\,,\\
\Delta_{\phi}^+(x_1,x_2)&=\frac{1}{2}\langle\{\Phi(x_1),\Phi(x_2)\}\rangle_c=\frac{1}{2}\left(\Delta_{\phi}^{-+}(x_1,x_2)+\Delta_{\phi}^{+-}(x_1,x_2)\right)\,.
\end{align}
\end{subequations}
The spectral correlation function encodes the information about the spectrum of the theory and the statistical correlation function gives the information about occupation numbers of different modes for the fluctuations.

The Schwinger-Dyson equations for the one- and two-point functions can be derived from the one-particle-irreducible (1PI) or two-particle-irreducible (2PI) effective actions~\cite{Jackiw:1974cv,Cornwall:1974vz}.\footnote{The $n$PI effective actions are very powerful and can also be used to study higher-order corrections to false vacuum decay~\cite{Baacke:2006kv,Garbrecht:2015oea,Garbrecht:2015yza,Ai:2018guc,Ai:2020sru,Ai:2023yce}.} Here we use the 2PI effective action. If the effective action is not truncated, the 1PI and 2PI give the same EoMs. For a given order in the loop expansion, the 2PI effective action includes resummation in the propagators compared to the 1PI effective action. The definition of the 2PI effective action is based on the generating functional with local and non-local sources,
\begin{align}
   & Z[J_\Phi,J_\chi,K_\Phi,K_\chi]
       =\int\mathcal{D}\Phi^+\mathcal{D}\Phi^-\mathcal{D}\chi^+\mathcal{D}\chi^-\,\notag\\ &\times\exp\left(\i\left[S[\Phi^+,\chi^+]-S[\Phi^-,\chi^-]+ J^A_{\Phi, x}\Phi^A_x+ J^A_{\chi,x} \chi^A_x+\frac{1}{2}\Phi^A_x K^{AB}_{\Phi, xy}\Phi^B_y+\frac{1}{2} \chi^A_xK^{AB}_{\chi,xy}\chi^B_y\right]\right).
\end{align}
where $A,B=\pm$, and we have used the DeWitt notation
\begin{subequations}
\begin{align}
    &J^A_{\Phi,x}\Phi^A_x=\int\d^4 x\, J^A_{\Phi}(x)\Phi^A(x)\,,\\
    &\Phi^A_x K^{AB}_{\Phi,xy}\Phi^B_y=\int\d^4 x\,\d^4 y\, \Phi^A(x)K^{AB}_{\Phi}(x,y)\Phi^B(y)\,.
\end{align}
\end{subequations}
The Einstein summation convention is understood for repeated indices. One can easily check that
\begin{subequations}
\begin{align}
    \frac{\delta\ln Z[J_\Phi,J_\chi,K_\Phi,K_\chi]}{\i\delta  J^A_\Phi(x)}&=\langle\Phi^A(x)\rangle\equiv \varphi^A(x)\,,\\
    \frac{\delta \ln Z[J_\Phi,J_\chi,K_\Phi,K_\chi]}{\i\delta K^{AB}_{\Phi,xy}}&=\frac{1}{2}\left(\Delta^{AB}_\phi(x_1,x_2)+\varphi^A(x)\varphi^B(y)\right)\,,
\end{align}
\end{subequations}
and similarly for quantities involving $\chi$. Note that in the above equations, the one- and two-point functions on the RHS are defined with corresponding sources. 
The 2PI effective action is then defined as the Legendre transform,
\begin{align}
    \Gamma_{\rm 2PI}[\varphi,\Bar{\chi},\Delta_\phi,\Delta_\chi]=&-\i\, \ln Z[J_\Phi,J_\chi,K_\Phi,K_\chi]-J^A_{\Phi,x}\varphi^A_x- J^A_{\chi,x}\Bar{\chi}^A_x\,,\\
    &-\frac{1}{2}K^{AB}_{\Phi,xy}(\varphi^A_x\varphi^B_y+\Delta^{AB}_{\phi,xy})-\frac{1}{2}K^{AB}_{\chi,xy}(\Bar{\chi}_x^A\Bar{\chi}_y^B+\Delta_{\chi,xy}^{AB})\,,
\end{align}
where $\bar{\chi}^A=\langle \chi^A\rangle$. The Schwinger-Dyson equations for the one- and two-point functions read
\begin{subequations}
\begin{align}
\label{eq:eom-varphi}
    &\frac{\delta\Gamma_{\rm 2PI}[\varphi,\Bar{\chi},\Delta_\phi,\Delta_\chi]}{\delta\varphi^A(x)}=-J^A_\Phi(x)-K^{AB}_{\Phi,xy}\varphi^B_y\,,\\
    \label{eq:eom-Delta-phi}
    &\frac{\delta\Gamma_{\rm 2PI}[\varphi,\Bar{\chi},\Delta_\phi,\Delta_\chi]}{\delta\Delta^{AB}_{\phi/\chi}(x,y)}=-\frac{1}{2}K^{AB}_{\Phi/\chi}(x,y)\,.
\end{align}
\end{subequations}
Taking all the $J$ and $K$ to be vanishing, we obtain the EoMs in the absence of external sources
\begin{align}
     &\left.\frac{\delta\Gamma_{\rm 2PI}[\varphi,\Bar{\chi},\Delta_\phi,\Delta_\chi]}{\delta\varphi^+(x)}\right|_{\varphi^-=\varphi^+=\varphi}=0\,,\\
    \label{eq:eom-Delta-phi2}
    &\left.\frac{\delta\Gamma_{\rm 2PI}[\varphi,\Bar{\chi},\Delta_\phi,\Delta_\chi]}{\delta\Delta^{AB}_{\phi/\chi}(x,y)}\right|_{\varphi^-=\varphi^+=\varphi}=0\,.
\end{align}

In the Feynman-diagram representation of the $2$PI effective action, we now have two types of vertices: the + and - types. Depending on the vertices, one would have four different types of propagators: $\Delta_{\phi/\chi}^{++}$, $\Delta_{\phi/\chi}^{+-}$, $\Delta_{\phi/\chi}^{-+}$, and $\Delta_{\phi/\chi}^{--}$. For example, a propagator connecting two + type vertices is of type $\Delta^{++}$, etc. Solving first on-shell for the two-point functions and assuming that $\Bar{\chi}=0$, one gets $\Delta_\phi[\varphi]$ and $\Delta_\chi[\varphi]$ as functionals of the condensate $\varphi$. Plugging them back into the 2PI effective action gives the EoM for the condensate
\begin{equation}
\label{eq:eom-general}
\left.\frac{ \delta \Gamma_\text{2PI}[\varphi,\Delta_\phi,\Delta_\chi] }{ \delta \varphi(x) }\right|_{\Delta_\phi=\Delta_\phi[\varphi],\,\Delta_\chi=\Delta_\chi[\varphi]}= 0 \, .
\end{equation}
Arriving here, one can follow Section 2 of Ref.~\cite{Ai:2021gtg} to perform the small-field expansion of the on-shell two-point functions $\Delta_\phi[\varphi]$ and $\Delta_\chi[\varphi]$. To the lowest order, they are independent of $\varphi$ and become the free thermal equilibrium propagators. Properly truncating the effective action, one ends up with Eq.~\eqref{eq:condensate eom}. 

\section{Self-energies}
\label{app:self-energy_and_four-vertex}

\begin{figure}[ht]
    \centering
    \includegraphics[scale=1]{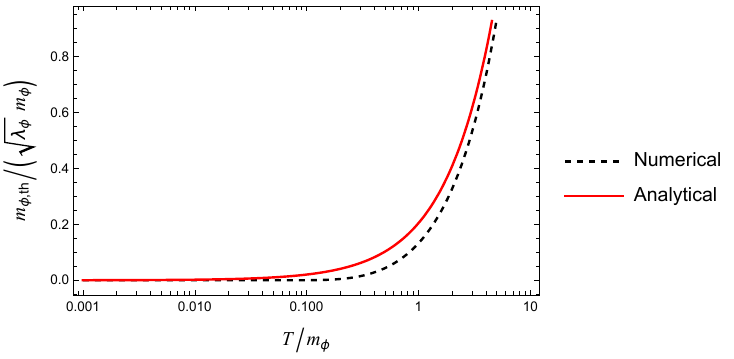}
    \caption{$m_{\phi,\rm th}/\sqrt{\lambda_\phi}m_\phi$ as a function of $T/m_\phi$ for the self-interactions. The red solid curve is obtained from the analytical expression in the high-temperature approximation, i.e., \eqref{eq:thermal-masses-highTa} with vanishing $g$. The black dashed curve is obtained from numerically integrating \eqref{eq:themrmal-masses-ex-a} with the simplification \eqref{eq:replacement}. Note that the ratio $m_{\phi,\rm th}/m_\phi$ is further suppressed by a factor of $\sqrt{\lambda_\phi}$ compared with what has been plotted. This plot shows that, away from the high-temperature regime, the thermal corrections to the masses could be simply neglected for perturbatively small couplings.}
    \label{fig:thermalmass}
\end{figure}

The Feynman-diagram rules for the free thermal equilibrium propagators with thermal corrections to the masses taken into account are
\begin{equation}
\begin{tikzpicture}[baseline={0cm-0.5*height("$=$")}]
\draw[thick] (-0.6,0) -- (0.6,0) ;
\filldraw (-0.6,0) circle (1.5pt) node[below] {$\scriptstyle(x,A)$} ;
\filldraw (0.6,0) circle (1.5pt) node[below] {$\scriptstyle(x',B)$} ;
\end{tikzpicture}
\equiv G_\phi^{AB}(x-x') \, ,
\qquad \qquad
\begin{tikzpicture}[baseline={0cm-0.5*height("$=$")}]
\draw[thick,densely dashed] (-0.6,0) -- (0.6,0) ;
\filldraw (-0.6,0) circle (1.5pt) node[below] {$\scriptstyle(x,A)$} ;
\filldraw (0.6,0) circle (1.5pt) node[below] {$\scriptstyle(x',B)$} ;
\end{tikzpicture}
\equiv
G_\chi^{AB}(x-x') \, ,
\label{thermal propagators}
\end{equation}
where
\begin{subequations}
\label{eq:propagators}
\begin{align}
G_{\phi/\chi}^{++}(x-x')
=&\int\frac{\d^4 k}{(2\pi)^4}\,
\e^{-\i k(x-x')}
\left[\frac{\i}{k^2 -M_{\phi/\chi}^2+\i\varepsilon}+2\pi f_{\rm B}(|k_0|)\delta\left( k^2 - M_{\phi/\chi}^2 \right) \right] \, ,\label{D++ thermal}\\
G_{\phi/\chi}^{+-}(x- x')
=&\int\frac{\d^4 k}{(2\pi)^4 } \,
\e^{-\i k (x-x')}\left[2\pi f_{\rm B}(k_0) \sign(k_0) \,\delta\left(k^2 - M_{\phi/\chi}^2 \right) \right] \, ,\label{D+- thermal}
\end{align}
\end{subequations}
with the remaining components being complex
conjugates,~$G^{--}_{\phi/\chi}(x-x') = \left[ G^{++}_{\phi/\chi}(x-x') \right]^*$, $G^{-+}_{\phi/\chi}(x-x') = \left[ G^{+-}_{\phi/\chi}(x-x') \right]^*$.
Here, the thermally corrected masses are
\begin{subequations}
\begin{align}
M_\phi^2 &=m_\phi^2-\frac{\i \lambda_\phi}{2}
\begin{tikzpicture}[baseline={0cm-0.5*height("$=$")}]
\draw[thick] (0,0) circle (0.5) ;
\filldraw (0.5,0) circle (1.5pt) node[right] {$\scriptstyle (x,+)$} ;
\end{tikzpicture}
-\frac{\i g}{2}
\begin{tikzpicture}[baseline={0cm-0.5*height("$=$")}]
\draw[thick, dashed] (0,0) circle (0.5) ;
\filldraw (0.5,0) circle (1.5pt) node[right] {$\scriptstyle (x,+)$} ;
\end{tikzpicture}
\, ,\label{thermal Mphi}\\
M_\chi^2 & = m_\chi^2-\frac{\i \lambda_\chi}{2}
\begin{tikzpicture}[baseline={0cm-0.5*height("$=$")}]
\draw[thick,dashed] (0,0) circle (0.5) ;
\filldraw (0.5,0) circle (1.5pt) node[right] {$\scriptstyle (x,+)$} ;
\end{tikzpicture}
-\frac{
\i g}{2}
\begin{tikzpicture}[baseline={0cm-0.5*height("$=$")}]
\draw[thick] (0,0) circle (0.5) ;
\filldraw (0.5,0) circle (1.5pt) node[right] {$\scriptstyle (x,+)$} ;
\end{tikzpicture}+({\rm other\ diagrams\ with\ SM\ fields})
\, .
\label{thermal Mchi}
\end{align}
\end{subequations}
Writing $M_{\phi/\chi}^2=m_{\phi/\chi}^2+m_{\phi/\chi,\rm th}^2$, we then have~\cite{Bellac:2011kqa} 
\begin{subequations}
\label{eq:themrmal-masses-ex}
    \begin{align}
        &m^2_{\phi,\rm th}=\frac{\lambda_\phi}{2}\int\frac{\d^3\veck}{(2\pi)^3}\frac{f_{\rm B}(\omega_{\phi,\veck})}{\omega_{\phi,\veck}}+\frac{g}{2}\int\frac{\d^3\veck}{(2\pi)^3}\frac{f_{\rm B}(\omega_{\chi,\veck})}{\omega_{\chi,\veck}}\,,\label{eq:themrmal-masses-ex-a}\\
        &m^2_{\chi,\rm th}=\frac{\lambda_\chi}{2}\int\frac{\d^3\veck}{(2\pi)^3}\frac{f_{\rm B}(\omega_{\chi,\veck})}{\omega_{\chi,\veck}}+\frac{g}{2}\int\frac{\d^3\veck}{(2\pi)^3}\frac{f_{\rm B}(\omega_{\phi,\veck})}{\omega_{\phi,\veck}}+({\rm contributions\ from\ SM\ fields})\,,
    \end{align}
\end{subequations}
where $\omega_{\phi/\chi,\veck}=\sqrt{M^2_{\phi/\chi} +\veck^2}$ and we have subtracted the divergent $T=0$ part. Note that the above equations are self-consistent equations as $m^2_{\phi/\chi,\rm th}$ appear on both sides. This is due to that resumed propagators are used in the 2PI effective action. In particular, with the assumed form for the propagators given in Eqs.~\eqref{eq:propagators}, the famous daisy resummation has been taken into account. In the high-temperature limit $m_{\phi,\chi}\ll T$, we get the familiar results~\cite{Bellac:2011kqa}
\begin{subequations}
\label{eq:thermal-masses-highT}
\begin{align}
    m^2_{\phi,\rm th} &= (\lambda_\phi+g)\frac{T^2}{24}+\O(\lambda_\phi^{3/2})+\O(g^{3/2})\,,\label{eq:thermal-masses-highTa}\\
    m_{\chi,\rm th}^2 &=(\lambda_\chi+g)\frac{T^2}{24}+\O(\lambda_\chi^{3/2})+\O(g^{3/2})+\kappa^2 T^2\,,
\end{align}
\end{subequations}
where $\kappa^2 T^2$ is due to possible interactions between $\chi$ and the SM fields. We shall take $\kappa^2\approx 0.1$ in our numerical calculations, which corresponds to roughly the value of the thermal corrections to the Higgs mass in the SM. Beyond the high-temperature limit $m_{\phi,\chi}\ll T$, one in principle needs to solve Eqs.~\eqref{eq:themrmal-masses-ex} iteratively. At the leading order, we simply have Eqs.~\eqref{eq:themrmal-masses-ex} with the following replacements on the RHS
\begin{align}
\label{eq:replacement}
    \omega_{\phi,\veck}\rightarrow \sqrt{m^2_\phi+\veck^2}\,,\qquad \omega_{\chi,\veck}\rightarrow \sqrt{m^2_\chi+\veck^2}\,,
\end{align}
i.e., by neglecting the thermal corrections on the RHS. Taking the self-interaction as an example, in Fig.~\ref{fig:thermalmass} we plot $m_{\phi,\rm th}/(\sqrt{\lambda_\phi} m_\phi)$ as a function of $T/m_\phi$ with and without the high-temperature approximation. It shows that away from the high-temperature regime, i.e., for $T\lsim m_\phi$, the thermal corrections to the mass are small anyway compared with the zero-temperature mass $m_\phi$. Therefore, it is not material whether or not one uses the high-temperature approximation for the thermal corrections to the mass throughout. Note $m_{\phi,\rm th}/m_\phi$ is further suppressed by a factor of $\sqrt{\lambda_\phi}$ compared with what has been plotted in Fig.~\ref{fig:thermalmass}.

The various self-energies and proper four-vertex functions read
\begin{subequations}
\begin{align}
\label{PiDef}
\Pi^{AB}(x-x')
	&=
\
-
\,
\frac{\i (AB)\lambda_\phi^2}{6}
\,
\begin{tikzpicture}[baseline={0cm-0.5*height("$=$")}]
\draw[thick] (0,0) circle (0.5) ;
\draw[thick] (-0.5,0) -- (0.5,0) ;
\filldraw (-0.5,0) circle (1.5pt) node[left] {$\scriptstyle(x,A)$} ;
\filldraw (0.5,0) circle (1.5pt) node[right] {$\scriptstyle(x',B)$} ;
\end{tikzpicture}
\
-
\
\frac{\i (AB)g^2}{2}
\,
\begin{tikzpicture}[baseline={0cm-0.5*height("$=$")}]
\draw[thick,dashed] (0,0) circle (0.5) ;
\draw[thick] (-0.5,0) -- (0.5,0) ;
\filldraw[fill=black] (0.5,0) circle (1.5pt) node[right] {$\scriptstyle(x',B)$} ;
\filldraw[fill=black] (-0.5,0) circle (1.5pt) node[left] {$\scriptstyle(x,A)$} ;
\end{tikzpicture}
\,,\\ 
\label{VDef}
V^{AB}(x-x')
&=
\
-
\
\frac{3\i (AB) \lambda_\phi^2}{2}
\,
\begin{tikzpicture}[baseline={0cm-0.5*height("$=$")}]
\draw[thick] (0,0) circle (0.5) ;
\filldraw[fill=black] (0.5,0) circle (1.5pt) node[right] {$\scriptstyle(x',B)$} ;
\filldraw[fill=black] (-0.5,0) circle (1.5pt) node[left] {$\scriptstyle(x,A)$} ;
\end{tikzpicture}
\
-
\
\frac{3\i (AB) g^2}{2}
\,
\begin{tikzpicture}[baseline={0cm-0.5*height("$=$")}]
\draw[thick, dashed] (0,0) circle (0.5) ;
\filldraw[fill=black] (0.5,0) circle (1.5pt) node[right] {$\scriptstyle(x'\!,B)$} ;
\filldraw[fill=black] (-0.5,0) circle (1.5pt) node[left] {$\scriptstyle(x,A)$} ;
\end{tikzpicture}
\,.
\end{align}
\end{subequations}
The retarded self-energy and retarded proper four-vertex function are then defined as
\begin{subequations}
\label{piR+vR def}
\begin{align}
&
\Pi^{\rm R}(x-x')
    = \Pi^{++}(x-x') + \Pi^{+-}(x-x')
	= \theta(t-t')
	    \left[ - \Pi^{-+}(x-x')
	        + \Pi^{+-}(x-x')  \right] \, ,
\label{piR def}
\\
&
V^{\rm R}(x-x')
    = V^{++}(x-x') + V^{+-}(x-x')
	= \theta(t-t')
	    \left[ - V^{-+}(x-x')
	        + V^{+-}(x-x')  \right] \, .
\label{vR def}
\end{align}
\end{subequations}
And $\pi^{\rm R}$ and $v^{\rm R}$ are defined as
\begin{equation}
\pi^{\rm R}(t-t')
	= \int \d^{3}x \, \Pi^{\rm R}(x-x') \, ,
\qquad
v^{\rm R}(t-t')
	= \int \d^{3}x \, V^{\rm R}(x-x') \, .
\end{equation}
Important quantities that would appear in the solutions for the condensate evolution are Fourier transforms of the retarded self-energy and proper four-vertex function, which are defined as
\begin{equation}
\label{PiTildeDef2}
\widetilde{\pi}^{\rm R}(\omega) = \int_{-\infty}^{\infty} \,
	\d t' \, \e^{\i\omega(t-t')} \, \pi^{\rm R}(t-t') \,,
\qquad \quad
\widetilde{v}^{\rm R}(\omega) = \int_{-\infty}^{\infty} \,
	\d t' \, \e^{\i\omega(t-t')} \, v^{\rm R}(t-t') \, .
\end{equation}
Note that 
\begin{align}
    \widetilde{\pi}^{\rm R}(\omega)=\widetilde{\Pi}^{\rm R}(\omega,\vec{p}=0)\,,\qquad \quad \widetilde{v}^{\rm R}(\omega)=\widetilde{V}^{\rm R}(\omega,\vec{p}=0)\,,
\end{align}
where $\omega=p_0$ and $\widetilde{\Pi}^{\rm R}$ and $\widetilde{V}^{\rm R}$ are Fourier transforms of $\Pi^{\rm R}$ and $V^{\rm R}$, respectively.

In terms of the free thermal propagators, the retarded self-energy and proper four-vertex function in position space read 
\begin{subequations}
\label{piR+vR exp}
\begin{align}
\Pi^{\rm R}(x-x')\equiv &\Pi_\phi^{\rm R}(x-x')+\Pi_\chi^{\rm R}(x-x')
    = -\frac{\i \lambda_\phi^2}{6}\left[\left(G^{++}_\phi(x-x')\right)^3-\left(G_\phi^{+-}(x-x')\right)^3\right]\notag\\
    &\quad -\frac{\i g^2}{2}\left[\left(G_\chi^{++}(x-x')\right)^2 G_\phi^{++}(x-x')-\left(G_\chi^{+-}(x-x')\right)^2 G_\phi^{+-}(x-x')\right]\,,
\label{piR exp}
\\
V^{\rm R}(x-x')\equiv & V_\phi^{\rm R}(x-x')+V_\chi^{\rm R}(x-x')
    =  -\frac{3\i \lambda_\phi^2}{2}\left[\left(G_\phi^{++}(x-x')\right)^2-\left(G_\phi^{+-}(x-x')\right)^2\right]\notag\\
	 &\quad -\frac{3\i g^2}{2}\left[\left(G_\chi^{++}(x-x')\right)^2-\left(G_\chi^{+-}(x-x')\right)^2\right]\,.
\label{vR exp}
\end{align}
\end{subequations}

\subsection{Proper four-vertex function}

As a warm-up, in this subsection we present the calculation of ${\rm Im} \widetilde{V}^{\rm R}(p)$ which gives $\sigma$ when evaluated at $p=(2M_\phi,\vec{0})$. In the next subsection, we will present the more complicated calculation for $\gamma$. 

\subsubsection{Summary of the results}

Let us first introduce a quantity $V^-$ that in position space reads 
\begin{align}
V^-(x-x')\equiv V_\phi^-(x-x')+V_\chi^-(x-x')&=\sum_{a=\phi,\chi} c_a\left[\left(G^{-+}_a(x-x')\right)^2-\left(G^{+-}_a(x-x')\right)^2\right]\notag\\
&=-2\i\sum_{a=\phi,\chi} c_a G_a^+(x-x') G_a^-(x-x')\,,
\label{eq:piDeltaplusDeltaminus}
\end{align}
where $c_\phi=-3\i \lambda_\phi^2/2$, $c_\chi=-3\i g^2/2$. The imaginary part of the retarded proper four-vertex function in momentum space, ${\rm Im}\widetilde{V}^{\rm R}(p)$, is related to $\widetilde{V}^-(p)$ through (see, e.g., Ref.~\cite{Anisimov:2008dz})
\begin{align}
\label{eq:PiRPiminus}
{\rm Im} \widetilde{V}^{\rm R}(p)=-\frac{\i}{2}\widetilde{V}^-(p)\,,
\end{align}
where $\widetilde{V}^-(p)$ is the Fourier transform of $V^-(x-x')$,
\begin{align}
\label{eq:piDeltaplusDeltaminus2}
\widetilde{V}^-(p)=\sum_{a=\phi,\chi} \widetilde{V}^-_{a}(p)=-2\i\sum_{a=\phi,\chi} c_a\int\frac{\d^4 k}{(2\pi)^4}\left(G^{+}_a(k) G^{-}_a(p-k)\right)\,.
\end{align}
The real part, ${\rm Re} \widetilde{V}^{\rm R}(p)$, is related to the imaginary part via the Kramers-Kronig relation
\begin{align}
{\rm Re}\widetilde{V}^{\rm R}(p)=\mathcal{P}\int\frac{\d\omega'}{\pi}\frac{{\rm Im}\widetilde{V}^{\rm R}(\omega',\vec{p})}{\omega'-p_0}\,.
\end{align}

Substituting the Kubo-Martin-Schwinger (KMS) relation
\begin{align}
\label{eq:KMS}
G_a^+(q)=-\i\left(\frac{1}{2}+f_{\rm B}(q_0)\right) G_a^-(q)
\end{align}
into Eq.~\eqref{eq:piDeltaplusDeltaminus2} and using the definition for the spectral function, $G^-_a(k)=\i \rho_{a,\vec{k}}(k_0)$, one gets
\begin{align}
\widetilde{V}^-_{a}(p) &= c_a\int\frac{\d^4 k_1}{(2\pi)^4}\frac{\d^4 k_2}{(2\pi)^4} (2\pi)^4 \delta^{(4)}(p-k_1-k_2) [1+f_{\rm B}(k_{1,0})+f_{\rm B}(k_{2,0})]\rho_{a,\vec{k}_1}(k_0)\rho_{a,\vec{k}_2}(k_{2,0})\notag\\
&\equiv c_a I_a(p)\,,
\end{align}
where we have defined the integral $I_a(p)$. Using Eq.~\eqref{eq:PiRPiminus}, one finally arrives at
\begin{align}
{\rm Im}\widetilde{V}^R_{a}(p)\equiv -\frac{\i c_a}{2} I_a(p)\,.
\end{align}

Substituting into the integral $I_a(p)$ the free thermal spectral function 
\begin{align}\label{eq:appB_rho_k of delta}
\rho_{a,\vec{k}}(k_0)=2\pi {\rm sign}(k_0)\delta(k_0^2-\omega_{a,\vec{k}}^2)\,,
\end{align}
where $\omega^2_{a,\vec{k}}=M_a^2+\vec{k}^2$, one obtains (see below for the detailed derivation)
\begin{align}
\label{eq:imaginary_fish1}
   I_a(p)&=\frac{\theta(p^2-4M_a^2)}{8\pi|\vec{p}|}\left[\left(\Omega_{a,-}-\Omega_{a,+}\right)+2T\log\left(\frac{f_{\rm B}(\Omega_{a,-})}{f_{\rm B}(\Omega_{a,+})}\right)\right]\notag\\
   &\qquad\qquad\qquad\qquad\qquad+\frac{\theta(-p^2)}{8\pi|\vec{p}|}\left[-2\Omega_{a,+}+2T\log\left|\frac{f_{\rm B}(\Omega_{a,-})}{f_{\rm B}(\Omega_{a,+})}\right|\right]\,,
\end{align} 
where 
\begin{align}
\label{eq:Omega+-}
   \Omega_{a,\pm} = \frac{p_0}{2}  \pm \frac{|{\vec p}|}{2}  \sqrt{1 - \frac{(2M_a)^2}{p^2}}\,. 
\end{align}
Taking $\vec p \to \vec 0$, one obtains
\begin{equation}
I_a(p_0, \vec{0}) = \frac{1}{ 8 \pi} \sqrt{1- \frac{4M_a^2}{p_0^2}} \theta\left(p_0^2 - 4 M_a^2\right)\left(1+2f_{\rm B}\left(\frac{p_0}{2}\right)\right)\,.
\end{equation}
Therefore 
\begin{align}
    {\rm Im}\widetilde{v}_a^{\rm R}(p_0)={\rm Im}\widetilde{V}^{\rm R}_{a}(p_0,\vec{0})=-\frac{\i c_a}{16\pi}\sqrt{1- \frac{4M_a^2}{p_0^2}} \theta\left(p_0^2 - 4 M_a^2\right)\left(1+2f_{\rm B}\left( \frac{p_0}{2}\right)\right)\,.
\end{align}
Taking further $p_0=2 M_\phi$ and substituting the above into the definition of $\sigma$, one obtains Eq.~\eqref{Im vR omega T}.

\subsubsection{Computation of \texorpdfstring{$I_a$}{TEXT}}

Below for clarity we suppress the subscript ``a''. Doing the integral over $k_{1,0}$ and $k_{2,0}$ and using 
\begin{align}
\label{eq:app-delta function-1}
\delta(f(x))=\sum_{\rm roots}\frac{\delta(x-x_i)}{|f'(x_i)|}\,,
\end{align}
we obtain
\begin{align}
    I(p)=\int\frac{\d^3 \vec{k}_1}{(2\pi)^3}\int\frac{\d^3\vec{k}_2}{(2\pi)^3}\frac{(2\pi)^4\delta^{(3)}(\vec{p}-\vec{k}_1-\vec{k}_2)}{4\omega_{\vec{k}_1}\omega_{\vec{k}_2}} \G(p_0,\omega_{\vec{k}_1},\omega_{\vec{k}_2})\,,
\end{align}
where
\begin{align}
    \G(p_0,\omega_{\vec{k}_1},\omega_{\vec{k}_2})=\sum_{\alpha_1,\alpha_2=\pm }(\alpha_1\alpha_2)\delta(p_0-\alpha_1\omega_{\vec{k}_1}-\alpha_2\omega_{\vec{k}_2})\left[1+f_{\rm B}(\alpha_1\omega_{\vec{k}_1})+f_{\rm B}(\alpha_2\omega_{\vec{k}_2})\right]\,.
\end{align}
Now using the property
\begin{align}
\label{eq:property}
    f_{\rm B}(-\omega_\vec{k})=-1-f_{\rm B}(\omega_\vec{k})\,,
\end{align}
one can rewrite $\G$ as
\begin{align}
\label{eq:G}
    \G(p_0,\omega_{\vec{k}_1},\omega_{\vec{k}_2})&=\delta(p_0-\omega_{\vec{k}_1}-\omega_{\vec{k}_2})\left\{\left[ 1+f_{\rm B}(\omega_{\vec{k}_1})\right]\left[1+f_{\rm B}(\omega_{\veck_2})\right]-f_{\rm B}(\omega_{\veck_1})f_{\rm B}(\omega_{\veck_2})\right\}\notag\\
    &+\delta(p_0+\omega_{\vec{k}_1}+\omega_{\vec{k}_2})\left\{f_{\rm B}(\omega_{\veck_1})f_{\rm B}(\omega_{\veck_2})-\left[ 1+f_{\rm B}(\omega_{\vec{k}_1})\right]\left[1+f_{\rm B}(\omega_{\veck_2})\right]\right\}\notag\\
    &+\delta(p_0+\omega_{\vec{k}_1}-\omega_{\vec{k}_2})\left\{f_{\rm B}(\omega_{\veck_1})\left[1+f_{\rm B}(\omega_{\veck_2})\right]-\left[ 1+f_{\rm B}(\omega_{\vec{k}_1})\right]f_{\rm B}(\omega_{\veck_2})\right\}\notag\\
    &+\delta(p_0-\omega_{\vec{k}_1}+\omega_{\vec{k}_2})\left\{\left[ 1+f_{\rm B}(\omega_{\vec{k}_1})\right]f_{\rm B}(\omega_{\veck_2})-f_{\rm B}(\omega_{\veck_1})\left[1+f_{\rm B}(\omega_{\veck_2})\right]\right\}\,.
\end{align}
It is easy to see the antisymmetry $p_0\leftrightarrow -p_0$ in $\G$ and also in the integral $I(p)$, $I(p_0,\vec{p})=-I(-p_0,\vecp)$. Without loss of generality, in the following calculations we take $p_0>0$.

Equation~\eqref{eq:G} has a very clear interpretation in terms of microscopic particles processes. Assuming that we consider $I_\phi(p)$ and $p_0>0$, the first line then describes the process $(\varphi\varphi)\leftrightarrow \phi_1\phi_2$ where $(\varphi\varphi)$ are the external condensate di-quanta while $\phi_1$, $\phi_2$ are two particles for the fluctuation field $\phi$. Parent particles or daughter particles can be distinguished by the factors from the Boltzmann distributions multiplied with the $\delta$-functions. For example, the factor $1+f_{\rm B}(\omega_{\veck_1})$ indicates that the particle $\phi_1$ is a daughter particle while $f_{\rm B}(\omega_{\vec{k}_1})$ indicates that $\phi_1$ is a parent particle. Therefore, when multiplied with $\left[ 1+f_{\rm B}(\omega_{\vec{k}_1})\right]\left[1+f_{\rm B}(\omega_{\veck_2})\right]$, the first $\delta$-function describes $(\varphi\varphi)\rightarrow \phi_1\phi_2$, while when multiplied with $f_{\rm B}(\omega_{\vec{k}_1})f_{\rm B}(\omega_{\veck_2})$ it describes $\phi_1\phi_2\rightarrow (\varphi\varphi)$. The quanta of the condensate are not associated with thermal distribution functions. Note that the ``inverse'' process makes a negative contribution to the imaginary part of the retarded proper four-vertex compared with the ``forward'' process, as expected. The $\delta$-function in the second line cannot be satisfied for $p_0>0$.  Similarly, one can interpret the third and fourth lines in terms of the processes $(\varphi\varphi)\phi_1\leftrightarrow \phi_2$ and $(\varphi\varphi)\phi_2\leftrightarrow \phi_1$. One can also interpret the terms in Eq.~\eqref{eq:G} for $p_0<0$. In this case, for the same term, the condensate di-quanta $(\varphi\varphi)$ must be moved from the LHS to RHS in the ``reaction'' equation. For example, the third line describes the process $\phi_1\leftrightarrow (\varphi\varphi)\phi_2$ for $p_0<0$. For the particular case $p=(2M_\phi,\vec{0})$, we are left with only the $\sigma$ channel given in Eq.~\eqref{eq:decaychannel-two}~\cite{Wang:2022mvv}.

The third line is equal to the fourth line with the exchange $\veck_1\leftrightarrow \veck_2$. Further using $1+f_{\rm B}(\omega_{\veck})=\exp(\beta \omega_\vec{k})f_{\rm B}(\omega_\vec{k})$ where $\beta=1/T$, one arrives at $I=I_1+I_2$ where
\begin{subequations}
\begin{align}
    &I_1(p)=\left(\e^{\beta p_0}-1\right)\int\frac{\d^3\veck_1}{(2\pi)^3}\int\frac{\d^3\veck_2}{(2\pi)^3}\frac{(2\pi)^4\delta^{(3)}(\vecp-\veck_1-\veck_2)}{4\omega_{\veck_1}\omega_{\veck_2}} \delta(p_0-\omega_{\veck_1}-\omega_{\veck_2})f_{\rm B}(\omega_{\veck_1})f_{\rm B}(\omega_{\veck_2})\,,\\
    &I_2(p)=2\left(\e^{\beta p_0}-1\right)\int\frac{\d^3\veck_1}{(2\pi)^3}\int\frac{\d^3\veck_2}{(2\pi)^3}\frac{(2\pi)^4\delta^{(3)}(\vecp-\veck_1-\veck_2)}{4\omega_{\veck_1}\omega_{\veck_2}} \delta(p_0+\omega_{\veck_1}-\omega_{\veck_2})\notag\\
    &\qquad\qquad\qquad\qquad\qquad\qquad\qquad\qquad\qquad\qquad\qquad\qquad\qquad\times \left[1+f_{\rm B}(\omega_{\veck_1})\right]f_{\rm B}(\omega_{\veck_2})\,.
\end{align}
\end{subequations}
Performing the integral over $\veck_2$ gives 
\begin{subequations}
\begin{align}
    &I_1(p)=2\pi\left(\e^{\beta p_0}-1\right)\int\frac{\d^3\veck_1}{(2\pi)^3}\frac{\delta(p_0-\omega_{\veck_1}-\omega_{\vecp-\veck_1})}{4\omega_{\veck_1}\omega_{\vecp-\veck_1}} f_{\rm B}(\omega_{\veck_1})f_{\rm B}(\omega_{\vecp-\veck_1})\,,\\
    &I_2(p)=4\pi\left(\e^{\beta p_0}-1\right)\int\frac{\d^3\veck_1}{(2\pi)^3}\frac{\delta(p_0+\omega_{\veck_1}-\omega_{\vecp-\veck_1})}{4\omega_{\veck_1}\omega_{\vecp-\veck_1}} \left[1+f_{\rm B}(\omega_{\veck_1})\right]f_{\rm B}(\omega_{\vecp-\veck_1})\,.
\end{align}
\end{subequations}
To do the integral over $\veck_1$, we use the spherical coordinates and let $\theta=0$ be along the direction of $\vec{p}$. 

Let us first focus on $I_1$, which now reads
\begin{align}
    I_1(p)=2\pi\left(\e^{\beta p_0}-1\right)\int_0^{\infty}\frac{\d|\veck_1|\, |\veck_1|^2}{(2\pi)^2}\int_{-1}^1 \d (\cos\theta)\, \frac{\delta(p_0-\omega_{\veck_1}-\omega_{\vecp-\veck_1})}{4\omega_{\veck_1}\omega_{\vecp-\veck_1}}f_{\rm B}(\omega_{\veck_1})f_{\rm B}(\omega_{\vecp-\veck_1})\,.
\end{align}
The $\delta$-function can be eliminated by the integral over $\cos\theta$. However, since $\cos\theta\in [-1,1]$, the $\delta$-function cannot be satisfied by any $|\veck_1|\in [0,\infty)$, leading to a constraint for the integral range of $|\veck_1|$ after the elimination of the $\delta$-function. The $\delta$-function can be fulfilled when
\begin{align}
\label{eq:four-vertex-condition}
    \sqrt{M^2+|\vecp|^2+|\veck_1|^2-2|\vecp||\veck_1|}\leq p_0-\omega_{\veck_1}\leq \sqrt{M^2+|\vecp|^2+|\veck_1|^2+2|\vecp||\veck_1|}\,.
\end{align}
The above constraint is equivalent to 
\begin{subnumcases}{}
\label{eq:fourvertex_omegak1_cond1}
\omega _{\vec{k}_1} \leq  p_0\,, \\
\label{eq:fourvertex_omegak1_cond2}
 |p^2-2p_0\omega_{\veck_1}|\leq 2 |\vecp||\veck_1|\,. 
\end{subnumcases}
Taking the square of Eq.~\eqref{eq:fourvertex_omegak1_cond2} gives
\begin{align}
\label{eq:fourvertex-omegak1_cond22}
    4p^2\omega^2_{\veck_1}-4p_0 p^2 \omega_{\veck_1}+p^4+4|\vecp|^2 M^2\leq 0\,.
\end{align}
The discriminant of the quadratic polynomial on the LHS is $\Delta=16 p^4|\vecp|^2 C$ where $C=1-\frac{4M^2}{p^2}$.

Now we have two cases. Case $(i)$: if $p^2>0$, condition~\eqref{eq:fourvertex-omegak1_cond22} requires that $\Delta>0$ which gives $p^2> 4 M^2$. We then have two real roots for the quadratic polynomial, $\omega_{\veck_1}^{(\pm)}=\Omega_{\pm}$ where we have used the definition~\eqref{eq:Omega+-} for $\Omega_{\pm}$. Therefore, Eq.~\eqref{eq:fourvertex-omegak1_cond22} leads to
\begin{align}
\label{eq:ineq-omegak}
\Omega_-\leq \omega_{\vec{k}}\leq \Omega_+\,.
\end{align}
The above condition, together with $p^2>4M^2$, ensures that Eq.~\eqref{eq:fourvertex_omegak1_cond1} is satisfied automatically. Case $(ii)$: if $p^2<0$, then $C>1$. We thus have 
\begin{align}
    \omega_{\veck_1}\leq \Omega_-\,\quad {\rm or}\quad  \omega_{\veck_1} \geq \Omega_2\,.
\end{align}
However, for $p^2<0$, one can show that $\Omega_-<0$ which makes the first condition impossible, and $\Omega_+>p_0$ which makes the second condition incompatible with Eq.~\eqref{eq:fourvertex_omegak1_cond1}. In conclusion, the condition in Eq.~\eqref{eq:four-vertex-condition} can be satisfied when $p^2>4M^2$ and $\Omega_-\leq \omega_{\veck_1}\leq \Omega_+$.

Using the formula~\eqref{eq:app-delta function-1} to perform the integral over $\cos\theta$ and further using $|\veck_1|\d |\veck_1|=\omega_{\veck_1}\d  \omega_{\veck_1}$, we finally get
\begin{align}
    I_1(p)&=\frac{\theta(p^2-4M^2)}{8\pi|\vecp|}\left(\e^{\beta p_0}-1\right)\int_{\Omega_-}^{\Omega_+}\d\omega_{\veck_1} f_{\rm B}(\omega_{\veck_1}) f_{\rm B}(p_0-\omega_{\veck_1})\,,\notag\\
    &=\frac{\theta(p^2-4M^2)}{8\pi|\vecp|}\left[(\Omega_- -\Omega_+) +2T\log\left(\frac{f_{\rm B}(\Omega_-)}{f_{\rm B}(\Omega_+)}\right)\right]\,.
\end{align}

We can repeat the analysis for $I_2$ which in the polar coordinates reads
\begin{align}
    I_2(p)=4\pi \left(\e^{\beta p_0}-1\right)\int_0^{\infty}\frac{\d |\veck_1|\, |\veck_1|^2}{(2\pi)^2}\int_{-1}^1\d(\cos\theta)\, \frac{\delta(p_0+\omega_{\veck_1}-\omega_{\vecp-\veck_1})}{4\omega_{\veck_1}\omega_{\vecp-\veck_1}}\left[1+f_{\rm B}(\omega_{\veck_1})\right]f_{\rm B}(\omega_{\vecp-\veck_1})\,.
\end{align}
The difference is that in this case the $\delta$-function can be satisfied when $p^2<0$ and $\omega_{\veck_1}\geq -\Omega_-$ (note that $\Omega_-<0$ for $p^2<0$). Then we obtain
\begin{align}
    I_2(p)&=\frac{\theta(-p^2)}{4\pi|\vecp|}\left(\e^{\beta p_0}-1\right)\int_{-\Omega_-}^\infty\d\omega_{\veck_1}\left[1+f_{\rm B}(\omega_{\veck_1})\right]f_{\rm B}(p_0+\omega_{\veck_1})\,,\notag\\
    &=\frac{\theta(-p^2)}{8\pi|\vecp|}\left[-2\Omega_+ + 2T\log\left|\frac{f_{\rm B}(\Omega_-)}{f_{\rm B}(\Omega_+)}\right|\right]\,.
\end{align}

Adding up $I_1$ and $I_2$, one obtains~\eqref{eq:imaginary_fish1}.
This expression is anti-symmetric in $p_0$ and so is valid also for $p_0<0$.

\subsection{Sunset self-energy}
\label{app:sefl-energy}

Similarly, we first introduce $\Pi^-$,
\begin{align}
\label{eq:Pi-}
    \Pi^-(x-x')&\equiv \Pi^-_{\phi}(x-x')+\Pi^-_\chi(x-x')\notag\\
    &=\sum_{a=\phi,\chi}d_a\left[G^{-+}_\phi(x-x')(G^{-+}_a(x-x'))^2-G^{+-}_\phi(x-x')(G^{+-}_a(x-x'))^2\right]\,,
\end{align}
where
\begin{align}
\label{eq:dphi-dh}
    d_\phi=-\frac{\i\lambda_\phi^2}{6}\,,\quad d_\chi=-\frac{\i g^2}{2}\,.
\end{align}
One can re-express the RHS of Eq.~\eqref{eq:Pi-} in terms of $G^-_a(x-x')$ using Eqs.~\eqref{eq:DeltaplusDeltaminus} and the KMS relation~\eqref{eq:KMS}. In momentum space, one has
\begin{align}\label{eq:appPi_def}
    \widetilde{\Pi}^-_{a}(p)=\i d_a J_a(p)\equiv &\i d_a\int\frac{\d^4 k_1}{(2\pi)^4}\int\frac{\d^4 k_2}{(2\pi)^4}\int\frac{\d^4 k_3}{(2\pi)^4}(2\pi)^4 \delta^{(4)}(p-k_1-k_2-k_3)G^-_\phi(k_1)G^-_a(k_2)G^-_a(k_3)\notag\\
    &\times \left(1+f_{\rm B}(k_{1,0})+f_{\rm B}(k_{2,0})+f_{\rm B}(k_{3,0})+f_{\rm B}(k_{1,0})f_{\rm B}(k_{2,0})+f_{\rm B}(k_{1,0})f_{\rm B}(k_{3,0})\right.\notag\\
    &\left.+f_{\rm B}(k_{2,0})f_{\rm B}(k_{3,0})\right)\,,
\end{align}
where we have defined the integrals $J_a(p)$. We can obtain the imaginary part of $\widetilde{\Pi}^{\rm R}_{a}$ through the relation
\begin{align}
\label{eq:PiR-Ja}
    {\rm Im}\widetilde{\Pi}_a^{\rm R}(p)=-\frac{\i}{2}\widetilde{\Pi}_a^{-}(p)=\frac{d_a}{2} J_a(p) \,.
\end{align}
Below, we shall calculate $J_\phi(M_\phi,\vec{0})$ and $J_h(M_\phi,\vec{0})$ separately.

\subsubsection{Computation of \texorpdfstring{$J_{\phi}$}{TEXT}}
\label{app:sec Jphi}

The $J_{\phi}(p)$ defined in Eq.~\eqref{eq:appPi_def} can be written as 
\begin{align}
\label{eq:appB_Jphi}
J_\phi(p)=  &-\i \int\frac{\d^4 k_1}{(2\pi)^4}\int\frac{\d^4 k_2}{(2\pi)^4}\int\frac{\d^4 k_3}{(2\pi)^4}(2\pi)^4 \delta^{(4)}(p-k_1-k_2-k_3)  \rho_{\phi,\vec{k}_1}(k_{1,0}) \rho_{\phi,\vec{k}_2}(k_{2,0}) \rho_{\phi,\vec{k}_3}(k_{3,0})\notag\\
    &\times \left\{\left[1+f_{\rm B}(k_{1,0})\right] \left[1+f_{\rm B}(k_{2,0})\right] \left[1+f_{\rm B}(k_{3,0})\right]-f_{\rm B}(k_{1,0})f_{\rm B}(k_{2,0})f_{\rm B}(k_{3,0})\right\}\,.
\end{align}
Since there are only the $\phi$-field quantities involved in $J_\phi$, we suppress the subscript $\phi$ below. With Eq.~\eqref{eq:appB_rho_k of delta}, we could first perform the integrals over $k_{i,0}$ (with $i=1,2,3$) in Eq.~\eqref{eq:appB_Jphi} and obtain 
\begin{align}
\label{eq:appB_Jphi_2}
 J_\phi(p)=  &-\i \int\frac{\d^3 \vec{k}_1}{(2\pi)^3}\int\frac{\d^3 \vec{k}_2}{(2\pi)^3}\int\frac{\d^3 \vec{k}_3}{(2\pi)^3} \frac{(2\pi)^4 \delta^{(3)}(\vec{p}-\vec{k}_1-\vec{k}_2-\vec{k}_3) }{8\, \omega _{\vec{k}_1}\omega _{ \vec{k}_2}\omega _{\vec{k}_3}}\, \mathcal{F}_1(p_0,\omega _{\vec{k}_1},\omega _{\vec{k}_2},\omega _{\vec{k}_3}) \,,
\end{align}
where
\begin{align}
    &\mathcal{F}_1(p_0,\omega _{\vec{k}_1},\omega _{\vec{k}_2},\omega _{\vec{k}_3})=\sum_{\alpha_1,\alpha_2,\alpha_3=\pm}(\alpha_1\alpha_2\alpha_3)\times\delta (p_0-\alpha_1\omega _{\vec{k}_1}-\alpha_2\omega _{\vec{k}_2}-\alpha_3\omega _{\vec{k}_3})\notag\\
    &\times\left\{\left[1+f_{\rm B}(\alpha_1\omega _{\vec{k}_1})\right] \left[1+f_{\rm B}(\alpha_2\omega _{\vec{k}_2})\right] \left[1+f_{\rm B}(\alpha_3\omega _{\vec{k}_3})\right]-f_{\rm B}(\alpha_1\omega _{\vec{k}_1})f_{\rm B}(\alpha_2\omega _{\vec{k}_2})f_{\rm B}(\alpha_3\omega _{\vec{k}_3})\right\}\,.
\end{align}
Making use of Eq.~\eqref{eq:property} and relabeling the integration variables in Eq.~\eqref{eq:appB_Jphi_2}, $\mathcal{F}$ can be simplified to 
\begin{align}
\label{eq:appB_Jphi_F}
&\mathcal{F}_1(p_0,\omega _{\vec{k}_1},\omega _{\vec{k}_2},\omega _{\vec{k}_3})
=\Big[\delta (p_0-\omega _{\vec{k}_1}-\omega _{\vec{k}_2}-\omega _{\vec{k}_3})-\delta (p_0+\omega _{\vec{k}_1}+\omega _{\vec{k}_2}+\omega _{\vec{k}_3})\Big] \notag\\ 
&\qquad \times \bigg\{\left[1+f_{\rm B}(\omega _{\vec{k}_1})\right] \left[1+f_{\rm B}(\omega _{\vec{k}_2})\right] \left[1+f_{\rm B}(\omega _{\vec{k}_3})\right]- f_{\rm B}(\omega _{\vec{k}_1})f_{\rm B}(\omega _{\vec{k}_2})f_{\rm B}(\omega _{\vec{k}_3})\bigg\}\notag\\
&\qquad+3 \Big[\delta (p_0-\omega _{\vec{k}_1}+\omega _{\vec{k}_2}+\omega _{\vec{k}_3})-\delta (p_0+\omega _{\vec{k}_1}-\omega _{\vec{k}_2}-\omega _{\vec{k}_3})\Big] \notag\\ 
&\qquad \times \bigg\{\left[1+f_{\rm B}(\omega _{\vec{k}_1})\right] f_{\rm B}(\omega _{\vec{k}_2}) f_{\rm B}(\omega _{\vec{k}_3})- f_{\rm B}(\omega _{\vec{k}_1})[1+f_{\rm B}(\omega _{\vec{k}_2})][1+f_{\rm B}(\omega _{\vec{k}_3})]\bigg\}\,.
\end{align}

Again, Eq.~\eqref{eq:appB_Jphi_F} has a clear interpretation in terms of particle production processes. The first $\delta$-function can only be satisfied for $p_0>0$ and describes the process $ \varphi \leftrightarrow  \phi_1 \phi_2 \phi_3$. The second $\delta$-function can only be satisfied for $p_0<0$ and describes $\phi_1\phi_2\phi_3\leftrightarrow \varphi$. The third $\delta$-function describes $\varphi\phi_2\phi_3\leftrightarrow \phi_1$ for $p_0>0$ and $\phi_2\phi_3\leftrightarrow \varphi\phi_1$ for $p_0<0$. The last $\delta$-function describes $\varphi\phi_1\leftrightarrow \phi_2\phi_3$ for $p_0>0$ and $\phi_1\leftrightarrow \varphi\phi_2\phi_3$ for $p_0<0$. 

From Eq.~\eqref{eq:appB_Jphi_2} and Eq.~\eqref{eq:appB_Jphi_F}, we observe that $
   J_\phi(p_0,\vec{p}) = - J_\phi(-p_0,\vec{p})$.
Therefore, we will also restrict the following calculation to the case of $p_0 >0$. Then the second $\delta$-function in Eq.~\eqref{eq:appB_Jphi_F} cannot be fulfilled. Further, we shall take $\vec{p}=0$ which is the relevant case for the condensate decay. In this case, we have 
\begin{align}
    \vec{k}_1+\vec{k}_2+\vec{k}_3=0\,,
\end{align}
which leads to
\begin{align}\label{eq:app_k1_k2}
    |\vec{k}_2|+|\vec{k}_1+\vec{k}_2|\ge |\vec{k}_1|\,,
\end{align}
and thus (for $M^2_\phi\neq 0$)
\begin{align}
    \omega _{\vec{k}_2}+\omega _{\vec{k}_3} > \omega _{\vec{k}_1}\,.
\end{align}
Therefore, for $p_0>0$ the third $\delta$-function in Eq.~\eqref{eq:appB_Jphi_F} also gives zero. Taking these into account and after some simple algebra, one obtains
\begin{align}\label{eq:appB JA and JB}
    J_\phi(p_0,\vec{0})=3\,J^A _\phi(p_0,\vec{0}) +J^B _\phi(p_0,\vec{0})\,,
\end{align}
where 
\begin{align}
\label{eq:appb_JA}
   J^A _\phi(p_0,\vec{0})
   =&-\i \left(\e ^{\beta p_0}-1\right) \int\frac{\d^3 \vec{k}_1}{(2\pi)^3}\int\frac{\d^3 \vec{k}_2}{(2\pi)^3}\int\frac{\d^3 \vec{k}_3}{(2\pi)^3}(2\pi)^4 \delta^{(3)}(\vec{k}_1+\vec{k}_2+\vec{k}_3)  \frac{1}{8\, \omega _{\vec{k}_1}\omega _{ \vec{k}_2}\omega _{\vec{k}_3}} \notag\\ 
   & \times \delta (p_0+\omega _{\vec{k}_1}-\omega _{\vec{k}_2}-\omega _{\vec{k}_3})\times \left[1+f_{\rm B}(\omega _{\vec{k}_1})\right] f_{\rm B}(\omega _{\vec{k}_2}) f_{\rm B}(\omega _{\vec{k}_3}) \,,
\end{align}
and 
\begin{align}\label{eq:appb_JB}
    J^B _\phi(p_0,\vec{0})
   =&-\i \left(\e ^{\beta p_0}-1\right) \int\frac{\d^3 \vec{k}_1}{(2\pi)^3}\int\frac{\d^3 \vec{k}_2}{(2\pi)^3}\int\frac{\d^3 \vec{k}_3}{(2\pi)^3}(2\pi)^4 \delta^{(3)}(\vec{k}_1+\vec{k}_2+\vec{k}_3)  \frac{1}{8\, \omega _{\vec{k}_1}\omega _{ \vec{k}_2}\omega _{\vec{k}_3}} \notag\\ 
   & \times \delta (p_0-\omega _{\vec{k}_1}-\omega _{\vec{k}_2}-\omega _{\vec{k}_3})\times f_{\rm B}(\omega _{\vec{k}_1}) f_{\rm B}(\omega _{\vec{k}_2}) f_{\rm B}(\omega _{\vec{k}_3}) \,. 
\end{align}
$J_\phi^A(p_0>0,\vec{0})$ corresponds to the process $\varphi\phi_1\leftrightarrow\phi_2\phi_2$ while $J_\phi^B(p_0>0,\vec{0})$ to $\varphi\leftrightarrow\phi_1\phi_2\phi_3$.

With $p_0=M_\phi$, the $\delta$-function in $J_\phi^B(p_0,\vec{0})$ can never be satisfied. Hence, only the process
\begin{align}
    \varphi\phi\leftrightarrow\phi\phi
\end{align}
is allowed. Below we thus focus on $J^A _\phi(p_0,\vec{0})$. Performing the integrals over $\vec{k}_3$ in Eq.~\eqref{eq:appb_JA}, we obtain
\begin{align}
    J^A _\phi(p_0,\vec{0})=&-2\pi \i \left(\e ^{\beta p_0}-1\right) \int\frac{\d^3 \vec{k}_1}{(2\pi)^3}\int\frac{\d^3 \vec{k}_2}{(2\pi)^3}\frac{1}{8\, \omega _{\vec{k}_1}\omega_{ \vec{k}_2}\omega_{\vec{k}_1+\vec{k}_2}} \\ \notag
   & \times \delta (p_0+\omega _{\vec{k}_1}-\omega _{\vec{k}_2}-\omega_{\vec{k}_1+\vec{k}_2})\times \left[1+f_{\rm B}(\omega _{\vec{k}_1})\right] f_{\rm B}(\omega _{\vec{k}_2}) f_{\rm B}(\omega _{\vec{k}_1+\vec{k}_2}) \,.
\end{align}
To perform the $\vec{k}_2$ integration, we can use spherical coordinates with $\theta =0$ being the direction of $\vec{k}_1$. Then, 
\begin{align}
\label{eq:appb JA-cos theta}
    J^A _\phi(p_0,\vec{0})=&-2\pi \i \left(\e ^{\beta p_0}-1\right) \int\frac{\d^3 \vec{k}_1}{(2\pi)^3}\int _0 ^{\infty}\frac{\d |\vec{k}_2| \, |\vec{k}_2|^2}{(2\pi)^2} \int^{1} _{-1} \d(\cos\theta)\,  \frac{1}{8\, \omega _{\vec{k}_1}\omega _{ |\vec{k}_2|}\omega _{\vec{k}_1+\vec{k}_2}} \\ \notag
   & \times \delta (p_0+\omega _{\vec{k}_1}-\omega _{\vec{k}_2}-\omega _{\vec{k}_1+\vec{k}_2})\times \left[1+f_{\rm B}(\omega _{\vec{k}_1})\right] f_{\rm B}(\omega _{\vec{k}_2}) f_{\rm B}(\omega _{\vec{k}_1+\vec{k}_2}) \,.
\end{align}
Again, since $-1\leq\cos\theta\leq 1$, after eliminating the $\delta$-function with the integral over $\cos\theta$ the integral range for $|\vec{k}_2|$ must be restricted to :
\begin{align}
    \sqrt{M^2 _{\phi} +\left(|\vec{k}_1|-|\vec{k}_2|\right)^2 } \le p_0+\omega _{\vec{k}_1}-\omega _{\vec{k}_2}  \le \sqrt{ M^2 _{\phi} +\left(|\vec{k}_1|+|\vec{k}_2|\right)^2 }\,.
\end{align}
The above constraint is equivalent to 
\begin{subnumcases}{}
\label{eq:Jphi_omegak2_cond1}
\omega _{\vec{k}_2} \leq p_0 + \omega _{\vec{k}_1}\,, \\
\label{eq:Jphi_omegak2_cond2}
\Big|(p_0+\omega _{\vec{k}_1}-\omega _{\vec{k}_2})^2-(\omega _{\vec{k}_1}^2+\omega _{\vec{k}_2}^2-M_{\phi}^2)\Big|\leq 2 |\vec{k}_1||\vec{k}_2| \,.
\end{subnumcases}

To simplify the conditions further, take the square of Eq.~\eqref{eq:Jphi_omegak2_cond2}, giving
\begin{align}
\label{eq:appb-condtion1}
    \Sigma_1  \omega _{\vec{k}_2}^2 + \Sigma_2  \omega _{\vec{k}_2} +\Sigma_3\le 0\,,
\end{align}
where
\begin{subequations}
\begin{align}
    \Sigma_1 &= 4 \left(M^2 _{\phi} +p_0 ^2+2 p_0 \omega _{\vec{k}_1} \right)\,,\\
    \Sigma_2 &=-4 (p_0+\omega _{\vec{k}_1})\left(M^2 _{\phi} +p_0 ^2+2 p_0 \omega _{\vec{k}_1} \right)\,, \\
    \Sigma_3 &=-3 M_{\phi}^4+2 M_{\phi}^2 \left(p_0^2+2 p_0 \omega _{\vec{k}_1}+2 \omega _{\vec{k}_1}^2\right)+p_0^2 (p_0+2 \omega _{\vec{k}_1})^2\,.
\end{align}
\end{subequations}
Since $\Sigma_1 >0$, the condition~\eqref{eq:appb-condtion1} is equivalent to
\begin{subnumcases}{}
\label{eq:discrimnant}
p^2_0 +2p_0\omega_{\vec{k}_1}-3M^2_\phi>0\,, \\
\label{eq:Jphi_omegak2_cond22}
\widetilde{\Omega}^{-} _{2}\leq \omega_{\vec{k}_2} \le \widetilde{\Omega}^{+}_{2} \,, 
\end{subnumcases}
where 
\begin{align}
    \widetilde{\Omega}^{\pm} _{2} = \frac{1}{2} \left[p_0+\omega_{\vec{k}_1} \pm \frac{\sqrt{(p_0 ^2+2p_0 \omega_{\vec{k}_1}-3M_{\phi}^2)(\omega_{\vec{k}_1} ^2-M^2_{\phi})(p_0 ^2+2p_0 \omega_{\vec{k}_1}+M_{\phi}^2)}}{p_0 ^2+2p_0 \omega_{\vec{k}_1}+M_{\phi}^2} \right]\,.
\end{align}
The condition~\eqref{eq:discrimnant} ensures that the discriminant of the quadratic polynomial in Eq.~\eqref{eq:appb-condtion1} is positive. Taking into account condition~\eqref{eq:Jphi_omegak2_cond1} and the physical conditions $\{\omega_{\vec{k}_1},\omega_{\vec{k}_2}\}\geq M_\phi$, we finally have 
\begin{subnumcases}{}
\omega_{\vec{k}_1}\geq \Omega_1^- \,, \\
\label{eq:app_omegak2_cond22}
\Omega ^{-} _{2}\leq \omega _{\vec{k}_2} \leq  \Omega ^{+} _{2} \,, 
\end{subnumcases}
where 
\begin{align}
    \Omega_1^-={\rm max}\{M_\phi, (3M_\phi^2 - p_0^2)/2 p_0\}\,,\  \Omega_2^-={\rm max}\{M_\phi,\widetilde{\Omega}^{-}_{2}\}\,,\ \Omega_{2}^+={\rm min}\{p_0+\omega_{\vec{k}_1},\widetilde{\Omega}^{+}_{2}\}\,.
\end{align}

One can also introduce spherical coordinates for $\veck_1$ (the direction $\theta'=0$ can be chosen arbitrarily) and perform the integral over $\cos\theta'$ with the help of Eq.~\eqref{eq:app-delta function-1}. Then, we have 
\begin{align}
    J^A _\phi(p_0,\vec{0})=\frac{\left(\e ^{\beta p_0}-1\right)}{\i\, 32\pi^3} \int ^{\infty} _{\Omega^-_1}\d \omega _{\vec{k}_1}\int _{\Omega^{-} _{2}} ^{{\Omega^{+} _{2}}}\d \omega _{\vec{k}_2} \left[1+f_{\rm B}(\omega _{\vec{k}_1})\right] f_{\rm B}(\omega _{\vec{k}_2}) f_{\rm B}(p_0+\omega _{\vec{k}_1}-\omega _{\vec{k}_2})  \,.
\end{align}
Doing the integral over $\omega _{\vec{k}_2}$, we obtain
\begin{align}
\label{eq:Jphi_A_p0}
   J^A _\phi(p_0,\vec{0})=& \frac{\e^{\beta p_0}-1}{\i\,32 \pi ^3 \beta} \int ^{\infty} _{\Omega^-_1} \d \omega_{\vec{k}_1} [1+f_{\rm B}(\omega _{\vec{k}_1})]f_{\rm B}(p_0+\omega _{\vec{k}_1})\,\log \left[ \frac{\left(\e ^{\beta \Omega^+ _2}-1\right)\left(\e^{\beta(p_0+\omega_{\vec{k}_1})}-\e^{\beta \Omega_{2}^-} \right)}{\left(\e^{\beta(p_0+\omega_{\vec{k}_1})}-\e^{\beta \Omega_{2}^+} \right)\left(\e ^{\beta \Omega^- _2}-1\right)}\right] \,. 
\end{align}
For general $p_0$, the integral over $\omega_{\vec{k}_1}$ is too complicated to give a closed form of $J^A _\phi(p_0)$. However, for $p_0=M_{\phi}$, which is the case we are interested in, \emph{analytic} results can be obtained.

\begin{figure}[ht]
    \centering
    \includegraphics[scale=0.7]{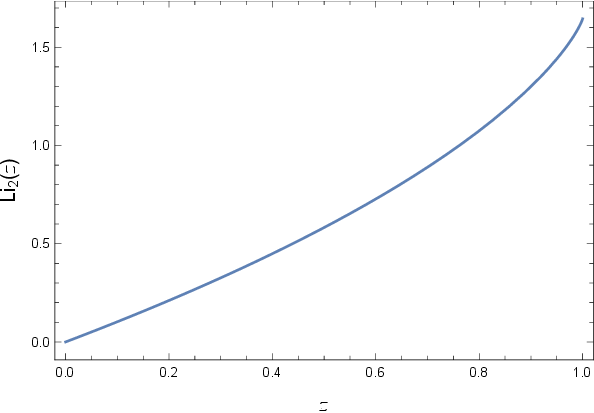}
    \caption{The dilogarithm function $\text{Li}_2(z)$}
    \label{fig:di-log}
\end{figure}

With $p_0=M_{\phi}$, we have $\Omega^-_{1} = M_{\phi}$, $\Omega^{-}_{2}=M_{\phi}$, $\Omega ^{+} _{2}=\omega_{\vec{k}_1}$. Substituting them into the integral in Eq.~\eqref{eq:Jphi_A_p0} gives\footnote{Using Mathematica to do the integral in  Eq.~\eqref{eq:appB_integral omega1}, one may obtain \begin{align*}
    \frac{v}{\beta(v-1)}\left[2 \text{Li}_2\left(\frac{v}{v-1}\right)+\text{Li}_2(v)+[\ln (1-v)]^2\right]\,,
\end{align*} 
which can be simplified to $\frac{v}{\beta(1-v)} \text{Li}_2(v)$ with the Landen's identity~\cite{Gordon1997}
\begin{align*}
    \text{Li}_2(-z) + \text{Li}_2\left(\frac{z}{1+z}\right) = -\frac{1}{2} [\ln (1+z) ]^2\,.
\end{align*}}
\begin{align}
\label{eq:appB_integral omega1}
\frac{1}{\beta}\int ^{v} _{0} \frac{\d u}{1-u} \frac{v}{1-u\cdot v} \left( \log \left[\frac{ (1-u)^2}{u }\right]+\log \left[\frac{v }{ (1-v)^2}\right]\right) = \frac{v\, \text{Li}_2(v)}{\beta(1-v)}\,, 
\end{align}
where
\begin{align}
    u&=\e ^{-\beta \omega_{\vec{k}_1} }\,,\quad v=\e ^{-\beta M_{\phi}}\,,
\end{align}
and the dilogarithm function $\text{Li}_2(z)$ is defined as 
\begin{align}
    \text{Li}_2(z) =-\int ^{1} _{0} \frac{\d x}{x} \log (1-z\,x)= \sum ^{\infty} _{n=1} \frac{z^n}{n^2}\,,
\end{align}
for $z\in (-\infty, 1)$. In the high-temperature limit ($T\gg M_{\phi}$) it approaches a constant,
\begin{align}
    \text{Li}_2(\e ^{- M_{\phi}/T}) = \frac{\pi ^2}{6} + \O\left[\frac{M_{\phi}}{T}\log \left(\frac{M_{\phi}}{T}\right)\right]\,.
\end{align}
In the low-temperature limit ($T\ll M_{\phi}$), one has
\begin{align}
    \text{Li}_2(\e ^{- M_{\phi}/T}) = \e ^{- M_{\phi}/T} + \O\left(\e ^{-2 M_{\phi}/T}\right)\,.
\end{align}
The behaviour of ${\rm Li}_2(\e^{-M_\phi/T})$ for the full region $T\in[0,\infty)$ is shown in Fig.~\ref{fig:di-log}.

As we mentioned above, $J_\phi^B(M_\phi,\vec{0})=0$. Therefore, we finally have 
\begin{align}
  J _\phi(M_{\phi},\vec{0}) = 3 J_\phi^A(M_\phi,\vec{0})= \frac{3 T^2}{\i\,32 \pi ^3 } \,  \text{Li}_2(\e ^{- M_{\phi}/T})\,.
\end{align}  
Substituting the above into Eq.~\eqref{eq:PiR-Ja} and using Eq.~\eqref{eq:dphi-dh}, one obtains
\begin{align}
\label{eq:B118}
  {\rm Im}\widetilde{\Pi}^{\rm R}_\phi(M_\phi,\vec{0})= {\rm Im} \widetilde{\pi}(M_\phi) = - \frac{\lambda _{\phi} ^2\,T^2}{128 \pi ^3 } \,  \text{Li}_2(\e ^{- M_{\phi}/T}) \,.
\end{align}
Substituting the above into the definition of $\gamma$, one obtains Eq.~\eqref{eq:Im-pi-phi}. Eq.~\eqref{eq:B118} is the result obtained in Refs.~\cite{Jeon:1994if,Wang:1995qg,Wang:1995qf}.
For other related discussions, see, e.g., Refs.~\cite{Parwani:1991gq,Berera:1998gx}.

\subsubsection{Computation of \texorpdfstring{$J_\chi$}{TEXT}}

Now we consider $J_{\chi}(p)$ which can be written as 
\begin{align}
\label{eq:appB_Jchi}
J_\chi(p)=  &-\i \int\frac{\d^4 k_1}{(2\pi)^4}\int\frac{\d^4 k_2}{(2\pi)^4}\int\frac{\d^4 k_3}{(2\pi)^4}(2\pi)^4 \delta^{(4)}(p-k_1-k_2-k_3)  \rho_{\phi,\vec{k}_1}(k_{1,0}) \rho_{\chi,\vec{k}_2}(k_{2,0}) \rho_{\chi,\vec{k}_3}(k_{3,0})\notag\\
    &\times \left\{\left[1+f_{\rm B}(k_{1,0})\right] \left[f_{\rm B}(k_{2,0})+1\right] \left[f_{\rm B}(k_{3,0})+1\right]-f_{\rm B}(k_{1,0})f_{\rm B}(k_{2,0})f_{\rm B}(k_{3,0})\right\}\,.
\end{align}
Similar to what we have done earlier, performing the integrals over $k_{i,0}$ and relabeling the integration variables lead to 
\begin{align}
 J_\chi(p)=  &-\i \int\frac{\d^3 \vec{k}_1}{(2\pi)^3}\int\frac{\d^3 \vec{k}_2}{(2\pi)^3}\int\frac{\d^3 \vec{k}_3}{(2\pi)^3}(2\pi)^4  \delta^{(3)}(\vec{p}-\vec{k}_1-\vec{k}_2-\vec{k}_3)  \frac{\mathcal{F}_2(p_0,\omega_{\phi,\vec{k}_1},\omega_{\chi,\vec{k}_2},\omega_{\chi,\vec{k}_3})}{8\, \omega _{\phi,\vec{k}_1}\omega _{\chi,\vec{k}_2}\omega _{\chi,\vec{k}_3}}\,, 
\end{align}
where
\begin{align}
\label{eq:F2}
 &\mathcal{F}_2(p_0,\omega_{\phi,\vec{k}_1},\omega_{\chi,\vec{k}_2},\omega_{\chi,\vec{k}_3})= \Big[\delta (p_0-\omega _{\phi,\vec{k}_1}-\omega _{\chi,\vec{k}_2}-\omega _{\chi,\vec{k}_3})-\delta (p_0+\omega _{\phi,\vec{k}_1}+\omega _{\chi,\vec{k}_2}+\omega _{\chi,\vec{k}_3})\Big] \notag\\
 &\quad  \times \bigg(\left[1+f_{\rm B}(\omega _{\phi,\vec{k}_1})\right] \left[f_{\rm B}(\omega _{\chi,\vec{k}_2})+1\right] \left[f_{\rm B}(\omega _{\chi,\vec{k}_3})+1\right]- f_{\rm B}(\omega _{\phi,\vec{k}_1})f_{\rm B}(\omega _{\chi,\vec{k}_2})f_{\rm B}(\omega _{\chi,\vec{k}_3})\bigg)\notag\\
&\quad + \Big[\delta (p_0-\omega _{\phi,\vec{k}_1}+\omega _{\chi,\vec{k}_2}+\omega _{\chi,\vec{k}_3})-\delta (p_0+\omega _{\phi,\vec{k}_1}-\omega _{\chi,\vec{k}_2}-\omega _{\chi,\vec{k}_3})\Big] \notag\\ 
&\quad \times \bigg(\left[1+f_{\rm B}(\omega _{\phi,\vec{k}_1})\right] f_{\rm B}(\omega _{\chi,\vec{k}_2}) f_{\rm B}(\omega _{\chi,\vec{k}_3})- f_{\rm B}(\omega _{\phi,\vec{k}_1})[1+f_{\rm B}(\omega _{\chi,\vec{k}_2})][1+f_{\rm B}(\omega _{\chi,\vec{k}_3})]\bigg) \notag \\
&\quad +2\times \Big[\delta (p_0+\omega _{\phi,\vec{k}_1}-\omega _{\chi,\vec{k}_2}+\omega _{\chi,\vec{k}_3})-\delta (p_0-\omega _{\phi,\vec{k}_1}+\omega_{\chi,\vec{k}_2}-\omega _{\chi,\vec{k}_3})\Big] \notag\\ 
&\quad \times \bigg(\left[1+f_{\rm B}(\omega _{\chi,\vec{k}_2})\right] f_{\rm B}(\omega _{\phi,\vec{k}_1}) f_{\rm B}(\omega _{\chi,\vec{k}_3})- f_{\rm B}(\omega _{\chi,\vec{k}_2})[1+f_{\rm B}(\omega _{\phi,\vec{k}_1})][1+f_{\rm B}(\omega _{\chi,\vec{k}_3})]\bigg)\,.
\end{align}
One can similarly interpret every $\delta$-function in terms of particle production processes. 
 
The calculation of $J_\chi(p)$ is much more involved. We will restrict ourselves to $p=(M_{\phi},\vec{0})$ from the beginning. For $p_0=M_\phi>0$, the second $\delta$-function in Eq.~\eqref{eq:F2} cannot be fulfilled. Further, since $\omega_{\phi,\vec{k}_1} \geq M_\phi$, the first $\delta$-function cannot be fulfilled either. Using Eq.~\eqref{eq:app_k1_k2} and after some simple algebra, one can show that the third and fifth $\delta$-functions cannot be fulfilled. Therefore, the only processes that can be on-shell are
\begin{align}
    \varphi\phi_1 \leftrightarrow \chi_2 \chi_3 \;\; \text{and} \;\; \varphi \chi_2 \leftrightarrow \phi_1 \chi_3\,,
\end{align}
corresponding to the fourth and sixth $\delta$-functions.

Using $1+f_{\rm B}(\omega_\veck)=\exp(\beta\omega_\veck)f_{\rm B}(\omega_\veck)$, one can write $J_\chi(M_{\phi},\vec{0})$ as
\begin{align}
\label{eq:appB chi JA and JB}
    J_\chi(M_{\phi},\vec{0})=J^A _\chi(M_{\phi},\vec{0}) +2J^B _\chi(M_{\phi},\vec{0})\,,
\end{align}
where 
\begin{align}\label{eq:appb chi_JA}
   J^A_\chi(M_{\phi},\vec{0})
   =&-\i \left(\e ^{\beta p_0}-1\right) \int\frac{\d^3 \vec{k}_1}{(2\pi)^3}\int\frac{\d^3 \vec{k}_2}{(2\pi)^3}\int\frac{\d^3 \vec{k}_3}{(2\pi)^3} \frac{(2\pi)^4 \delta^{(3)}(\vec{k}_1+\vec{k}_2+\vec{k}_3)}{8\, \omega _{\phi,\vec{k}_1}\omega _{\chi,\vec{k}_2}\omega _{\chi,\vec{k}_3}} \notag\\ 
   & \times \delta (M_{\phi}+\omega _{\phi,\vec{k}_1}-\omega _{\chi,\vec{k}_2}-\omega _{\chi,\vec{k}_3})\times \left[1+f_{\rm B}(\omega _{\phi,\vec{k}_1})\right] f_{\rm B}(\omega _{\chi,\vec{k}_2}) f_{\rm B}(\omega _{\chi,\vec{k}_3}) \,,
\end{align}
and 
\begin{align}
\label{eq:appb chi_JB}
    J^B _\chi(M_{\phi},\vec{0})
   =&-\i \left(\e ^{\beta p_0}-1\right) \int\frac{\d^3 \vec{k}_1}{(2\pi)^3}\int\frac{\d^3 \vec{k}_2}{(2\pi)^3}\int\frac{\d^3 \vec{k}_3}{(2\pi)^3} \frac{(2\pi)^4 \delta^{(3)}(\vec{k}_1+\vec{k}_2+\vec{k}_3) }{8\, \omega _{\phi,\vec{k}_1}\omega _{\chi,\vec{k}_2}\omega _{\chi,\vec{k}_3}} \notag\\ 
   & \times \delta (M_{\phi}+\omega _{\chi,\vec{k}_2}-\omega _{\phi,\vec{k}_1}-\omega _{\chi,\vec{k}_3})\times \left[1+f_{\rm B}(\omega _{\chi,\vec{k}_2})\right] f_{\rm B}(\omega _{\phi,\vec{k}_1}) f_{\rm B}(\omega _{\chi,\vec{k}_3})\,. 
\end{align}
In the following, we calculate $ J^A _\chi(M_{\phi},\vec{0})$ and $ J^B _\chi(M_{\phi},\vec{0})$, separately.

\subsubsection*{\underline{\texorpdfstring{$J^A_\chi(M_{\phi},\vec{0})$}{TEXT}}}

Performing the integrals over $\vec{k}_3$ in Eq.~\eqref{eq:appb chi_JA}, we obtain
\begin{align}
    J^A _\chi(M_{\phi},\vec{0})=&-2\pi \i \left(\e ^{\beta M_{\phi}}-1\right) \int\frac{\d^3 \vec{k}_1}{(2\pi)^3}\int\frac{\d^3 \vec{k}_2}{(2\pi)^3}\frac{\delta (M_{\phi}+\omega _{\phi,\vec{k}_1}-\omega _{\chi,\vec{k}_2}-\omega _{\chi,\veck_1+\veck_2})}{8\, \omega _{\phi,\vec{k}_1}\omega _{\chi,\vec{k}_2}\omega _{\chi,\veck_1+\veck_2}} \\ \notag
   &  \times \left[1+f_{\rm B}(\omega _{\phi,\vec{k}_1})\right] f_{\rm B}(\omega _{\chi,\vec{k}_2}) f_{\rm B}(\omega _{\chi,\veck_1+\veck_2}) \,.
\end{align}
To perform the $\vec{k}_2$ integration, we use spherical coordinates with $\vec{k}_1$ being the polar axis ($\theta =0$). Then, 
\begin{align}
\label{eq:appb chi_JA-cos theta}
    J^A _\chi(M_{\phi},\vec{0})=&-2\pi \i \left(\e ^{\beta M_{\phi}}-1\right) \int\frac{\d^3 \vec{k}_1}{(2\pi)^3}\int _0 ^{\infty}\frac{\d |\vec{k}_2| \, |\vec{k}_2|^2}{(2\pi)^2} \int^{1} _{-1} \d \cos\theta \\ \notag
   & \times\frac{\delta (M_{\phi}+\omega _{\phi,\vec{k}_1}-\omega _{\chi,\vec{k}_2}-\omega _{\chi,\veck_1+\veck_2})}{8\, \omega _{\phi,\vec{k}_1}\omega _{\chi,\vec{k}_2}\omega _{\chi,\veck_1+\veck_2}}  \left[1+f_{\rm B}(\omega _{\phi,\vec{k}_1})\right] f_{\rm B}(\omega _{\chi,\vec{k}_2}) f_{\rm B}(\omega_{\chi,\veck_1+\veck_2}) \,.
\end{align}
Since $-1\le\cos \theta\le 1$, the $\delta$-function requires that
\begin{align}
    \sqrt{M^2 _{\chi} +\left(|\vec{k}_1|-|\vec{k}_2|\right)^2 } \le M_{\phi}+\omega_{\phi,\vec{k}_1}-\omega _{\chi,\vec{k}_2}  \leq \sqrt{M^2_{\chi} +\left(|\vec{k}_1|+|\vec{k}_2|\right)^2 }\,,
\end{align}
which can be reduced to 
\begin{subnumcases}{}
\label{eq:JhA_condition1}
    \omega_{\chi,\vec{k}_2}\leq M_{\phi} + \omega_{\phi,\vec{k}_1} \,, \\
\label{eq:JhA_condition2}
  \Big|(M_{\phi}+\omega _{\phi,\vec{k}_1}-\omega _{\chi,\vec{k}_2})^2-(\omega _{\phi,\vec{k}_1}^2+\omega _{\chi,\vec{k}_2}^2-M_{\phi}^2) \Big| \leq 2 |\vec{k}_1||\vec{k}_2| \,. 
\end{subnumcases}
Taking the square of Eq.~\eqref{eq:JhA_condition2} gives
\begin{align}
\label{eq:JhA_condtion22}
    \Sigma_1  \omega _{\chi,\vec{k}_2}^2 + \Sigma_2  \omega _{\chi,\vec{k}_2} +\Sigma_3\le 0\,,
\end{align}
where 
\begin{align}
    \Sigma_1 &= 8 M _{\phi} \left(M _{\phi} +\omega _{\phi,\vec{k}_1} \right)\,,\\
    \Sigma_2 &=-8 M _{\phi} \left(M _{\phi} +\omega _{\phi,\vec{k}_1} \right)^2\,, \\
    \Sigma_3 &=4 \left(M _{\phi} +\omega _{\phi,\vec{k}_1} \right)  \left(M^3_{\phi} -M_{\phi}M^2 _{\chi}+M_{\phi}^2\omega _{\phi,\vec{k}_1}+M_{\chi}^2\omega _{\phi,\vec{k}_1} \right)\,.
\end{align}
Since $\Sigma_1 >0$, the condition~\eqref{eq:JhA_condtion22} is equivalent to 
\begin{subnumcases}{}
\label{eq:JhA_omegak2_cond1}
-2M_{\chi}^2+M_{\phi}^2+M_{\phi}\omega_{\phi,\vec{k}_1} > 0\,,\\
\label{eq:JhA_omegak2_cond2}
\Omega^{-}_{2}\le \omega _{\chi,\vec{k}_2} \le \Omega^{+} _{2}\,,
\end{subnumcases}
where 
\begin{align}
    \Omega^{\pm} _{2} =\frac{M_{\phi}(M_{\phi}+\omega _{\phi,\vec{k}_1})\pm \sqrt{M_{\phi}(\omega _{\phi,\vec{k}_1}-M_{\phi})(-2M_{\chi}^2+M_{\phi}^2+M_{\phi}\omega_{\phi,\vec{k}_1}) }}{2M_{\phi} }\,.
\end{align}

To have $\Omega^{+}_{2}>M_{\chi}$, we require 
\begin{align}
\label{eq:Omega2+>Mh}
   M_{\phi}(M_{\phi}+\omega _{\phi,\vec{k}_1}-2M_\chi)+\sqrt{M_{\phi}(\omega _{\phi,\vec{k}_1}-M_{\phi})(-2M_{\chi}^2+M_{\phi}^2+M_{\phi}\omega_{\phi,\vec{k}_1})}> 0\,.
\end{align}
If
\begin{align}
\label{eq:Cond_Omega2+>Mh}
 \omega _{\phi,\vec{k}_1}> 2M_{\chi}-M_{\phi}\,,
\end{align}
The condition~\eqref{eq:Omega2+>Mh} is satisfied automatically. If $\omega _{\phi,\vec{k}_1}<2M_{\chi}-M_{\phi}$, then the condition~\eqref{eq:Omega2+>Mh} requires
\begin{align}
    M_{\phi}(\omega _{\phi,\vec{k}_1}-M_{\phi})(-2M_{\chi}^2+M_{\phi}^2+M_{\phi}\omega_{\phi,\vec{k}_1})> \left[M_{\phi}(M_{\phi}+\omega _{\phi,\vec{k}_1}-2M_\chi)\right]^2\,.
\end{align}
The above condition can be simplified to $(M_\chi - M_\phi)^2 M_\phi (M_\phi + \omega_{\phi,\veck_1}) < 0$ which cannot be fulfilled. Therefore, we have the requirement~\eqref{eq:Cond_Omega2+>Mh} to ensure $\Omega_2^+>M_h$. It also can be shown that
\begin{align}
    \Omega^{+} _{2}< M_{\phi}+\omega _{\phi,\vec{k}_1}-M_{\chi}\,,
\end{align}
such that the condition~\eqref{eq:JhA_condition1} is automatically satisfied when  Eq.~\eqref{eq:JhA_omegak2_cond2} is fulfilled. On the other hand, one can show that 
\begin{align}
    \Omega^-_2=\frac{M_\phi+\omega_{\phi,\veck_1}}{2}-\frac{1}{2}\sqrt{\left(\omega_{\phi,\veck_1}-\frac{M_\chi^2}{M_\phi}\right)^2-\left(\frac{M^2_\chi}{M_\phi}-M_\phi\right)^2}\geq \frac{1}{2}\left(M_\phi+\frac{M^2_\chi}{M_\phi}\right)\geq M_\chi\,.
\end{align}

\begin{figure}[H]
    \centering
    \includegraphics[scale=0.8]{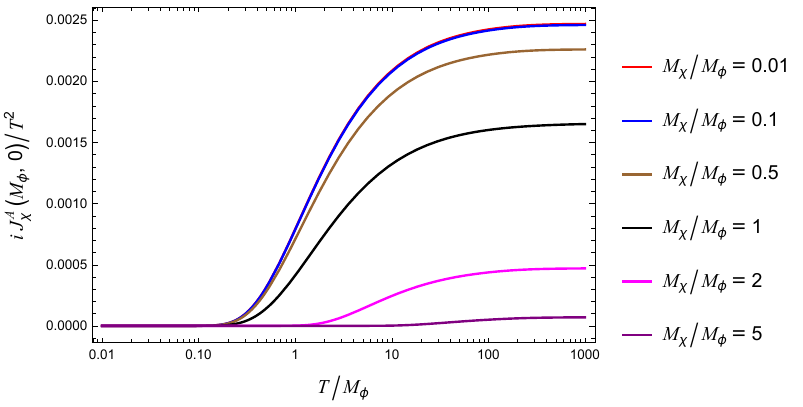}
    \caption{Numerical results for $J^A _\chi(M_{\phi},\vec{0})$. Note that the red and blue curves are overlapped, indicating that for $M_\chi\lsim 0.1 M_\phi$ the integral depends very weekly on $M_\chi/M_\phi$.}
    \label{fig:JhA_numericalsol1}
\end{figure}

In conclusion, the  integral over $\cos \theta $ in Eq.~\eqref{eq:appb chi_JA-cos theta} gives the following constraint 
\begin{subnumcases}{}
\omega _{\phi,\vec{k}_1} \geq \Omega^-_{1} \,,\\
    \Omega^{-}_{2}\leq \omega _{\chi,\vec{k}_2} \le \Omega ^{+}_{2}\,,
\end{subnumcases}
where 
\begin{align}
 \Omega^-_{1} = \max[M_{\phi}, (2M_{\chi}^2-M_{\phi}^2)/M_{\phi}, 2M_{\chi}-M_{\phi}]=\max[M_{\phi}, (2M_{\chi}^2-M_{\phi}^2)/M_{\phi}] \,.
\end{align}
Similarly, introducing spherical coordinates for $\veck_1$ and doing the integral over the angular coordinates gives
\begin{align}
    J^A _\chi(M_{\phi},\vec{0})=&\frac{\left(\e ^{\beta M_{\phi}}-1\right)}{\i\,32\pi^3} \int ^{\infty} _{\Omega^-_1}\d \omega _{\phi,\vec{k}_1}\int _{\Omega^{-} _{2}} ^{{\Omega^{+} _{2}}}\d \omega _{\chi,\vec{k}_2} \left[1+f_{\rm B}(\omega _{\phi,\vec{k}_1})\right] f_{\rm B}(\omega _{\chi,\vec{k}_2}) f_{\rm B}(M_{\phi}+\omega _{\phi,\vec{k}_1}-\omega _{\chi,\vec{k}_2})  \,.
\end{align}
Doing the integral over $\omega _{\vec{k}_2}$, we obtain
\begin{align}
\label{eq:JhA_final}
   J^A _\chi(M_{\phi},\vec{0})=& \frac{\e ^{\beta M_{\phi}}-1}{\i\,32 \pi ^3 \beta} \int ^{\infty} _{\Omega^-_1} \d \omega_{\phi,\vec{k}_1} [1+f_{\rm B}(\omega _{\phi,\vec{k}_1})]f_{\rm B}(M_{\phi}+\omega_{\phi,\vec{k}_1})\notag\\
   &\qquad\qquad\qquad\times\ln \left[ \frac{\left(\e ^{\beta \Omega^+ _2}-1\right)\left(\e^{\beta(M_{\phi}+\omega_{\phi,\vec{k}_1})}-\e^{\beta \Omega_{2}^-} \right)}{\left(\e^{\beta(M_{\phi}+\omega_{\phi,\vec{k}_1})}-\e^{\beta \Omega_{2}^+} \right)\left(\e ^{\beta \Omega^- _2}-1\right)}\right] \,. 
\end{align}
For $M_{\phi} =M_{\chi}$, we have $J^A_\chi(M_{\phi},\vec{0})=J^A_\phi(M_{\phi},\vec{0})$.
For general $M_{\chi}$, there is no a closed form of $J^A _\chi(M_{\phi},\vec{0})$. Numerical results of $J^A _\chi(M_{\phi},\vec{0})$ for different values of $M_{\chi}/M_{\phi}$ are shown in Fig.~\ref{fig:JhA_numericalsol1}.

\begin{figure}[ht]
    \centering
    \includegraphics[scale=0.9]{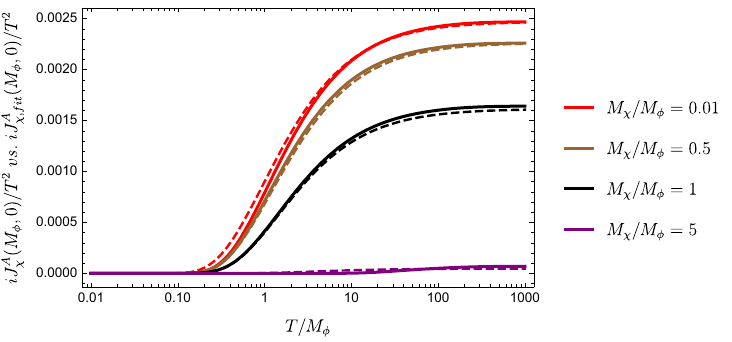}
    \caption{Comparison between the fit function $J_{\chi,\rm fit}^A(M_\phi,\vec{0})$ (dashed lines) and exact numerical results of $J_\chi^A(M_\phi,\vec{0})$ (solid lines).}
    \label{fig:JhA_fitting}
\end{figure}

In order to have an analytic expression for $J_\chi^{A}(M_\phi,\vec{0})$, we do the fitting using the known function $J^A_\phi(M_\phi,\vec{0})$,
\begin{align}
    J_\chi^A(M_\phi,\vec{0})\sim J_{\chi,\rm fit}^A(M_\phi,\vec{0})\equiv 1.49\times \frac{6}{\pi^2}\times {\rm Li}_2\left(\frac{4}{(M_\chi/M_\phi)^3+4}\right) J_\phi^A\left(\frac{M_\chi+2 M_\phi}{3},\vec{0}\right)\,.
\end{align}
The comparison between this fit function and the exact numerical results is shown in Fig.~\ref{fig:JhA_fitting}.

\subsubsection*{\underline{$ J^B _\chi(M_{\phi},\vec{0})$}}

Performing the integrals over $\vec{k}_3$ in Eq.~\eqref{eq:appb chi_JB}, we obtain
\begin{align}
    J^B _\chi(M_{\phi},\vec{0})=&-2\pi \i \left(\e ^{\beta M_{\phi}}-1\right) \int\frac{\d^3 \vec{k}_2}{(2\pi)^3}\int\frac{\d^3 \vec{k}_1}{(2\pi)^3}\frac{ \delta (M_{\phi}+\omega _{\chi,\vec{k}_2}-\omega _{\phi,\vec{k}_1}-\omega _{\chi,\veck_1+\veck_2})}{8\, \omega _{\phi,\vec{k}_1}\omega _{\chi,\vec{k}_2}\omega _{\chi,\veck_1+\veck_2}} \\ \notag
   & \times \left[1+f_{\rm B}(\omega _{\chi,\veck_2})\right] f_{\rm B}(\omega_{\phi,\vec{k}_1}) f_{\rm B}(\omega_{\chi,\veck_1+\veck_2}) \,.
\end{align}
Different from the earlier calculations, we now perform the integral over $\veck_1$ first. To do that, we introduce spherical coordinates for $\veck_1$ with $\vec{k}_2$ being the polar axis ($\theta =0$). Then, 
\begin{align}
\label{eq:JhB_costheta}
    J^B _\chi(M_{\phi},\vec{0})=&-2\pi \i \left(\e ^{\beta M_{\phi}}-1\right) \int\frac{\d^3 \vec{k}_2}{(2\pi)^3}\int _0 ^{\infty}\frac{\d |\vec{k}_1| \, |\vec{k}_1|^2}{(2\pi)^2} \int^{1} _{-1} \d \cos \theta\notag\\
   &\times \frac{\delta (M_{\phi}+\omega _{\chi,\vec{k}_2}-\omega_{\phi,\vec{k}_1}-\omega _{\chi,\veck_1+\veck_2})}{8\, \omega _{\phi,\vec{k}_1}\omega_{\chi,\vec{k}_2}\omega_{\chi,\veck_1+\veck_2}}  \left[1+f_{\rm B}(\omega _{\chi,\vec{k}_2})\right] f_{\rm B}(\omega _{\phi,\vec{k}_1}) f_{\rm B}(\omega_{\chi,\veck_1+\veck_2}) \,.
\end{align}
The $\delta$-function is satisfied when 
\begin{align}
\label{eq:JhB_delta_contraint}
   M_{\phi}+\omega_{\chi,\vec{k}_2}-\omega_{\phi,\vec{k}_1}- \sqrt{M^2_{\chi} + |\vec{k}_1|^2+|\vec{k}_2|^2+2|\vec{k}_1||\vec{k}_2|\cos \theta} =0\,.
\end{align}
Since $-1\leq\cos \theta\leq 1$, Eq.~\eqref{eq:JhB_delta_contraint} requires that
\begin{align}
    \sqrt{M^2 _{\chi} +\left(|\vec{k}_1|-|\vec{k}_2|\right)^2 } \le M_{\phi}+\omega_{\chi,\vec{k}_2}-\omega _{\phi,\vec{k}_1}  \le \sqrt{ M^2 _{\chi} +\left(|\vec{k}_1|+|\vec{k}_2|\right)^2 }\,,
\end{align}
which can be reduced to 
\begin{subnumcases}{}
\label{eq:JhB_omega1_cond1}
    \omega_{\phi,\vec{k}_1} \leq M_{\phi} + \omega_{\chi,\vec{k}_2} \,,\\
\label{eq:JhB_omega1_cond2}
   \Big|(M_{\phi}+\omega _{\chi,\vec{k}_2}-\omega_{\phi,\vec{k}_1})^2-(\omega _{\phi,\vec{k}_1}^2+\omega _{\chi,\vec{k}_2}^2-M_{\phi}^2) \Big|\leq 2 |\vec{k}_1||\vec{k}_2|\,.
\end{subnumcases}
Taking the square of Eq.~\eqref{eq:JhB_omega1_cond2} gives
\begin{align}
\label{eq:JhB_omega1_cond22}
    \Sigma_1  \omega _{\phi,\vec{k}_1}^2 + \Sigma_2  \omega _{\phi,\vec{k}_1} +\Sigma_3\le 0\,,
\end{align}
where 
\begin{align}
    \Sigma_1 &= 4 \left(M _{\phi}^2 +M _{\chi}^2+ 2 M _{\phi} \omega_{\chi,\vec{k}_2} \right)\,,\\
    \Sigma_2 &=-8 M _{\phi} \left(M _{\phi} +\omega_{\chi,\vec{k}_2} \right)^2\,, \\
    \Sigma_3 &=4M _{\phi}^2  \left(M^2_{\phi} -M_{\chi}^2 +2M_{\phi}\omega _{\chi,\vec{k}_2}+2\omega^2_{\chi,\vec{k}_2} \right)\,.
\end{align}

Since $\Sigma_1 >0$, condition Eq.~\eqref{eq:JhB_omega1_cond22} gives
\begin{align}
    \Omega^{-}_{1}\le \omega_{\phi,\vec{k}_1} \le \Omega^{+}_{1}\,,
\end{align}
where 
\begin{align}
     \Omega^{-} _{1}& =M_{\phi}\,,\\
     \Omega^{+} _{1}&= \frac{M_{\phi}(M_{\phi}^2-M_{\chi}^2+2M_{\phi} \omega _{\chi,\vec{k}_2}+ 2\omega _{\chi,\vec{k}_2}^2)}{ M_{\phi}^2+M_{\chi}^2+2M_{\phi}\omega _{\chi,\vec{k}_2}}\,.
\end{align}
It can be shown that 
\begin{align}
    \Omega^{+} _{1} \le M_{\phi} + \omega _{\chi,\vec{k}_2} -M_{\chi}\,,
\end{align}
so that Eq.~\eqref{eq:JhB_omega1_cond1} is automatically satisfied when $\omega_{\phi,\veck_1}\leq \Omega_1^+$.

In conclusion, the  integral over $\cos \theta $ in Eq.~\eqref{eq:JhB_costheta} gives the following constraint 
\begin{align}
    &\omega _{\chi,\vec{k}_2} \ge M_{\chi} \,,\qquad \Omega ^{-} _{1}\le \omega _{\phi,\vec{k}_1} \le \Omega ^{+}_{1}\,.
\end{align}
Then, we have  
\begin{align}
    J^B _\chi(M_{\phi},\vec{0})=&\frac{\left(\e ^{\beta M_{\phi}}-1\right)}{\i\, 32\pi^3} \int ^{\infty} _{M_{\chi}}\d \omega_{\chi,\vec{k}_2}\int _{\Omega^{-}_{1}} ^{{\Omega^{+}_{1}}}\d \omega _{\phi,\vec{k}_1} \left[1+f_{\rm B}(\omega _{\chi,\vec{k}_2})\right] f_{\rm B}(\omega _{\phi,\vec{k}_1}) f_{\rm B}(M_{\phi}+\omega _{\chi,\vec{k}_2}-\omega_{\phi,\vec{k}_1})  \,.
\end{align}
Doing the integral over $\omega_{\phi,\veck_1}$, one arrives at
\begin{align}
\label{eq:JhB_final}
   J^B _\chi(M_{\phi},\vec{0})=& \frac{\e ^{\beta M_{\phi}}-1}{\i\,32 \pi ^3 \beta} \int ^{\infty} _{M_{\chi}} \d \omega_{\chi,\vec{k}_2} [1+f_{\rm B}(\omega _{\chi,\vec{k}_2})]f_{\rm B}(M_{\phi}+\omega _{\chi,\vec{k}_2})\notag\\
   &\qquad\times \ln \left[ \frac{\left(\e ^{\beta \Omega^+_1}-1\right)\left(\e^{\beta(M_{\phi}+\omega_{\chi,\vec{k}_2})}-\e^{\beta \Omega_{1}^-} \right)}{\left(\e^{\beta(M_{\phi}+\omega_{\chi,\vec{k}_2})}-\e^{\beta \Omega_{1}^+} \right)\left(\e ^{\beta \Omega^- _1}-1\right)}\right] \,. 
\end{align}

\begin{figure}[H]
    \centering
    \includegraphics[scale=0.75]{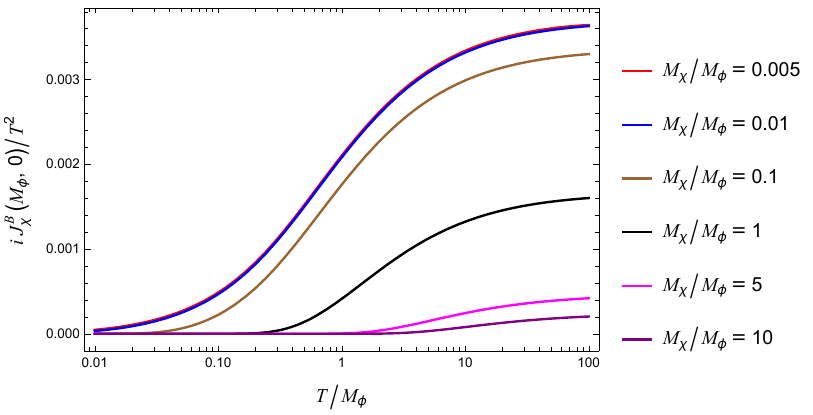}
    \caption{Numerical results of $J^B _\chi(M_{\phi},\vec{0})$. Note that the red and blue curves are almost overlapped, indicating that for $M_\chi\lsim 0.01 M_\phi$ the integral depends very weekly on $M_\chi/M_\phi$.}
    \label{fig:JhB_numericalsol1}
\end{figure}

For $M_{\phi} =M_{\chi}$, we have 
$J^B _\chi(M_{\phi},\vec{0})=J^A _\phi(M_{\phi},\vec{0})$. For general $M_{\chi}$, the expressions for $\Omega^{\pm}_1$ are too complicated to evaluate a closed form of $J^B _\chi(M_{\phi},\vec{0})$. Numerical results of $J^B _\chi(M_{\phi},\vec{0})$ for different values of $M_{\chi}/M_{\phi}$ are shown in Fig.~\ref{fig:JhB_numericalsol1}. 

\begin{figure}[ht]
    \centering
    \includegraphics[scale=0.85]{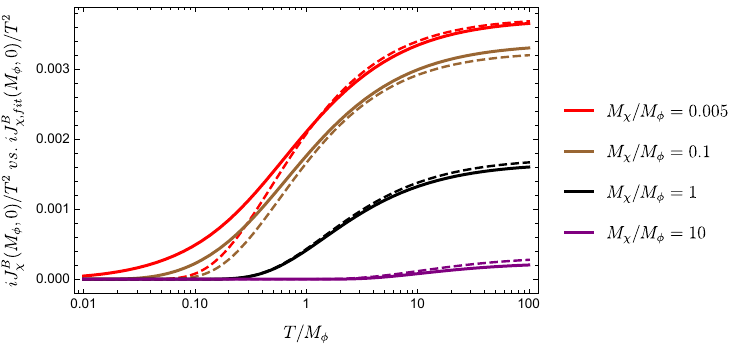}
    \caption{Comparison between the fit function $J_{\chi,\rm fit}^B(M_\phi,\vec{0})$ (dashed lines) and exact numerical results of $J_\chi^B(M_\phi,\vec{0})$ (solid lines). }
    \label{fig:JhB_fitting}
\end{figure}

We use the following function to approximate $J_h^B(M_\phi,\vec{0})$,
\begin{align}
    J_\chi^B(M_\phi,\vec{0})\sim J_{\chi,\rm fit}^B(M_\phi,\vec{0}) \equiv 2.28\times \frac{6}{\pi^2}\times {\rm Li}_2\left(\frac{1.6}{(M_\chi/M_\phi)+1.6}\right) J^A_\phi\left(\frac{2M_\chi+M_\phi}{3},\vec{0}\right)\,.
\end{align}
The comparison between this fit function and the exact numerical results is shown in Fig.~\ref{fig:JhB_fitting}.

\section{Solving the condensate equation of motion with a time-dependent mass term}
\label{app:sol_eom}
The condensate EoM~\eqref{eq:condensate eom} has been solved in Refs.~\cite{Ai:2021gtg,Wang:2022mvv} for constant $M_\phi$ using multiple-scale analysis~\cite{Bender,Holmes}. In this section, we solve it when $M_\phi$ is time-dependent but adiabatic, $\dot{M}_\phi/M_\phi^2\ll 1$.

As a standard procedure for perturbation theory, one puts a {\it bookkeeping} parameter $\varepsilon$ in front of all perturbatively small terms in the EoM. When doing the perturbative expansion, the higher the power of $\varepsilon$ a quantity is associated with, the smaller the quantity is. Therefore, we have
\begin{align}
  \ddot{\varphi}(t) +  M_\phi^2  \varphi(t) + 3 \varepsilon H\dot{\varphi}(t)
    + \varepsilon \frac{\lambda_\phi \varphi^3(t) }{6}
	+ \varepsilon \int_{t_i}^{t}
	    \d t' \, \pi_{\rm R}(t-t') \varphi(t')+ \varepsilon \frac{\varphi(t) }{6}
	\int_{t_i}^{t} \d t' \,
	    v_{\rm R}(t-t') \varphi^2(t') = 0 \, .
\label{eq:condensate eom4}  
\end{align}
There is a hierarchy in the time scales in the evolution of the condensate. The shorter time scale corresponds to the oscillation frequency, and the longer time scale corresponds to the damping rate. One can understand this by imaging a form of the condensate evolution $\varphi(t)=A(t)B(t)$ where $B(t)$ is an oscillating function and $A(t)$ is the envelope function. One can expect that 
\begin{align}
\label{eq:hierarchy-time-evolution}
   M_\phi \sim \frac{1}{B(t)}\frac{\d B(t)}{\d t} \gg  \frac{1}{A(t)}\frac{\d A(t)}{\d t}\,. 
\end{align}
Therefore, when taking time derivatives of $\varphi$, one can obtain quantities at different orders in magnitude. In order to trace these quantities, we thus formally introduce two different time variables. In the above example, one can write it as $\varphi(t,\tau)=A(\tau)B(t)$ with $\tau=\varepsilon t$ where $\varepsilon$, as in Eq.~\eqref{eq:condensate eom4}, is only a bookkeeping parameter. Then we have 
\begin{align}
    \frac{\d\varphi}{\d t}=\frac{\d\tau}{\d t}\frac{\partial\varphi(t,\tau)}{\partial\tau}+\frac{\partial\varphi(t,\tau)}{\partial t}=\varepsilon \frac{\d A(\tau)}{\d\tau} B(t)+A(\tau)\frac{\d B(t)}{\d t}\,.
\end{align}
This way, the smaller term is now associated with a factor $\varepsilon$. These two-time variables are used when solving the above equation for constant $M_\phi$~\cite{Ai:2021gtg,Wang:2022mvv}.

Here we consider a slowly evolving $M_\phi(\tau)$ where $\tau$ is still the slow time variable. It turns out that in this case the fast time variable now is~\cite{Holmes}
\begin{align}
\eta=f(t)=\int^t\d t' M_{\phi}(\varepsilon t')\,.
\end{align}
We shall see shortly the advantage of introducing the $\eta$ time variable. We will also assume $H=H(\tau)$.

Now let $\varphi$ take the following form
\begin{equation}
    \varphi(t) = 
    \varphi(t,\tau;\varepsilon)=\varphi_0 (t,\tau)+ \varepsilon \varphi_1 (t,\tau) + \varepsilon ^2 \varphi_2 (t,\tau)+...\,,
    \label{eq:varphi in powers of varepsilon}
\end{equation}
where 
\begin{equation}
    \varphi_n (t,\tau) =\frac{1}{n!}\frac{\partial ^n \varphi (t,\tau;\varepsilon)}{\partial \varepsilon ^n}\bigg|_{\varepsilon=0}
\end{equation}
are the coefficients. 
Below, we will still use the time variable $t$ inside the integrals which then should be understood as $t=f^{-1}(\eta)$. Now one can directly expand the first four terms of Eq.~\eqref{eq:condensate eom4} in powers of $\varepsilon$. 
To expand the non-local terms of Eq.~\eqref{eq:condensate eom4} in powers of $\varepsilon$, we need some simplification. 
Assuming that the solution for the condensate $\varphi$ varies only slightly in the slow time $\tau$ during the window provided by the kernel of the non-local terms, we can Taylor-expand  $\varphi (t',\tau';\varepsilon)$  at $\tau$ as
\begin{align}\label{eq:appassumption}
    \varphi (t',\tau';\varepsilon) 
    =\varphi (t',\tau;\varepsilon) +\varepsilon(t'-t)\frac{\partial \varphi (t',\tau;\varepsilon) }{\partial \tau} + \frac{\varepsilon^2}{2}(t'-t)^2 \frac{\partial^2 \varphi (t',\tau;\varepsilon) }{\partial \tau^2} +...\,.
\end{align} 
Indeed, from Eqs.~\eqref{piR+vR exp} and~\eqref{eq:propagators}, one can see that the scale of
the window is $1/M_\phi$ while $(\d A(\tau)/\d \tau)\times (1/M_\phi)\ll A(\tau)$ (cf. Eq.~\eqref{eq:hierarchy-time-evolution}).

Then, up to the second order in $\varepsilon$, the non-local terms could be written as
\begin{align}
    \varepsilon \int_{t_i}^{t}& \d t' \, \pi_{\rm R}(t-t') \varphi(t',\tau ';\varepsilon) \notag \\
    &=\varepsilon \int_{t_i}^{t} \d t' \, \pi_{\rm R}(t-t') \varphi(t',\tau ;\varepsilon)-\varepsilon ^2 \int_{t_i}^{t} \d t' \,(t-t') \pi_{\rm R}(t-t') \frac{\partial \varphi(t',\tau ;\varepsilon)}{\partial \tau}+\O(\varepsilon^3)\,,
\end{align}
and
\begin{align}
\label{eq:non-local terms in powers of varepsilon}
    &\varepsilon \frac{\varphi(\eta,\tau;\varepsilon) }{6} \int_{t_i}^{t} \d t' \,
	    v_{\rm R}(t-t')\varphi ^2(t',\tau ';\varepsilon) =\varepsilon \frac{\varphi(\eta,\tau;\varepsilon) }{6} \int_{t_i}^{t} \d t' \, v_{\rm R}(t-t')\varphi ^2(t',\tau ;\varepsilon) \notag\\
    &\qquad\qquad\qquad\qquad- \varepsilon^2 \frac{\varphi(\eta,\tau;\varepsilon) }{3} \int_{t_i}^{t} \d t' \, (t-t') v_{\rm R}(t-t')\varphi (t',\tau ;\varepsilon)\frac{\partial \varphi(t',\tau ;\varepsilon)}{\partial \tau}+\O(\varepsilon^3)\,.
\end{align}

Now, we are able to organise Eq.~\eqref{eq:condensate eom4} in powers of $\varepsilon$. At the leading order $\O(\varepsilon^0)$, we have 
\begin{equation}
\label{eq:eom varphi_0}
    \frac{\partial ^2 \varphi_0(\eta,\tau) }{\partial \eta^2} +   \varphi_0(t,\tau)=0\,,
\end{equation}
which has the solution
\begin{align}
    \varphi_0(\tau,t) = {\rm Re}\left[R(\tau) \e^{-\i\eta}\right]\,.
\end{align}
Here, we see that using the time variable $\eta$, the leading-order equation takes the simple form of an EoM of a harmonic oscillator.
$R(\tau)$ can only be determined when one looks into the equation at the order $\O(\varepsilon)$, which reads
\begin{align}
& M_\phi^2\left[\frac{\partial ^2 \varphi_1(\eta,\tau) }{\partial \eta^2} +  \varphi_1(\eta,\tau) \right]=-2 M_\phi(\tau) \frac{\partial^2 \varphi_0(\eta,\tau) }{\partial \eta \partial \tau}
-\frac{\d M_{\phi} (\tau)}{\d \tau}\frac{\partial \varphi_0(\eta,\tau) }{\partial \eta}
-3 H(\tau) M_\phi(\tau) \frac{\partial \varphi_0(\eta,\tau) }{\partial \eta} \notag\\
&- \frac{\lambda_\phi }{6} \varphi_0 ^3(\eta,\tau)-\int_{t_i}^{t} \d t' \, \pi_{\rm R}(t-t') \varphi_0(t',\tau)-\frac{\varphi_0(\eta,\tau) }{6} \int_{t_i}^{t} \d t' \, v_{\rm R}(t-t')\varphi_0 ^2(t',\tau)\,.
\label{eq:eom varphi_1_0}
\end{align}
The non-local terms in Eq.~\eqref{eq:eom varphi_1_0} can be written as
\begin{align}
\label{eq:non-local-fourier_11}
  -\int_{t_i}^{t} \d t' \, \pi_{\rm R}(t-t') \varphi_0(t',\tau) =&-\text{Re}\left[R(\tau) \e^{-\i \eta  } \int_{t_i}^{t} \d t' \, \pi_{\rm R}(t-t') \e^{\i \int ^{t} _{t'} d \tilde{t}\, M_\phi (\varepsilon \tilde{t}) }\right]\,,
\end{align}
and
\begin{align}
\label{eq:non-local-fourier_12}
&-\frac{\varphi_0(\eta,\tau) }{6} \int_{t_i}^{t} \d t' \, v_{\rm R}(t-t')\varphi_0 ^2(t',\tau) =-\frac{1}{12}\text{Re}\left[\e^{-\i \eta }[R(\tau)]^2 R^*(\tau)  \int_{t_i}^{t} \d t' \, v_{\rm R}(t-t') \right]\notag \\
&\qquad\qquad\qquad\qquad\qquad-\frac{1}{24}\text{Re}\left[\e^{-\i \eta} [R(\tau)]^2 R^*(\tau)  \int_{t_i}^{t} \d t' \, v_{\rm R}(t-t') \e^{2\i \int ^{t}_{t'} \d \tilde{t}\, M_\phi (\varepsilon \tilde{t})  }\right]\notag \\
&\qquad\qquad\qquad\qquad\qquad-\frac{1}{24}\text{Re}\left[\e^{-3\i \eta} [R(\tau)]^3  \int_{t_i}^{t} \d t' \, v_{\rm R}(t-t')\e^{2\i \int ^{t}_{t'} \d \tilde{t}\,M_\phi (\varepsilon \tilde{t})  }\right]\,.
\end{align}
Taking into account that the kernels of the non-local terms only extend over several oscillation periods, one can make further simplifications. First, one can neglect any early-time transient effects due to initial conditions in the non-local terms by taking $t_i \rightarrow -\infty$. Second, the following approximation could be made (also for the corresponding term with $v_{\rm R}$)
\begin{align}
    \int_{t_i}^{t} \d t' \, \pi_{\rm R}(t-t') \e^{\i \int ^{t} _{t'} \d \tilde{t} \, M_\phi (\varepsilon \tilde{t}) } &\approx \int_{t_i}^{t} \d t' \, \pi_{\rm R}(t-t') \e^{\i M_\phi (\varepsilon t) (t-t') }\,,
\end{align}
provided that $M(\tau)$ varies slightly within the extension of the kernels.  Note that we could also replace $t$ by $+\infty$ in the upper limit of the integrals of $t'$  because of the Heaviside step function in the definitions of the retarded self-energy and the retarded four-vertex function (cf. Eq.~\eqref{piR+vR def}).
Therefore, we could actually recognize the integrals on the RHS of Eqs.~\eqref{eq:non-local-fourier_11} and~\eqref{eq:non-local-fourier_12} as Fourier transforms 
\begin{subequations}
\label{eq:FourierTransf}
\begin{align}
\int_{t_i}^{t} \d t' \, \pi_{\rm R}(t-t') \e^{\i \int^{t} _{t'} \d \tilde{t}\, M_\phi (\varepsilon \tilde{t})} &\approx \int_{-\infty}^{+\infty} \d t' \, \pi_{\rm R}(t-t') \e^{\i M_\phi (\tau)  (t-t')}=\widetilde{\pi}_{\rm R} (M_\phi (\tau) )\,,\\
\int_{t_i}^{t} \d t' \, v_{\rm R}(t-t') &\approx \int_{-\infty}^{+\infty} \d t' \, v_{\rm R}(t-t') \e^{\i 0(t-t')}=\widetilde{v}_{\rm R} (0)\,,\\
\int_{t_i}^{t} \d t' \, v_{\rm R}(t-t') \e^{2\i \int ^{t} _{t'} \d \tilde{t}\, M_\phi (\varepsilon \tilde{t})  }&\approx \int_{-\infty}^{+\infty} \d t' \, v_{\rm R}(t-t') \e^{2\i M_\phi (\tau)  (t-t')}=\widetilde{v}_{\rm R} (2M_\phi(\tau)  )\,.
\end{align}
\end{subequations}

Using the definitions given in Eq.~\eqref{eq:important_quantities},
Eq.~\eqref{eq:eom varphi_1_0} can be written as 
\begin{align}
\frac{\partial ^2 \varphi_1(\eta,\tau) }{\partial \eta^2} +  \varphi_1(\eta,\tau) 
=&\frac{2}{M^2_{\phi}}\text{Re}\Bigg\{ \i M_\phi\e^{-\i \eta} \bigg[ \frac{\d R(\tau)}{\d \tau}+\Big(\frac{3}{2}H(\tau)+\frac{1}{2M_\phi}\frac{\d M_{\phi} (\tau)}{\d \tau}+ \i \mu +\gamma \Big)R(\tau)\notag\\
&+\Big(\frac{\i \lambda_\phi}{16M_\phi}+\i \alpha +\frac{\sigma}{2}\Big) [R(\tau)]^2 R^*(\tau) \bigg]\Bigg\}\notag\\
&+\frac{2}{M^2_{\phi}}\text{Re}\Bigg\{ \i M_\phi \e^{-3\i \eta}\bigg(\frac{\i \lambda_{\phi}}{48 M_\phi}+ \i \alpha_2 +\frac{\sigma}{2}\bigg)[R(\tau)]^3\Bigg\}\,,
\label{eq:eom varphi_1_2}
\end{align}
where we used  
\begin{equation}
    \text{Im}[\widetilde{v}_{\rm R}(0)]=0\,,
\end{equation}
which is due to that the imaginary part of the Fourier-transformed proper four-vertex is anti-symmetric. 

The EoM governing $\varphi_1(t,\tau)$ describes a harmonic oscillator that is subject to two external oscillatory forces. The first two lines on the RHS of Eq.~\eqref{eq:eom varphi_1_2} describe a force with a frequency of $M_\phi$, which aligns with the natural frequency of $\varphi_1(t,\tau)$. The second force is represented by the last line and possesses a frequency of $3 M_\phi$. This second force alters the initial oscillatory behaviour by adding different oscillations having constant amplitudes. It is known that the first force can induce a resonant behaviour, resulting in the final amplitude of $\varphi_1(t,\tau)$ being proportional to time $t$, causing an unbounded increase in amplitude. To circumvent the emergence of non-physical, spurious resonances in the solution, we require
\begin{equation}
\label{eq:eom R_1}
    \frac{\d R(\tau)}{\d \tau}+\Big(\frac{3}{2}H(\tau)+\frac{1}{2M_\phi}\frac{\d M_{\phi} (\tau)}{\d \tau}+ \i \mu +\gamma \Big)R(\tau) +\Big(\frac{\i \lambda_\phi}{16M_\phi}+\i \alpha +\frac{\sigma}{2}\Big) [R(\tau)]^2 R^*(\tau)=0\,.
\end{equation}
Making the following Ansatz 
\begin{equation}\label{eq:app_ansatz}
    R(\tau) = A (\tau) \,\e ^{-\i f(\tau)}\,,
\end{equation}
where $A (\tau)$ and $f(\tau)$ are real functions of $\tau$, we can split the complex equation~\eqref{eq:eom R_1} into the following two real equations,
\begin{subequations}
\label{eq:eom_A and f(tau)_1}
\begin{align}
\label{eq:eom_A(tau)_1}
    \frac{\d A (\tau)}{\d \tau}+\left(\gamma +\frac{3}{2}H(\tau)+\frac{1}{2M_\phi}\frac{\d M_{\phi} (\tau)}{\d \tau}\right) A(\tau) +\frac{\sigma}{2} [A(\tau)]^3&=0\,,\\
    \frac{\d f(\tau)}{\d \tau}-\mu -\left(\frac{\lambda_{\phi}}{16M_{\phi}(t)}+\alpha \right)[A(\tau)]^2&=0\,.
\end{align}
\end{subequations}
Arriving here, one can take $\tau=t$.

\end{appendix}


\bibliographystyle{utphys}
\bibliography{ref}{}

\end{document}